\documentclass[11pt,letterpaper]{article}
\pdfoutput=1
\usepackage{jheppub}
\usepackage{color}
\usepackage{graphicx}
\usepackage{tabularx}
\usepackage{xspace}

\usepackage{verbatim}
\usepackage{amsmath}
\usepackage{amssymb}
\usepackage[caption=false]{subfig}
\usepackage{url}
\usepackage{bbold}
\usepackage{slashed}
\usepackage{array}

\usepackage{multirow}
\usepackage{threeparttable}
\usepackage{paralist}

\newcommand{\ev}{\text{event}}

\newcommand{\cut}{\text{cut}}

\newcommand{\Ecut}{E_{{\rm cut}}}

\newcommand{\ptc}{p_{T{\rm cut}}}

\newcommand{\pythia}{\textsc{Pythia~8}\xspace}
\newcommand{\herwig}{\textsc{Herwig++}\xspace}
\newcommand{\eventtwo}{\textsc{Event2}\xspace}

\newcommand{\FastJet}{\textsc{FastJet}\xspace}

\newcommand{\df}{\text{d}}

\DeclareRobustCommand{\Sec}[1]{Sec.~\ref{#1}}
\DeclareRobustCommand{\Secs}[2]{Secs.~\ref{#1} and \ref{#2}}

\DeclareRobustCommand{\App}[1]{App.~\ref{#1}}
\DeclareRobustCommand{\Tab}[1]{Table~\ref{#1}}
\DeclareRobustCommand{\Tabs}[2]{Tables~\ref{#1} and \ref{#2}}
\DeclareRobustCommand{\Fig}[1]{Fig.~\ref{#1}}

\DeclareRobustCommand{\Eq}[1]{Eq.~(\ref{#1})}
\DeclareRobustCommand{\Eqs}[2]{Eqs.~(\ref{#1}) and (\ref{#2})}
\DeclareRobustCommand{\Eqss}[3]{Eqs.~(\ref{#1}), (\ref{#2}), and (\ref{#3})}
\DeclareRobustCommand{\Ref}[1]{Ref.~\cite{#1}}
\DeclareRobustCommand{\Refs}[1]{Refs.~\cite{#1}}

\newcommand{\be}{\begin{equation}}
\newcommand{\ee}{\end{equation}}
\newcommand{\nn}{\nonumber}

\usepackage{xspace}

\newcommand{\e}{\epsilon}

\newcommand{\bq}{\bar{q}}
\newcommand{\zcut}{z_{\rm cut}}

\newcommand{\cL}{\mathcal{L}}
\newcommand{\cR}{\mathcal{R}}
\newcommand{\cF}{\mathcal{F}}
\newcommand{\cI}{\mathcal{I}}
\newcommand{\cK}{\mathcal{K}}
\newcommand{\beq}{\begin{eqnarray}}
\newcommand{\eeq}{\end{eqnarray}}

\newcommand{\ord}[1]{\mathcal{O}\!\left(#1\right)}

\newcommand{\Njet}{\widetilde{N}_{\rm jet}}
\newcommand{\dN}[1]{\Delta_{#1}}
\newcommand{\dNpm}{\Delta_{2\pm}}
\newcommand{\dNp}{\Delta_{2+}}
\newcommand{\dNm}{\Delta_{2-}}
\newcommand{\dNtm}{\Delta_{3-}}

\newcommand{\cT}{\mathcal{T}}
\newcommand{\as}{\alpha_s}
\renewcommand{\angle}{\theta}

\begin{document}

\title{The First Calculation of Fractional Jets}
\author[a,b]{Daniele Bertolini,}
\author[c]{Jesse Thaler,}
\author[a,b]{and Jonathan R. Walsh}
\affiliation[a]{Ernest Orlando Lawrence Berkeley National Laboratory, University of California, \\ Berkeley, CA 94720, U.S.A.}
\affiliation[b]{Center for Theoretical Physics, University of California, \\ Berkeley, CA 94720, U.S.A.}
\affiliation[c]{Center for Theoretical Physics, Massachusetts Institute of Technology, \\ Cambridge, MA 02139, U.S.A.}

\emailAdd{dbertolini@lbl.gov}
\emailAdd{jthaler@mit.edu}
\emailAdd{jwalsh@lbl.gov}


\abstract{In collider physics, jet algorithms are a ubiquitous tool for clustering particles into discrete jet objects.  Event shapes offer an alternative way to characterize jets, and one can define a jet multiplicity event shape, which can take on fractional values, using the framework of ``jets without jets''.  In this paper, we perform the first analytic studies of fractional jet multiplicity $\Njet$ in the context of $e^+e^-$ collisions.  We use fixed-order QCD  to understand the $\Njet$ cross section at order $\as^2$, and we introduce a candidate factorization theorem to capture certain higher-order effects.  The resulting distributions have a hybrid jet algorithm/event shape behavior which agrees with parton shower Monte Carlo generators.   The $\Njet$ observable does not satisfy ordinary soft-collinear factorization, and the $\Njet$ cross section exhibits a number of unique features, including the absence of collinear logarithms and the presence of soft logarithms that are purely non-global.  Additionally, we find novel divergences connected to the energy sharing between emissions, which are reminiscent of rapidity divergences encountered in other applications.  Given these interesting properties of fractional jet multiplicity, we advocate for future measurements and calculations of $\Njet$ at hadron colliders like the LHC.}


\preprint{MIT-CTP 4632}

\maketitle

\section{Introduction}
\label{introduction}

For almost forty years, we have known that high energy particle collisions can produce jets \cite{Bjorken:1969wi,Hanson:1975fe}.  The term ``jets'' has two related but different meanings:  ``jet formation'' is the physical process by which quarks and gluons shower and fragment into collimated sprays of hadrons \cite{Field:1977fa}, while ``jet algorithms'' are an analysis technique used to cluster those hadrons into proxies for the underlying quarks and gluons \cite{Sterman:1977wj}.  Jet algorithms are a powerful way to categorize and organize collision events \cite{Ellis:2007ib,Salam:2009jx}, but event shapes (and jet shapes) offer a more sensitive probe of jet formation itself \cite{Dasgupta:2003iq}.  Indeed, though the observation of three-jet structure in $e^+e^-$ collisions strongly hinted at the existence of gluons \cite{Ellis:1976uc,Wiik:1979cq}, an unambiguous discovery at PETRA \cite{Brandelik:1979bd,Barber:1979yr,Berger:1979cj,Bartel:1979ut} required the use of event shapes like thrust \cite{Farhi:1977sg} and oblateness \cite{Barber:1979yr}.

Recently, the distinction between jet algorithms and event shapes was blurred through the ``jets without jets'' framework,  in which standard jet-based observables are mapped into jet-like event shapes~\cite{Bertolini:2013iqa}.  These observables incorporate a transverse momentum threshold $\ptc$ and a jet radius $R$ just like traditional jet algorithms, but they behave more like event shapes because they involve an inclusive sum over particles in an event and do not uniquely assign hadrons to jet objects.  A particularly interesting jets-without-jets observable is jet multiplicity,
\be
\label{eq:Njetintro}
\Njet(\ptc,R) = \sum_{i\in\ev}\frac{p_{T i}}{p_{T i,R}}\, \Theta(p_{T i,R}-\ptc),
\ee
where $p_{T i,R}$ is the transverse momentum contained in a cone of radius $R$ around particle $i$.  By design, this observable is highly correlated with the standard jet count (for the same $\ptc$ and $R$ values).  Crucially, $\Njet$ can yield fractional values, offering a new probe of jet formation.

In this paper, we initiate the analytic study of fractional jet multiplicity.  For simplicity, we treat the case of $e^+ e^- \to \text{jets}$, though we briefly mention how to adapt our calculational techniques to collisions at the Large Hadron Collider (LHC).  With two or three partons in the final state ($e^+ e^- \to q \overline{q}$ or $q \bar{q} g$), $\Njet$ always yields an integer value.   Fractional jets only start appearing with four or more partons (e.g.\ $e^+ e^- \to q \overline{q} g g$), so our analytic calculations will start at $\ord{\as^2}$.  This is a general feature of fractional jets: non-integer values only appear with three or more particles in a given jet region.\footnote{For $e^+e^-$ collisions in the rest frame, momentum conservation prevents all particles from being in the same jet region.  Hence four partons are required for non-integer values.}  Our main technical results will be obtained using fixed-order perturbative QCD, though we will also discuss connections to factorization properties in soft-collinear effective theory (SCET) \cite{Bauer:2000ew,Bauer:2000yr,Bauer:2001ct,Bauer:2001yt,Bauer:2002nz}.

We will focus on the behavior of $\Njet$ in the vicinity of dijet configurations,
\be
\dNp = \Njet - 2, \qquad \dNm = 2-\Njet,
\ee
though we do present some results for $\dNtm = 3-\Njet$ as well.  This near-integer behavior exhibits a number of curious analytic features, which are already visible at $\ord{\as^2}$.
\begin{itemize}
\item \textit{Hybrid jet algorithm/event shape behavior.}  Jet algorithms have a finite cross section at (integer) jet multiplicities whereas event shapes typically have form factors that suppress the cross section at singular values.  The $\Njet$ distribution has both types of behavior.   Even though the Born-level process $e^+ e^- \to q \overline{q}$ gives the integer value $\Njet = 2$, one might naively expect the corresponding spike at $\dN{2\pm} = 0$ to be completely smeared out by multiple emissions that generate finite values of $\dN{2\pm}$.   Instead, the spike at $\dN{2\pm} = 0$ is robust, as there is a finite region of the many-body phase space that still gives rise to integer values of $\Njet$.  At the same time, the appearance of logarithms in $\dN{2\pm}$ at every perturbative order generates a shoulder at finite values of $\dN{2\pm}$, which is suppressed as $\dN{2\pm} \to 0$.  Thus, the cross section at exact-integer values of $\Njet$ behave like a jet algorithm while near-integer values behave like an event shape (see \Sec{bumpspike}). 

\item \textit{Cancellation of single- and double-soft divergences.}   The first non-trivial contributions to non-integer behavior of $\Njet$ arise from configurations where three partons are within a mutual radius of $2R$ but not within $R$ (see \Sec{lowmult}).  For $\dNp$, this three-parton phase space has singularities when one or two of the partons goes soft.  These divergences arise because the observable receives parametrically equivalent contributions from the single- and double-soft regions, which are not regulated in dimensional regularization.  Interestingly, these divergences are structurally similar to rapidity divergences, and we will introduce the analog of rapidity regularization~\cite{Chiu:2011qc,Chiu:2012ir} to see that the single-soft and double-soft divergences do indeed cancel (see \Sec{rapidity}).  We note that soft emissions contribute in a non-linear way, and therefore $\Njet$ is a non-additive observable.

\item \textit{No collinear logarithms.}  Typical event shapes have singularities in both the soft and collinear limit, giving rise to both soft and collinear logarithms.  The resulting double-logarithmic structure appears as Sudakov form factors in the cross section.  By contrast, collinear emissions do not generate logarithms of $\Njet$, and only soft logarithms appear in the $\Njet$ distribution (see \Sec{lowmult}).\footnote{Mass-dropped observables~\cite{Butterworth:2008iy,Dasgupta:2013ihk,Larkoski:2014wba} have the reverse behavior of only having collinear logarithms.}  Thus, the suppression in the $\dN{2\pm} \to 0$ limit is only single-logarithmic.

\item \textit{Non-global yet local structure.}  The $\dNpm$ cross section does not satisfy ordinary soft-collinear factorization, because the coefficients of the soft logarithms depend on the collinear structure of the jets (despite the absence of collinear logarithms).  Furthermore, the soft logarithms in $\Njet$ are purely non-global~\cite{Dasgupta:2001sh,Dasgupta:2002bw}, in the sense that they arise from soft emissions in a restricted region of phase space (see \Sec{NGLs}).\footnote{As explained in \Sec{NGLs}, our use of the term ``non-global'' can refer to both non-Abelian and Abelian correlations.}  These features would seem to preclude any standard factorization theorem, especially given the non-additive nature of $\Njet$.  In order to change the value of $\Njet$, however, soft emissions must lie within $\lesssim 2R$ from the hard core of the jet.  Thus, color coherence ensures that the dependence on $\dNpm$, albeit non-global, is local to each jet region.  We will present a candidate factorization theorem that exploits this  local structure (see \Sec{Njet}).

\end{itemize}
In addition to analytic studies, we will test $\Njet$ using high-statistics Monte Carlo samples from \pythia~\cite{Sjostrand:2007gs} and \herwig~\cite{Bahr:2008pv}.  Within theoretical uncertainties, the Monte Carlo results confirm our analytic understanding.

Given its many features and potential applications, $\Njet$ would be very interesting to measure at the LHC.  As mentioned above, $\Njet$ is a purely non-global observable, with the near-integer behavior determined only by soft and not by collinear divergences.  To our knowledge, it is the only jet or event shape observable with this behavior.  As such, it is a unique probe of soft physics, and measurements of $\Njet$ can be used to test color coherence, underlying event models, and pileup mitigation strategies.  Furthermore, $\Njet$ is useful basis to compare parton shower predictions for jet substructure, and we present an initial comparison in this work.   For new physics searches involving high-multiplicity final states, fractional $\Njet$ values interpolate between different jet multiplicities, obviating the need for exclusive jet bins.  This interpolation also makes for an interesting version of the classic ``staircase'' plots for vector boson plus $N$ jet production~\cite{Ellis:1985vn,Berends:1989cf,Berends:1990ax,Aaltonen:2007ae,Abazov:2009pp,Aad:2013ysa,Khachatryan:2014uva,Khachatryan:2014zya,Aad:2014qxa}.  Finally, for the growing field of matrix element/parton shower matching/merging \cite{Bengtsson:1986hr,Catani:2001cc,Lonnblad:2001iq,Mrenna:2003if,Hoche:2006ph,Alwall:2007fs,Giele:2007di,Frixione:2002ik,Nason:2004rx,Lavesson:2008ah,Hamilton:2010wh,Alioli:2013hqa}, $\Njet$ has a continuous distribution unlike standard jet algorithms and a huge dynamic range compared to standard event shapes, so $\Njet$ can be used to test whether matching/merging procedures achieve a smooth interpolation, even in the soft regime. We note, of course, that theoretical calculations of $\Njet$ for hadronic collisions must contend with additional effects such as the underlying event and pileup contamination.  Moreover, it will be non-trivial to include non-perturbative effects, power-suppressed terms, and higher-order perturbative effects such as the resummation of non-global logarithms \cite{Dasgupta:2001sh,Dasgupta:2002bw,Banfi:2002hw,Appleby:2003ai,Dasgupta:2007wa,Dasgupta:2014yra}. Although a detailed study of $\Njet$ for the LHC is beyond the scope of this paper, we briefly discuss some of these issues and potential solutions in \Sec{sec:LHC}.

The rest of this paper is organized as follows.  In \Sec{kinematics}, we review the basic physics behind $\Njet$ and explain the kinematic regimes that give rise to fractional jets.  In \Sec{rapidity}, we discuss the structure of rapidity-like divergences and how they appear in the $\Njet$ calculation.  In \Sec{nearint}, we perform fixed-order calculations of $\dN{2\pm}$ at $\ord{\as^2}$, using both the full $e^+ e^- \to 4$ parton matrix element as well as a $1 \to 3$ splitting function approximation.  We then present a candidate factorization theorem for $\dN{2\pm}$ in \Sec{Njet}, which includes a renormalization-group-independent ``collinear function''. In \Sec{sec:LHC}, we briefly discuss how to extend our results to the LHC.  We compare our analytic calculations to \pythia and \herwig in \Sec{MC}, and we conclude in \Sec{conclusions}.  The appendices contain further calculational results and details.

\section{Aspects of Fractional Jets}
\label{kinematics}

Since we will be looking at electron-positron collisions, it is more natural to work with a variant of $\Njet$ based on energies and angles (instead of transverse momenta and azimuth-rapidity distances):
\be
\label{eq:NjetSec2}
\Njet(\Ecut,R) = \sum_{i\in\ev}\frac{E_i}{E_{i,R}}\, \Theta(E_{i,R}-\Ecut),
\ee
where $E_i$ is the energy of particle $i$,
\be
E_{i,R} = \sum_{j \in \ev} E_j \, \Theta(R - \theta_{ij}),
\ee
and $\theta_{ij}$ is the opening angle between particles $i$ and $j$.

For particles whose angular separation is always larger than $R$, $E_i / E_{i,R}$ reduces to $1$, and $\Njet$ simply counts the number of hard particles above the energy threshold $\Ecut$.  Because $\Njet$ is an infrared/collinear (IRC) safe observable, $\Njet$ always takes on integer values in the extreme soft and collinear limit.  We will use the notation
\be
\label{eq:dNpm}
\dN{n\pm} = \pm (\Njet - n) 
\ee
to indicate the near-integer behavior, with $\dN{n+}$ ($\dN{n-}$) parameterizing the $\Njet$ behavior just above (below) $n$.  Our calculations will focus on $\dN{2\pm}$ and $\dN{3-}$ in $e^+e^-$ collisions.

\subsection{Hybrid Jet Algorithm/Event Shape Behavior}
\label{bumpspike}

\begin{figure}
\begin{center}
\subfloat[]{
\label{fig:MChighstat:a}
\includegraphics[width=0.48\textwidth]{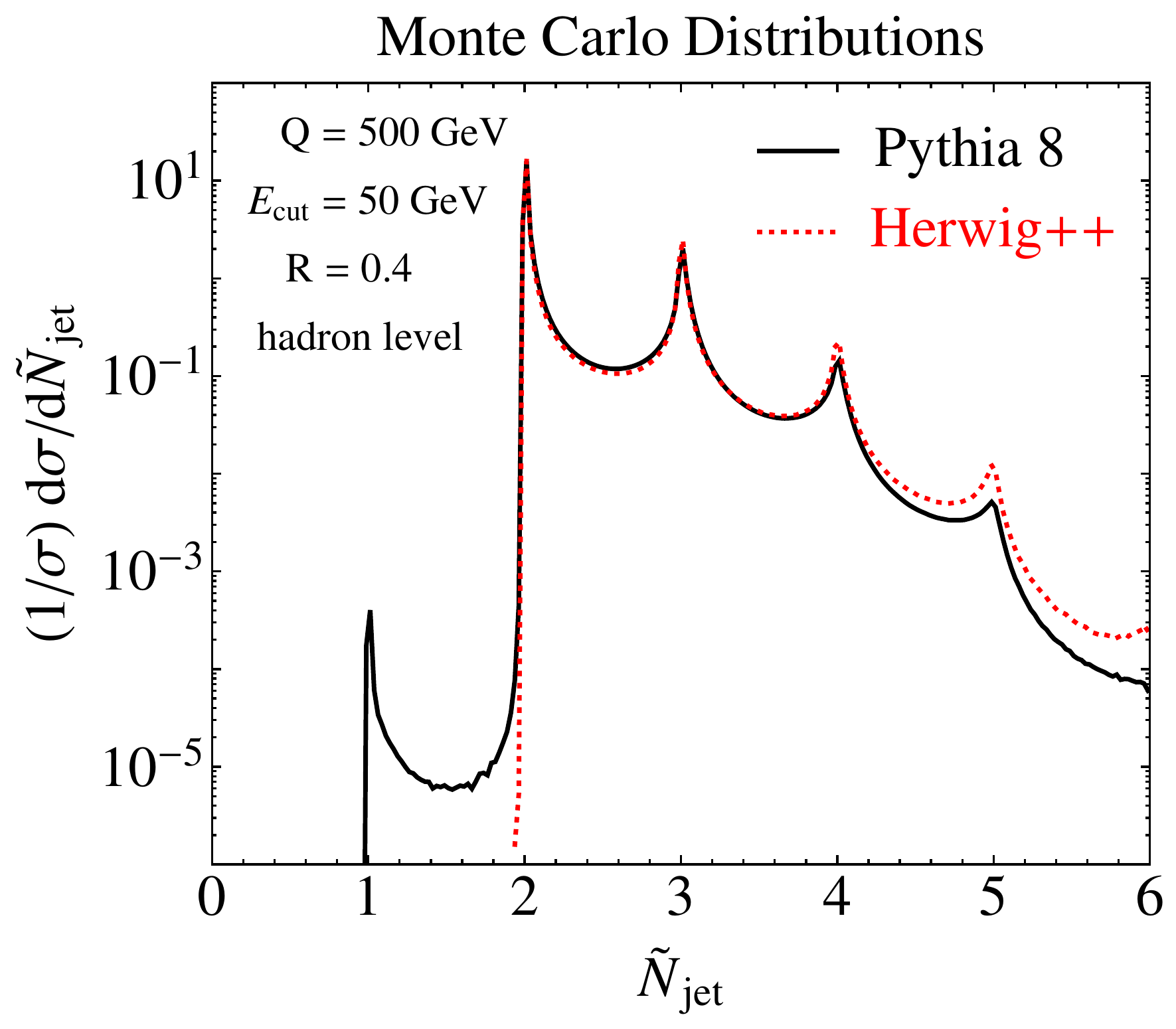}
}
\subfloat[]{
\label{fig:MChighstat:b}
\includegraphics[width=0.49\textwidth]{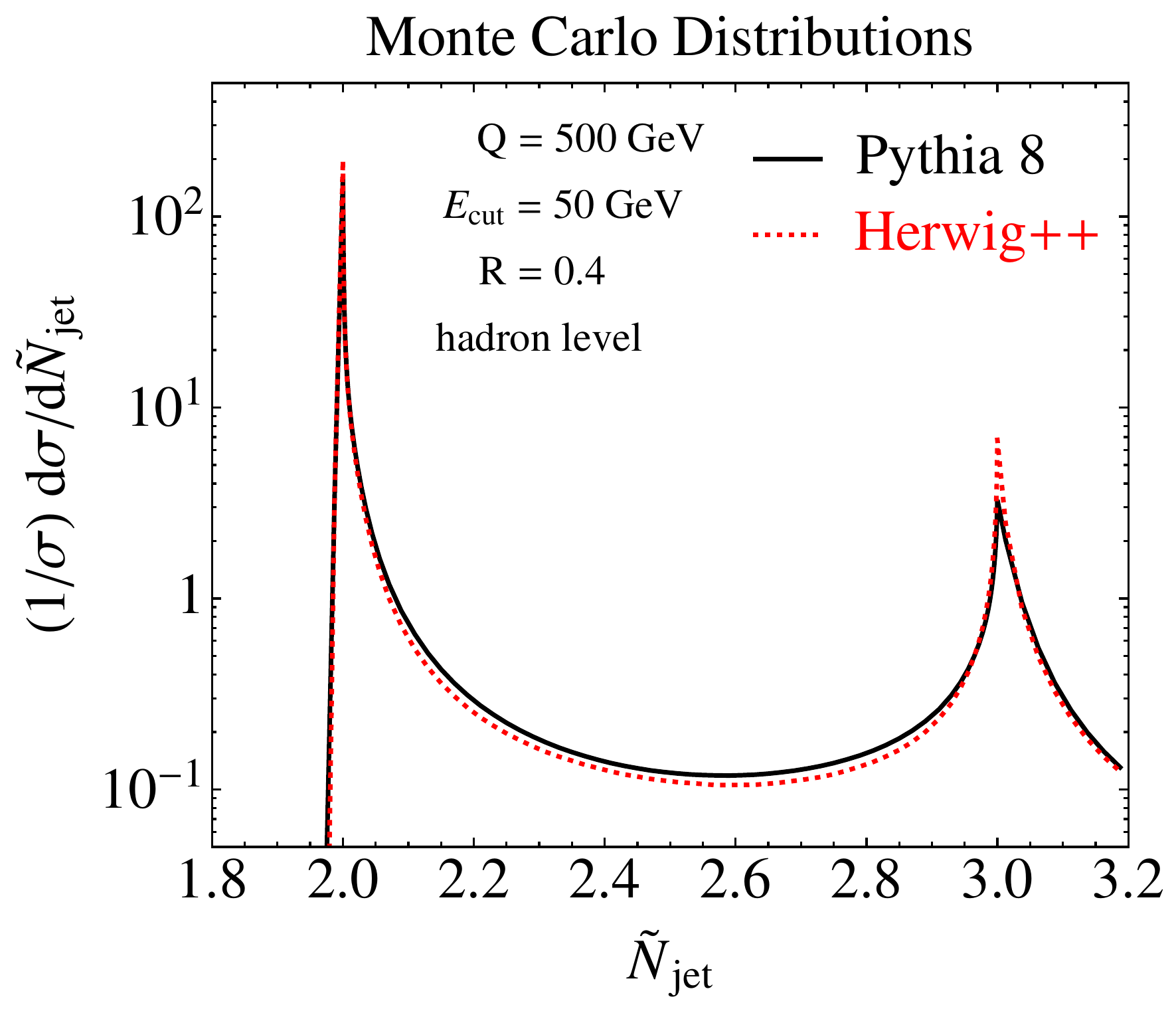}
}
\caption{The $\Njet$ distribution in $e^+e^-$ collisions from \pythia and \herwig, over a wide range of multiplicities (left plot) and for $2 \lesssim \Njet \lesssim 3$ (right plot).  The distribution peaks at integer values, and near-integer values are enhanced relative to intermediate values between integers.  It is this enhancement that we will understand quantitatively through our calculations.}
\label{fig:MChighstat}
\end{center}
\end{figure}

\begin{figure}
\begin{center}
\includegraphics[width=\textwidth]{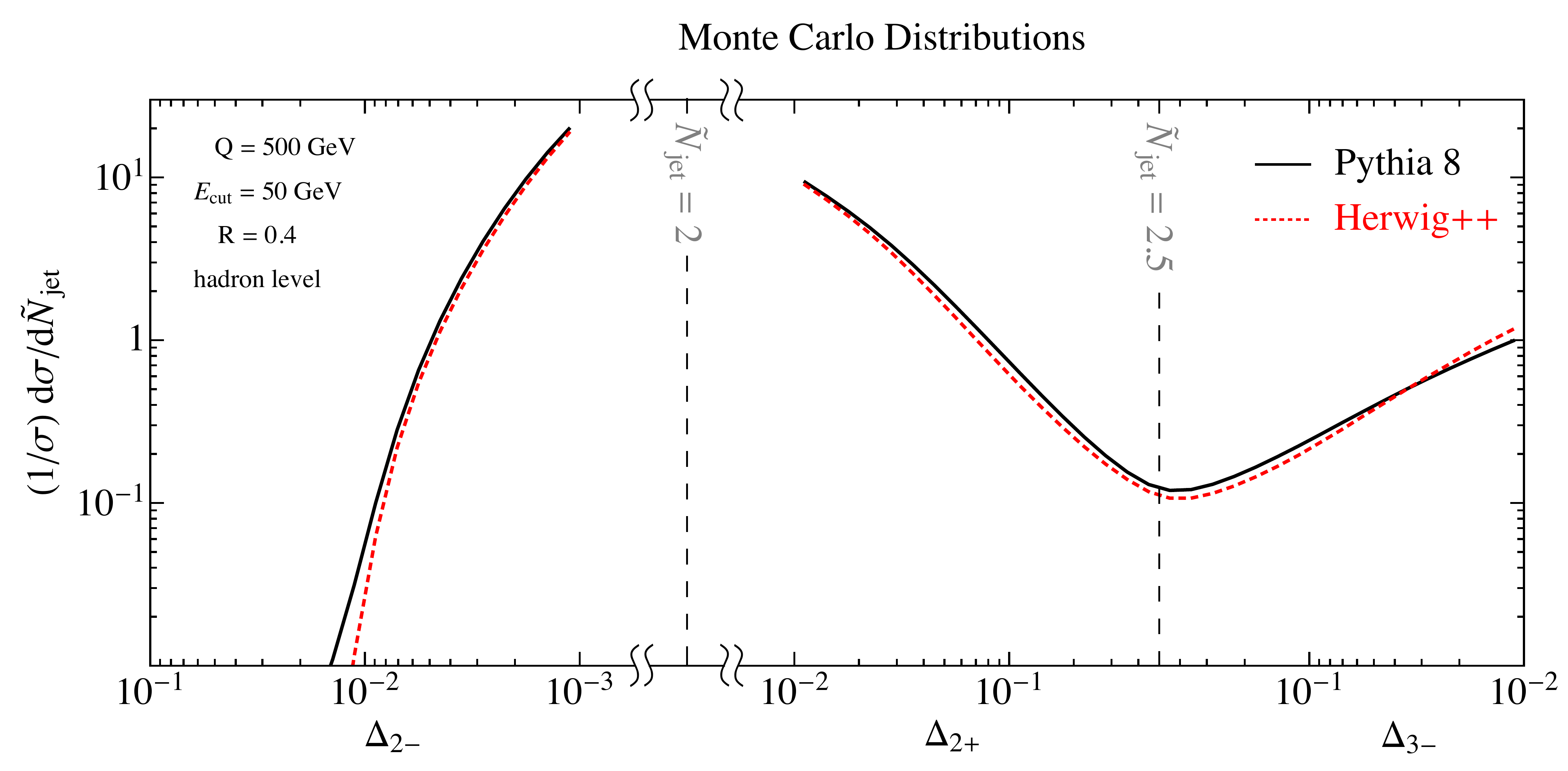}
\caption{The same $\Njet$ distributions as \Fig{fig:MChighstat}, now plotted in ``triptych'' form to show the near-integer behavior in $\dNm$, $\dNp$, and $\dNtm$.  Note that the $\dNm$ and $\dNtm$ axes run backwards.  The $\dNm$ and $\dNp$ distributions interface at $\Njet = 2$ (where $\dNpm = 0$), indicated by the left-hand dashed vertical line.  Since we do not plot $\dNpm$ all the way to 0, we put a gap over the region around $\Njet = 2$.  The $\dNp$ and $\dNtm$ distributions are connected at $\Njet = 2.5$ (where $\dNp = \dNtm = 0.5$), indicated by the right-hand dashed vertical line.}
\label{fig:MCtriptych}
\end{center}
\end{figure}

The $\Njet$ distribution is peaked at integer values, with substantial support in the near-integer regime.  These different parts of the $\Njet$ distribution can be qualitatively studied with Monte Carlo generators.\footnote{The original $\Njet$ variable in \Eq{eq:Njetintro} is available through the \textsc{JetsWithoutJets} package, an add-on to \FastJet~3 \cite{Cacciari:2011ma} contained in the \FastJet~\textsc{contrib} library (\url{http://fastjet.hepforge.org/contrib/}).  The variant in \Eq{eq:NjetSec2} is available from the authors upon request.}  We generate events for $e^+e^- \to$ hadrons at a center-of-mass energy of $Q = 500$ GeV in \pythia.183 \cite{Sjostrand:2007gs} and \herwig 2.7.1 \cite{Bahr:2008pv}, including showering and hadronization.  In \Fig{fig:MChighstat:a}, we plot the distribution of $\Njet$ across a wide range of values, showing the circus-tent-like peak and fall-off behavior of the cross section.\footnote{The peak at $\Njet = 1$ in \pythia seems to arise from events with $\tau$ leptons produced in hadron decays, where the resulting neutrinos carry away a substantial fraction of the jet momentum. The same feature is not visible in \herwig, nor is it visible when hadronization effects are turned off in \pythia.}  In \Fig{fig:MChighstat:b}, we focus on the distribution in the range $2 \lesssim \Njet \lesssim 3$, which is the region we aim to quantitatively describe in this paper.  In \Fig{fig:MCtriptych}, we plot the $\Njet$ distribution in a ``triptych'' form that shows in more detail the near-integer behavior in $\dNm$, $\dNp$, and $\dNtm$, in particular the logarithmic enhancement as $\dN{n\pm} \to 0$.

\begin{figure}
\begin{center}
\subfloat[]{
\label{fig:MChighstatFocus:a}
\includegraphics[width=0.49\textwidth]{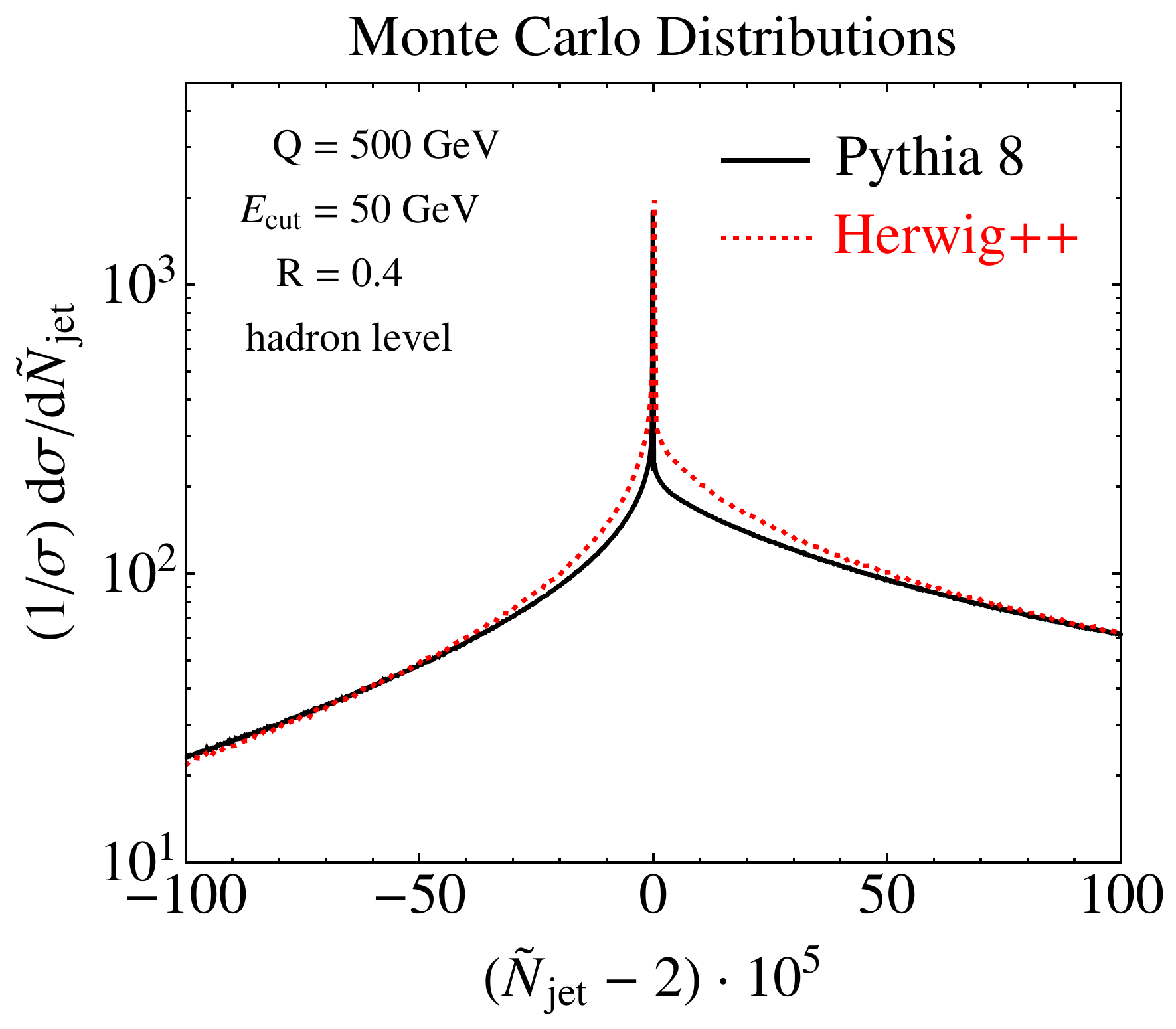}
}
\subfloat[]{
\label{fig:MChighstatFocus:b}
\includegraphics[width=0.48\textwidth]{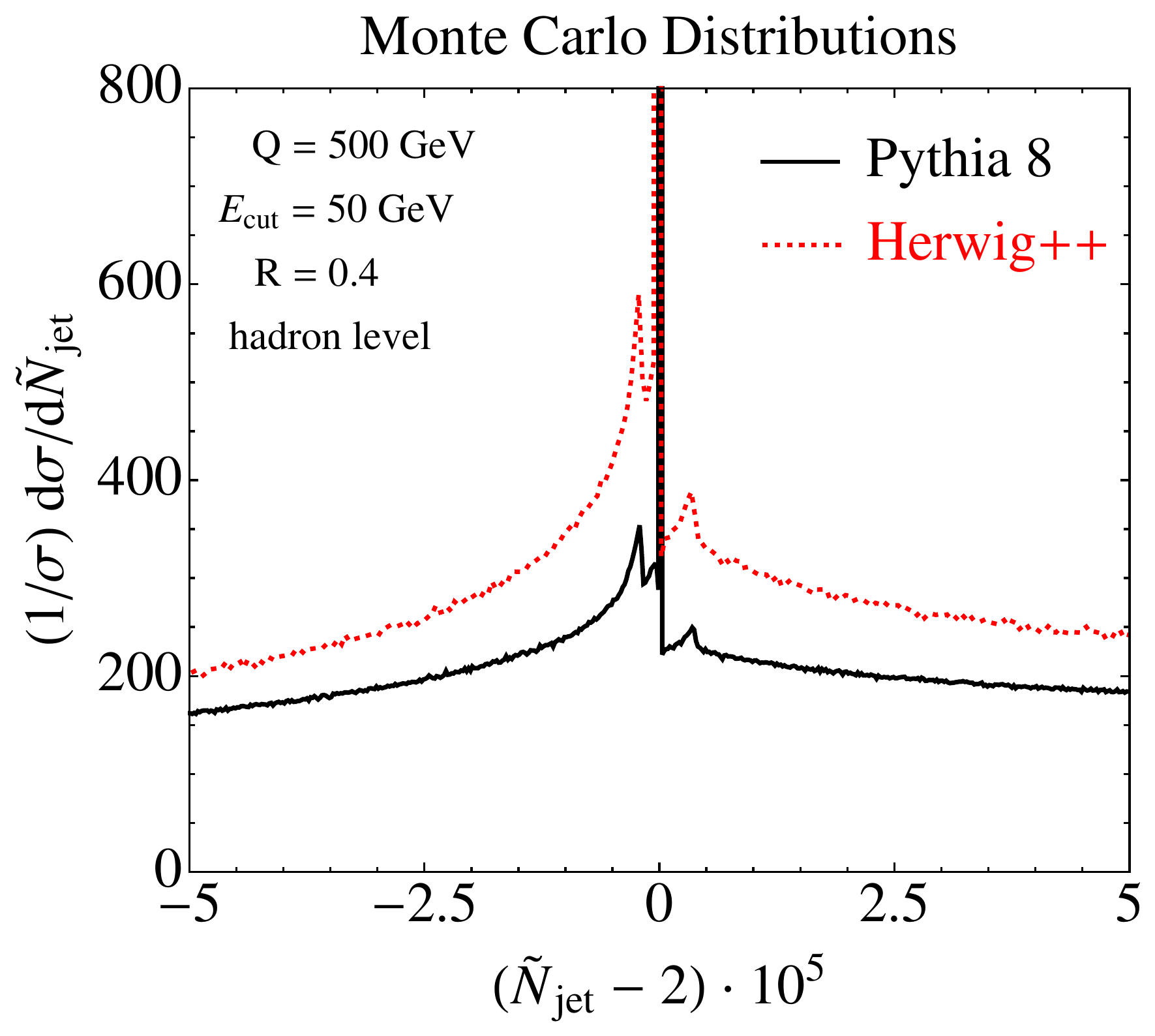}
}
\caption{The same $\Njet$ distributions as \Fig{fig:MChighstat}, now plotted in the very near vicinity of $\Njet = 2$.  Note the scale on the $x$-axis and the fact that the $y$-axis is logarithmic on the left plot but linear on the right one.  At $\Njet = 2$, there is a spike in the distribution from the jet algorithm behavior of the fractional jet multiplicity.  At non-integer values, the continuous distribution is more similar to event shapes.  We show the near-integer values in two different ranges, and for the closer range (right plot) one can see bumps in the Monte Carlo distributions from multiple emission and hadronization effects.}
\label{fig:MChighstatFocus}
\end{center}
\end{figure}

As mentioned in the introduction, the cross section at exact integer values has a different behavior than at near-integer values, a feature related to the hybrid jet algorithm/event shape nature of the observable.  Like a jet algorithm, the cross section at integer values $\Njet = n$ has a non-zero value.  Like an event shape, the non-integer cross section is suppressed by all-orders emissions as $\Njet \to n$.  The behavior can be seen in \Fig{fig:MChighstatFocus}, where we plot the distribution in the very near vicinity of $\Njet = 2$.  

The reason why integer values $\Njet = n$ have a finite cross section is that they receive contributions from regions of phase space with non-zero measure.  This can be seen easily at $\ord{\as}$ for $e^+e^- \to q \overline{q} g$, where the entire cross section lies at $\Njet = 2$ or $3$.  More generally, any collection of particles within a mutual radius of $R$ will give an integer contribution to $\Njet$.\footnote{Regions within mutual radius $R$ are bounded by curves of constant width $R$.  A circle of radius $R/2$ is the simplest example, but there are more general examples like the Reuleaux triangle or the Canadian loonie.}  This means that the differential distribution at integer values has the form
\be
\label{eq:deltafunctioninteger}
\frac{\df\sigma}{\df \Njet} (\Njet = n) \propto \delta(n) \,.
\ee
In contrast, the distribution for near-integer values $\dN{n\pm} > 0$ is dominated by logarithms of $\dN{n\pm}$.  We will later show that the logarithms of $\dN{n\pm}$ arise from soft emissions, and the most important terms scale single logarithmically as $(\as \ln \dN{n\pm})^k$.  These logarithms combine at all orders to suppress the cross section as $\dN{n\pm} \to 0$, leading to the disjoint behavior at integer $\Njet$.  Note that single-logarithmic suppression is not as strong as the more familiar double-logarithmic suppression, so there are no Sudakov peaks in the $\dN{n\pm}$ distributions. 

\subsection{Soft and Collinear Limits of Fractional Jets}
\label{lowmult}

To understand the leading near-integer behavior of $\Njet$, we can focus on the soft and collinear limits.  As discussed further in \Sec{Njet}, the all-orders structure of the cross section does not satisfy standard soft-collinear factorization, but a soft-collinear analysis is still fruitful at fixed order.  The essential physics appears already in the three-parton phase space, where the near-integer behavior is dominated by soft physics.  We will focus on a single jet region here; to describe $e^+ e^- \to \text{dijets}$, we sum over the contributions to $\Njet$ from both jet regions (see \Eq{factform}).

\begin{figure}
\begin{center}
\includegraphics[width=\textwidth]{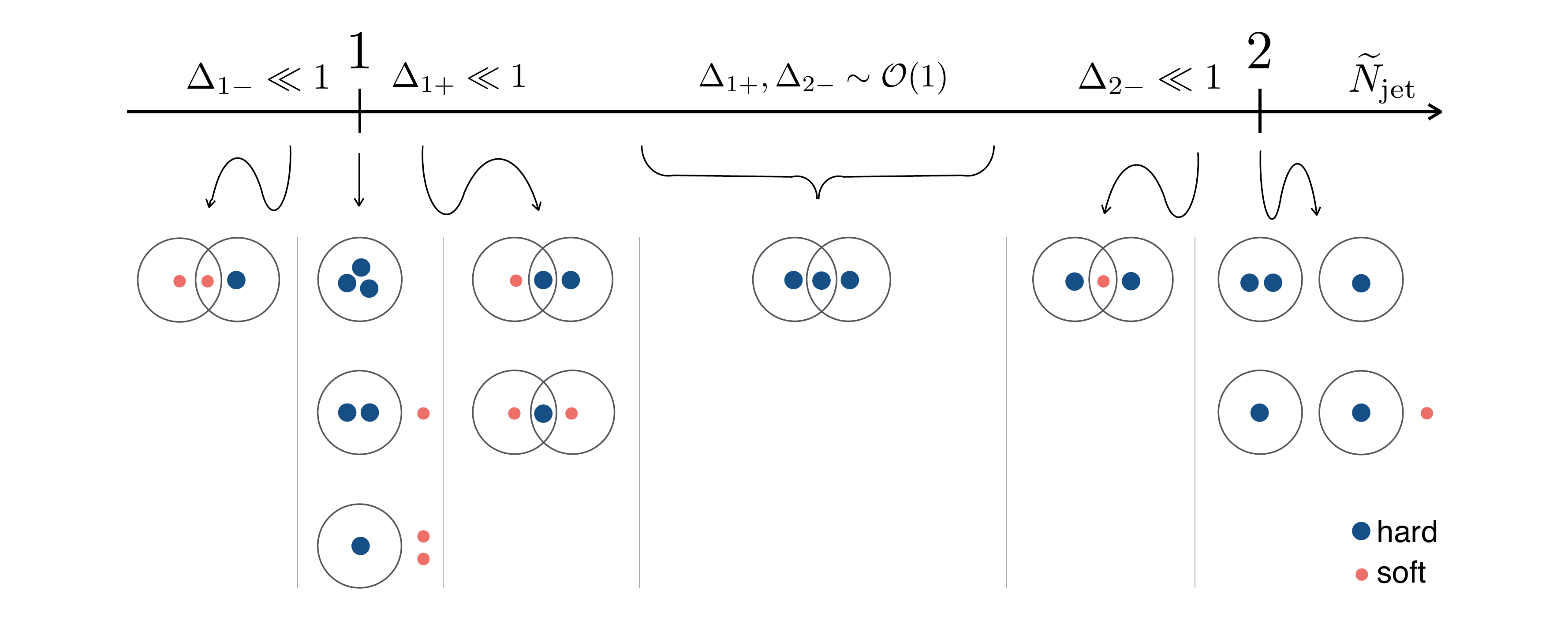}
\end{center}
\caption{Schematic representation of the different phase space configurations with three merged partons.  The numbered line shows the corresponding range of $\Njet$.  Here, we are considering just a single jet region; an $e^+ e^- \to \text{dijet}$ event has two such jet regions.  Circles represent cones of radius $R$, large blue dots represent energetic partons, and small red dots represent soft partons with $z<z_\text{cut}$. The near-integer regions $\Delta_{1-}$, $\Delta_{1+}$, and $\Delta_{2-}$ get contributions from single- and/or double-soft emissions.}
\label{fig:phasespace}
\end{figure}

We call a group of particles \emph{merged} if their contribution to $\Njet$ is connected, such that removing a subset of particles affects the contribution to $\Njet$ from other particles.  For a single isolated particle, its contribution to $\Njet$ is 1.  For a merged pair of particles with separation less than $R$, one still obtains an integer value of $\Njet$:
\be
\Njet = \frac{E_1}{E_1 + E_2} + \frac{E_2}{E_1 + E_2} = 1.
\ee
Because a single soft/collinear emission does not contribute to the value of $\Njet$, it is not linear in soft and collinear momenta in the singular region of phase space, and hence it is a non-additive observable.

Now consider three merged partons.  As shown in \Fig{fig:phasespace}, there are various different phase space configurations that lead to different values of $\Njet$.  To achieve non-integer values, one needs a phase space configuration with
\be
\theta_{12} > R \,, \qquad \theta_{13}, \theta_{23} < R,
\ee
or permutations of the parton labels. If all $E_i > \Ecut$, then the contribution to the jet multiplicity is
\be
\Njet = \frac{E_1}{E_1 + E_3} + \frac{E_2}{E_2 + E_3} + \frac{E_3}{E_1 + E_2 + E_3} \in (1,2) \,.
\ee
In terms of energy fractions
\be \label{zi}
z_i \equiv \frac{E_i}{E_1 + E_2 + E_3} \,, \qquad z_1 + z_2 + z_3 = 1 \,, \qquad \zcut \equiv \frac{\Ecut}{E_1 + E_2 + E_3} \,,
\ee
which removes the overall energy scale of the jet as a degree of freedom, the observable takes the value
\be
\Njet = \frac{z_1}{z_1 + z_3} + \frac{z_2}{z_2 + z_3} + z_3 \,.
\ee

For this merged triplet, near-integer behavior is obtained when one or two of the partons goes soft, as shown in \Fig{fig:phasespace}.  In these soft limits, the observable may depend on a single soft momentum or a product of soft momenta.  For example, if particle 1 is soft,
\be \label{Njet1soft}
\Njet \xrightarrow[\; z_1 \, \ll \, 1 \;]{} \; 1 + z_1 \frac{z_2 (2-z_2)}{1-z_2} + \ord{z_1^2} \,.
\ee
Or, if particles 1 and 2 are both soft,
\be \label{Njet12soft}
\Njet \xrightarrow[\; z_{1,2} \, \ll \, 1 \;]{} \; 1 + 2 z_1 z_2 + \ord{z_{1,2}^3} \,.
\ee
The first case is typical of event shape observables, as the near-integer behavior is linear in the soft particle's energy.  The second case, however, is unusual---it goes as the \emph{product} of soft momenta, another demonstration of the non-additive nature of $\Njet$.  This feature creates a novel structure in the perturbative series, reminiscent of rapidity divergences in SCET$_{\rm II}$~\cite{Chiu:2009yx,Chiu:2011qc}.  We discuss this further in \Sec{rapidity}, using a toy observable $\Delta = z_1 z_2$ that exhibits the same analytic features.

With more emissions, near-integer behavior requires the soft limit of one or more particles.\footnote{We are neglecting special configurations of energetic particles  where the value of $\Njet$ happens to be near-integer, since those regions of phase space are power suppressed.}  As mentioned above, an arbitrary collection of energetic particles will yield an integer $\Njet$ if the particles can be grouped into merged clusters where each particle in the cluster is within $R$ of all other particles in the cluster (such that each cluster has $\Njet = 1$).  Non-integer values are obtained when this is not satisfied (as in the merged triplet example above), though generically the resulting values are far from integers.  Near-integer values are obtained  by starting from an integer $\Njet$ configuration and then adding any number of soft particles which are not within the mutual radius $R$ of the cluster.  The contribution of these soft particles to $\Njet$ will be suppressed by
\be
\frac{E_{\rm soft}}{E_{{\rm soft},R}} \,,
\ee
and the deviation of $\Njet$ from integer values is similarly suppressed by this ratio as long as $E_{{\rm soft},R} > E_\cut$.  In this way, the near-integer behavior is determined by soft emissions in the vicinity of hard clusters, and soft emissions will generate logarithms of $\dN{n\pm}$. 

By contrast, collinear splittings cannot generate near-integer behavior of $\Njet$ and hence do not generate logarithms of $\dN{n\pm}$.  For collinear splittings of angle $R_c$, the only effect on $\Njet$ comes from particles who are within $R$ of either of the two daughters or the parent, but not within $R$ of all three.  For small $R_c$, this is a power-suppressed region of phase space and not logarithmically enhanced.  Said another way, the effect on $\Njet$ from a small-angle splitting is not smooth, as it either preserves the value of $\Njet$ or discretely changes it by including or excluding particles from the various $E_{i,R}$ terms; this behavior cannot generate logarithms of $\dN{n\pm}$.  So unlike standard jet shapes (like jet mass) which depend on both the softness and collinearity of a splitting, the near-integer $\Njet$ value depends only on the softness of emissions. 

\subsection{Non-Global yet Local Structure}
\label{NGLs}

We have argued that soft emissions contribute to near-integer values of $\Njet$, but this is only the case if the soft emissions lie within a restricted phase space of size $\lesssim 2R$ around the jet.  Wide-angle soft radiation does not contribute to $\Njet$, since those emissions yield $E_{i,R} < \Ecut$.\footnote{As discussed in \Sec{sec:local_fact}, wide-angle soft radiation does affect the cross section at a fixed value of $\Njet$, just not the $\Njet$ value itself.}  This angular restriction on soft radiation produces \emph{non-global} logarithms of $\dN{n\pm}$, which are logarithms that arise from emissions in a restricted angular region of phase space~\cite{Dasgupta:2001sh,Dasgupta:2002bw}.

At leading order, non-global logarithms are often associated with correlated soft emissions from non-Abelian matrix elements and are therefore proportional to $C_F C_A$. For the case of fractional jet multiplicity, however, the measurement itself introduces a correlation between different emissions, and this effect appears for both non-Abelian and Abelian matrix elements.   As we discuss in detail in \Sec{singular}, the allowed phase space for a soft emission to change $\Njet$ depends on the phase space of other soft emissions and on the phase space of the hard partons, similarly to what happens for clustering logarithms \cite{Banfi:2005gj,Delenda:2006nf,KhelifaKerfa:2011zu,Kelley:2012kj}.  In this regard, all soft logarithms of $\Njet$, including those proportional to $C_F^2$, can be considered non-global.

Moreover, the fact that the allowed phase space for soft emissions to change $\Njet$ depends on the phase space of the hard partons means that one cannot consider how $\Njet$ depends on soft emissions without also considering collinear emissions, and vice versa.  Because the contributions to $\Njet$ from soft and collinear emissions are inextricably linked, we will show in \Sec{softcollnonfact} that $\Njet$ does not obey standard soft-collinear factorization~\cite{Bauer:2001yt,Bauer:2008jx,Ellis:2010rwa,Walsh:2011fz}.    As mentioned above, collinear emissions by themselves do not generate logarithms of $\Njet$, but they do alter the allowed phase space for soft radiation within a jet region.  Thus, collinear emissions will modify the coefficients of soft logarithms of $\Njet$, which is a sign of collinear-soft non-factorization.

While the $\dN{n\pm}$ dependence is non-global, it is also local to each jet.  Emissions affecting $\Njet$ are restricted to an angular region of $\lesssim 2R$ around each jet direction.  Additional emissions away from all jets can also create their own jets, changing $\Njet$ by an integer or near-integer amount, but an emission far from all jets cannot change $\Njet$ by a small amount.  Because of color coherence, this implies that the contribution to $\Njet$ from a given jet is, to leading power, independent of all other jets in the event.\footnote{Color coherence states that emissions collinear to a given jet and well-separated from all other jets are only sensitive to the kinematics and color of that jet and the \emph{anti}-color of the jet, i.e., the color of all other jets in the event~\cite{Ellis:1991qj}.}  The $\Njet$ distribution is therefore a sum of contributions from jet regions that are to a good approximation independently determined.  We will formalize this ``local'' factorization structure in \Sec{sec:local_fact}, and see how it could simplify LHC calculations in \Sec{sec:LHC}.

\section{Rapidity-like Divergences}
\label{rapidity}

As mentioned in \Sec{lowmult}, the calculation of $\Njet$ at $\mathcal{O}(\as^2)$ features divergences not regulated by dimensional regularization (dim reg) when the calculation is divided into unique soft limits.  These divergences have a strong resemblance to rapidity divergences in SCET$_{\rm II}$, where the large rapidities of soft and collinear modes in the effective theory generate divergent integrals not regulated by dim reg~\cite{Chiu:2009yx,Chiu:2011qc,Chiu:2012ir}.  The similar divergences appearing in the $\Njet$ calculation are not from physical rapidities, but instead originate from the \emph{energy sharing} between particles and are easily cast in terms of a ``rapidity'' of this energy sharing:
\be
\label{eq:rapidity_definition}
y = \frac12 \ln \frac{z_1}{z_2} \,.
\ee
In this section, we show how these divergences arise in the $\Njet$ calculation and how we can adapt standard rapidity regulators to our case.

Instead of the complete $\Njet$ calculation, consider the simplified observable
\be
\Delta = z_1 z_2 \,,
\ee
where particles 1 and 2 are particles that may be soft.  $\Delta$ has the same soft scaling properties as the near-integer behavior of $\Njet$ (see \Eqs{Njet1soft}{Njet12soft}), and the other complications from the complete $\Njet$ calculation are irrelevant for the discussion here.  Like $\dNp$ to be calculated in \Sec{singular}, $\Delta$ receives contributions from both single-soft and double-soft regions of phase space.  Also like $\dNp$, $\Delta$ is a non-additive observable that does not get a linear contribution from each soft emission.

\begin{figure}
\begin{center}
\includegraphics[width=0.6\textwidth]{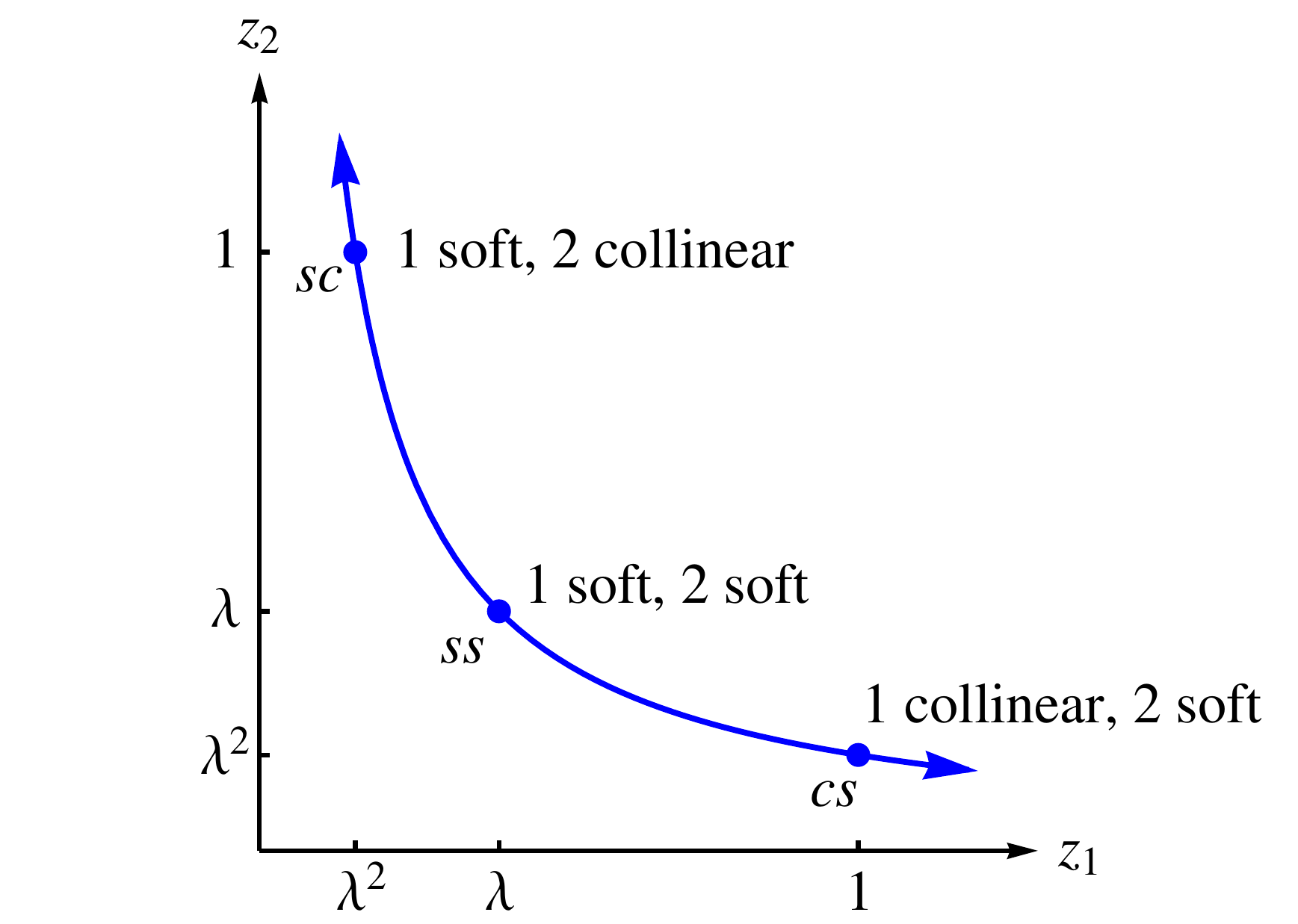}
\caption{The single- and double-soft modes depicted in the $z_1$-$z_2$ plane for the simplified observable $\Delta = z_1 z_2$ (similar to $\dNp$).  We assign the power counting $\Delta \sim \lambda^2$.  In the case with only one soft particle, the soft energy scales as $\lambda^2$, whereas in the case with two soft particles, the soft energies scale as $\lambda$.  These modes are connected by the hyperbola shown, which can be thought of as tracing out the rapidity in the energies of the two partons (see \Eq{eq:rapidity_definition}).  In our calculation, we encounter divergences in this rapidity variable for each mode depicted.  This picture has a strong analog in the divergences in physical rapidity of single particles in SCET$_{\rm II}$.}
\label{fig:rapidityfactorization}
\end{center}
\end{figure}

The interplay between the single- and double-soft limits is illustrated in \Fig{fig:rapidityfactorization}.  In order to have a consistent power counting for $\Delta$, we need to consider two different scalings of the soft modes.  Let $\lambda \ll 1$ be the power counting parameter, with $\Delta \sim \lambda^2$.  The phase space regions with the same parametric contribution to $\Delta$ are: 
\begin{align}
\text{1 soft, 2 collinear ($sc$): } \; & z_1 \sim \lambda^2 \,, \; z_2 \sim 1 \,, \nn \\
\text{1 soft, 2 soft ($ss$): } \; & z_1 \sim \lambda \,, \; z_2 \sim \lambda \,, \nn \\
\text{1 collinear, 2 soft ($cs$): } \; & z_1 \sim 1 \,, \; z_2 \sim \lambda^2 \,.
\end{align}
Thus, there are two types of soft modes --- single-soft modes scale as $\lambda^2$, whereas double-soft modes scale as $\lambda$ --- and there is a unique soft power counting within each sector.  This relative scaling is the same as ultrasoft modes in SCET$_{\rm I}$ (which scale as $\lambda^2$) and soft modes in SCET$_{\rm II}$ (which scale as $\lambda$).  This indicates that the calculation has contributions to the observable from soft modes in both SCET$_{\rm I}$ and SCET$_{\rm II}$.  The ``energy rapidity'' variable $y$ in \Eq{eq:rapidity_definition} separates these phase space regions by measuring the relative energy sharing between particles 1 and 2.  Like an ordinary rapidity, the range is $y \in (-\infty,\infty)$.  The $sc$ region has $z_1 \ll z_2$ ($y \to -\infty$), the $ss$ region has $z_1 \simeq z_2$ ($\lvert y \rvert \simeq 1$), and the $cs$ region has $z_1 \gg z_2$ ($y \to \infty$).  Each of the $sc$/$cs$/$ss$ sectors gives an independent contribution to the $\Delta$ distribution at $\mathcal{O}(\as^2)$ which should be properly summed (i.e.~one has to remove potential double-counting).  As we will now see, unbounded integrals in $y$ produce divergences, not regulated by dim reg.

In full QCD with dim reg in $d = 4-2\epsilon$ dimensions, the soft divergences of $\Delta$ are encapsulated by the integral
\begin{align}
I_{\rm full} (\Delta) &= \int_0^1 \frac{\df z_1}{z_1} \frac{\df z_2}{z_2} \, (z_1 z_2)^{-2\epsilon} \, \delta (\Delta - z_1 z_2) \nn \\
&= \Delta^{-1-2\epsilon} \int_{-\infty}^{\infty} \df y \, \Theta\Bigl( -\frac12 \ln (1/\Delta) < y < \frac12 \ln (1/\Delta) \Bigr) \nn \\
&= \frac{1}{4\epsilon^2} \delta(\Delta) - \cL_1 (\Delta) \, \Theta(\Delta < 1) + \ord{\epsilon} \,,
\end{align}
where $\cL_1$ is a logarithmic plus-function defined in \Eq{eq:plusfunctiondef}.  If we take soft limits of the phase space, where the boundaries scale as $[0,1] \to [0,\infty]$, we get:
\begin{align} \label{simplesoftlimits}
\text{1 soft: } \; I_{sc} (\Delta) &= \int_0^\infty \frac{\df z_1}{z_1} \int_0^1 \frac{\df z_2}{z_2} \, (z_1 z_2)^{-2\epsilon} \, \delta(\Delta - z_1 z_2) \nn \\
&= \Delta^{-1-2\epsilon} \int_{-\infty}^{\infty} \df y \, \Theta\Bigl( -\frac12 \ln (1/\Delta) < y \Bigr) \,, \nn \\
\text{2 soft: } \; I_{cs} (\Delta) &= \int_0^1 \frac{\df z_1}{z_1} \int_0^\infty \frac{\df z_2}{z_2} \, (z_1 z_2)^{-2\epsilon} \, \delta(\Delta - z_1 z_2) \nn \\
&= \Delta^{-1-2\epsilon} \int_{-\infty}^{\infty} \df y \, \Theta\Bigl( y < \frac12 \ln (1/\Delta) \Bigr)  \,, \nn \\
\text{1, 2 soft: } \; I_{ss} (\Delta) &= \int_0^\infty \frac{\df z_1}{z_1} \int_0^{\infty} \frac{\df z_2}{z_2} \, (z_1 z_2)^{-2\epsilon} \, \delta(\Delta - z_1 z_2) \nn \\
&= \Delta^{-1-2\epsilon} \int_{-\infty}^{\infty} \df y \,.
\end{align}
Each integral is divergent and unregulated, but the $\Delta < 1$ regime of full QCD is reproduced by the combination\footnote{Note that outside this toy example, only the logarithms of $\Delta$ in full QCD will be reproduced by the soft limits, instead of the full $\Delta < 1$ result.}
\be \label{unregsum}
I_{\rm full} (\Delta) = \bigl[ I_{sc} (\Delta) + I_{cs} (\Delta) - I_{ss} (\Delta) \bigr] \Theta(\Delta < 1).
\ee
This result is as expected, since the double-soft limit should remove the double-counting of the single-soft limits, even if each contribution is not well-defined individually.  Note that the unregulated divergences appear only in $y$.

This structure of divergences in $y$ is identical to those in physical rapidities of single particles in SCET$_{\rm II}$.  These divergences may be handled in a number of ways, and we will use the rapidity renormalization group (rapidity RG)~\cite{Chiu:2011qc,Chiu:2012ir}.  In the rapidity RG, divergences are regulated analogously to dim reg, using a scale $\nu$ and infinitesimal parameter $\eta$ that correspond to the usual $\mu$ and $\epsilon$ in dim reg.  At one loop, these regulator factors are~\cite{Chiu:2011qc,Chiu:2012ir}
\begin{align}
\text{1 soft: } \; R^{sc}_{\eta} &= \Bigl( \frac{\nu}{E_J} \Bigr)^{\eta} z_2^{-\eta} = \Bigl( \frac{\nu}{E_J} \Bigr)^{\eta} s^{-\eta/2} e^{y\eta}  \,, \nn \\
\text{2 soft: } \; R^{cs}_{\eta} &= \Bigl( \frac{\nu}{E_J} \Bigr)^{\eta} z_1^{-\eta} = \Bigl( \frac{\nu}{E_J} \Bigr)^{\eta} s^{-\eta/2} e^{-y\eta} \,, \nn \\
\text{1, 2 soft: } \; R^{ss}_{\eta} &= \Bigl( \frac{\nu}{E_J} \Bigr)^{\eta} \lvert z_1 - z_2 \rvert^{-\eta} = \Bigl( \frac{\nu}{E_J} \Bigr)^{\eta} s^{-\eta/2} \lvert 2 \sinh y \rvert^{-\eta} \,,
\end{align}
where $s = z_1 z_2$, and $E_J$ is the total jet energy such that $z_i = E_i / E_J$.  These regulators give well-defined terms for each soft limit,
\begin{align}
\text{1 soft: } \; I_{sc} (\Delta) &= \int_0^\infty \frac{\df z_1}{z_1} \int_0^1 \frac{\df z_2}{z_2} \, (z_1 z_2)^{-2\epsilon} \, \delta(\Delta - z_1 z_2) \, R^{sc}_{\eta} \nn \\
&= -\Bigl( \frac{\nu}{E_J} \Bigr)^{\eta} \Delta^{-1-2\epsilon} \, \frac{1}{\eta} \,, \nn \\
\text{2 soft: } \; I_{cs} (\Delta) &= \int_0^1 \frac{\df z_1}{z_1} \int_0^\infty \frac{\df z_2}{z_2} \, (z_1 z_2)^{-2\epsilon} \, \delta(\Delta - z_1 z_2) \, R^{cs}_{\eta} \nn \\
&= -\Bigl( \frac{\nu}{E_J} \Bigr)^{\eta} \Delta^{-1-2\epsilon} \, \frac{1}{\eta} \,, \nn \\
\text{1, 2 soft: } \; I_{ss} (\Delta) &= \int_0^\infty \frac{\df z_1}{z_1} \int_0^{\infty} \frac{\df z_2}{z_2} \, (z_1 z_2)^{-2\epsilon} \, \delta(\Delta - z_1 z_2) \, R^{ss}_{\eta} \nn \\
&= \Bigl( \frac{\nu}{E_J} \Bigr)^{\eta} \Delta^{-1-2\epsilon - \eta/2} \biggl[ \frac{2}{\eta} + \ord{\eta} \biggr] \,.
\end{align}
Now the sum of contributions is independent of the rapidity regulator and matches the full QCD results for $\Delta < 1$:
\begin{align} \label{regsum}
\bigl[ I_{sc} (\Delta) + I_{cs} (\Delta) + I_{ss} (\Delta) \bigr] \Theta(\Delta < 1) = \frac{1}{4\epsilon^2} \delta(\Delta) - \cL_1 (\Delta) \, \Theta(\Delta < 1) + \ord{\epsilon, \eta} = I_{\rm full} (\Delta) \,.
\end{align}

Comparing \Eq{regsum} to \Eq{unregsum}, we see that with the rapidity regulator, the full result is reproduced by the \emph{sum} of single- and double-soft contributions, instead of the difference.  This happens because the single- and double-soft limits produce canceling poles in the large positive and negative rapidity regions.  For example, the large rapidity limit is allowed in both the $sc$ and $ss$ contributions, where the rapidity regulators scale as
\begin{align}
\text{1 soft: } \; R^{sc}_{\eta} (y \to \infty) &\sim e^{y\eta} \,, \nn \\
\text{1, 2 soft: } \; R^{ss}_{\eta} (y \to \infty) &\sim e^{-y\eta} \,,
\end{align}
which lead to canceling poles:
\begin{align}
\text{1 soft: } &\; \int_0^{\infty} \df y \, e^{y\eta} = -\frac{1}{\eta} \,, \nn \\
\text{1, 2 soft: } &\; \int_0^{\infty} \df y \, e^{-y\eta} = \frac{1}{\eta} \,.
\end{align}
Thus the cancellation of the large rapidity regimes happens at the level of the sum of contributions, instead of the difference.

Not only does the rapidity RG make the various soft contributions well defined, but it also separates the double-soft limit from the zero-bin subtraction (needed to remove double counting \cite{Manohar:2006nz}) of the single-soft limits.\footnote{In this case, the zero-bin limit is the overlap of the single- and double-soft limits, and is tantamount to removing the restriction on $y$ in the single-soft limits in \Eq{simplesoftlimits}.}  Because the rapidity regulators in the single- and double-soft limits are \emph{not} related by scaling in rapidities, the double-soft contribution is not the zero-bin of the single-soft limit.  In fact, the zero-bin limit is scaleless; the rapidity regulator is unchanged in the zero-bin limit but now all rapidities are allowed, leading to the scaleless integral~\cite{Chiu:2012ir}
\be
\int_{-\infty}^{\infty} \df y \, e^{\pm y\eta} = 0 \,.
\ee
In SCET$_{\rm II}$ applications where the collinear and soft contributions to the distribution can be factorized into separated jet/beam and soft functions, this factorization allows these rapidity divergences to be divided into the collinear and soft functions and the resulting logarithms to be resummed using standard techniques (see, e.g., \cite{Chiu:2011qc,Chiu:2012ir}).

The features of rapidity divergences in this simple example of $\Delta$ repeat themselves in the calculation of near-integer behavior of $\Njet$ (specifically, $\dNp$) in \Sec{singular} below.

\section{Calculating the Near-Integer Behavior}
\label{nearint}

As seen in \Fig{fig:phasespace}, fractional values of $\Njet$ arise when radiation in a jet region extends beyond a radius $R$. In particular, near-integer values of $\Njet$ come from soft radiation beyond $R$, and the deviation from integer values grows as emissions become harder.  In this section, we study the leading-order near-integer behavior of $\Njet$ in $e^+ e^- \to \text{jets}$ at a center-of-mass energy of $Q$, which first occurs for $e^+ e^- \to 4$ partons.

Let $\dNpm$ be the jet multiplicity near the 2-jet peak as defined in \Eq{eq:dNpm}; we take the power counting $\dNpm\sim\lambda^2$, with $\lambda \ll 1$. We will first derive an analytic expression for the cross section at leading order in $\lambda$ (i.e.~the singular contributions), including a discussion of the aforementioned rapidity-like divergences. We will then calculate the full $\ord{\as^2}$ result using the Monte Carlo program \eventtwo~\cite{Catani:1996jh,Catani:1996vz}, which allows us to include non-singular terms as well as cross check our results for the singular contributions.  We will also calculate the $\ord{\as^2}$ contributions to $\dNtm$, though this is not the focus of our studies.

\subsection{Singular Contributions using Splitting Functions}
\label{singular}

The cross section for $\dNpm$ can be written as
\be
\label{eq:sigmadefinition}
\frac{\df\sigma}{\df\dNpm}=\int \df\Phi_4 \, \mathcal{T}(e^+e^-\to 4\,\text{ partons})\,\mathcal{F}(\dNpm,\Phi_4),
\ee
where $\Phi_4$ represents four-body phase space, $\mathcal{T}$ is the matrix element for $e^+e^-\to 4\text{ partons}$, and $\mathcal{F}(\dNpm,\Phi_4)$ is the measurement function, which  projects out the slice of phase space corresponding to a constant value of $\dNpm$.  The allowed values of $\Njet$ for four partons are:\footnote{Here we assume that each isolated parton and group of partons are sufficiently energetic to pass the energy cut $\Ecut$.  If this does not hold, then the integer values of $\Njet =$ 3 or 4 may be reduced to 2 or 3.}
\begin{align}
\text{Four isolated partons: }&  \Njet = 4, \\
\text{Two isolated partons and one merged pair: }& \Njet = 3, \\
\label{eq:merged_triplet_phase_space}\text{One isolated parton and one merged triplet: }& \Njet \in \left( 2 - (\Ecut/Q)^2,3 \right), \\
\text{Two merged pairs: }& \Njet = 2.
\end{align}

The only non-integer behavior is obtained for the merged triplet in \Eq{eq:merged_triplet_phase_space}, which requires three particles to be within an angular distance $\lesssim 2R$. Thus, for sufficiently small $R$, we can use the matrix element in the limit where three partons are collinear, allowing us to take advantage of collinear factorization \cite{Ellis:1978sf,Ellis:1978ty,Catani:1996jh,Catani:1996vz,Campbell:1997hg}:\footnote{We use spin-averaged splitting functions, which are sufficient since the value of $\Njet$ for the collinear system is independent of its orientation relative to other jets in the event (see, e.g., \cite{Alioli:2013hba}).   This $1 \to 3$ splitting function approach also appears in \Ref{Field:2012rw} for calculating the jet substructure observable planar flow \cite{Thaler:2008ju,Almeida:2008yp,Almeida:2008tp}.}
\begin{align} \label{Tfact}
\cT (e^+ e^- \to 4\,\text{ partons}) &\simeq \cT(e^+ e^- \to q\bar{q}) \cdot \sum_k \cT_k^\text{coll} (1 \to 3) \\
&= \cT(e^+ e^- \to q\bar{q}) \cdot \frac{(8\pi\as\mu^{2\e})^2}{s_{123}^2} \sum_k \langle \hat{P}^k_{1\to 3} \rangle\,,
\end{align}
where $k$ labels one of the parton channels $q \to g g q \,, \bar{q} \to g g \bar{q} \,, q \to q' \bar{q}' q \,, \text{or } \bar{q} \to q' \bar{q}' \bar{q}$, $s_{123}$ is the squared invariant mass of the three parton system, and $\langle \hat{P}^k_{1\to 3} \rangle$ is the spin-averaged $1\to3$ splitting function~\cite{Catani:1998nv,Catani:1999ss}.  The factorization in \Eq{Tfact} implies that the relevant phase space for the collinear splitting is 3-body, meaning we decompose the 4-body phase space of the whole event into the 2-body $q\bq$ system of the hard interaction and the 3-body phase space of the $1\to3$ splitting.  In the collinear limit, the 4-body phase space factorizes as $\df\Phi_4 = \df\Phi_2 (q\bar{q}) \times \df s_{123} / (2\pi) \times \df\Phi_3 (1\to3)$.  

The collinear matrix elements can be further decomposed according to their color structure. The channel $q \to g g q$ (and $\bar{q} \to g g \bar{q}$) can be written as a sum of Abelian (``ab") and non-Abelian (``nab") contributions:
\be
 \cT^\text{coll}(q \to g g q)=C_F^2 \cT_{\rm ab}^{\rm coll} (q \to g g q)+C_FC_A  \cT_{\rm nab}^{\rm coll} (q \to g g q),
\ee
while the $q \to q' \bar{q}' q$ (and $\bar{q} \to q' \bar{q}' \bar{q}$) channels will give contributions proportional to the $C_F T_R$ color structure.\footnote{For same flavor final state quarks $q'=q$, this matrix element contains also terms proportional to $C_F^2$ and $C_FC_A$. However, these terms do not contribute to the cross section at leading power.}  At $\ord{\as^2}$ the cross section for $\dNpm$ is obtained by summing over all channels and color structures. In the rest of this subsection we will discuss the calculation of the Abelian piece of the cross section; results for the other color structures can be found in \App{deets}.

The Abelian contribution comes from a $q\to ggq$ or $\bar{q} \to gg \bar{q}$ splitting.  The two channels give identical contributions and we will focus on the $q\to ggq$ case for definiteness, labeling the final state as:
\be
g_1 g_2 q_3.
\ee
As discussed in \Sec{lowmult}, at leading power we can take the soft limit of our observable, which corresponds to either {\it one} or {\it both} gluons becoming soft.  As in the toy calculation in \Sec{rapidity}, we will label the limit where gluon 1 (gluon 2) is soft as $sc$ ($cs$) and the double-soft limit as $ss$.  We again use the energy sharing variables $z_i = E_i / E_J$ and scaled veto variable $\zcut = E_\cut / E_J$ in \Eq{zi}, where here $E_J$ is the energy of one jet region (with $E_J \simeq Q/2$ for a collision at center-of-mass energy $Q$).  We generally assume $\zcut \ll 1$, which allows us to neglect power-suppressed contributions from soft quarks, and if some $z_i < \zcut$ then that particle may be treated as soft.

\begin{table}[p]
\begin{center}
\begin{tabularx}{\textwidth}{  >{\centering\arraybackslash}m{0.15\textwidth} || >{\centering\arraybackslash}m{0.2\textwidth} | >{\centering\arraybackslash}m{0.2\textwidth} | >{\centering\arraybackslash}m{0.35\textwidth}}
\hline\hline
Observable & $\cR_A$ & $\cR_B$ & $\cR_C$\\
\hline\hline
$\dNm$ &\; \includegraphics[width=0.15\textwidth]{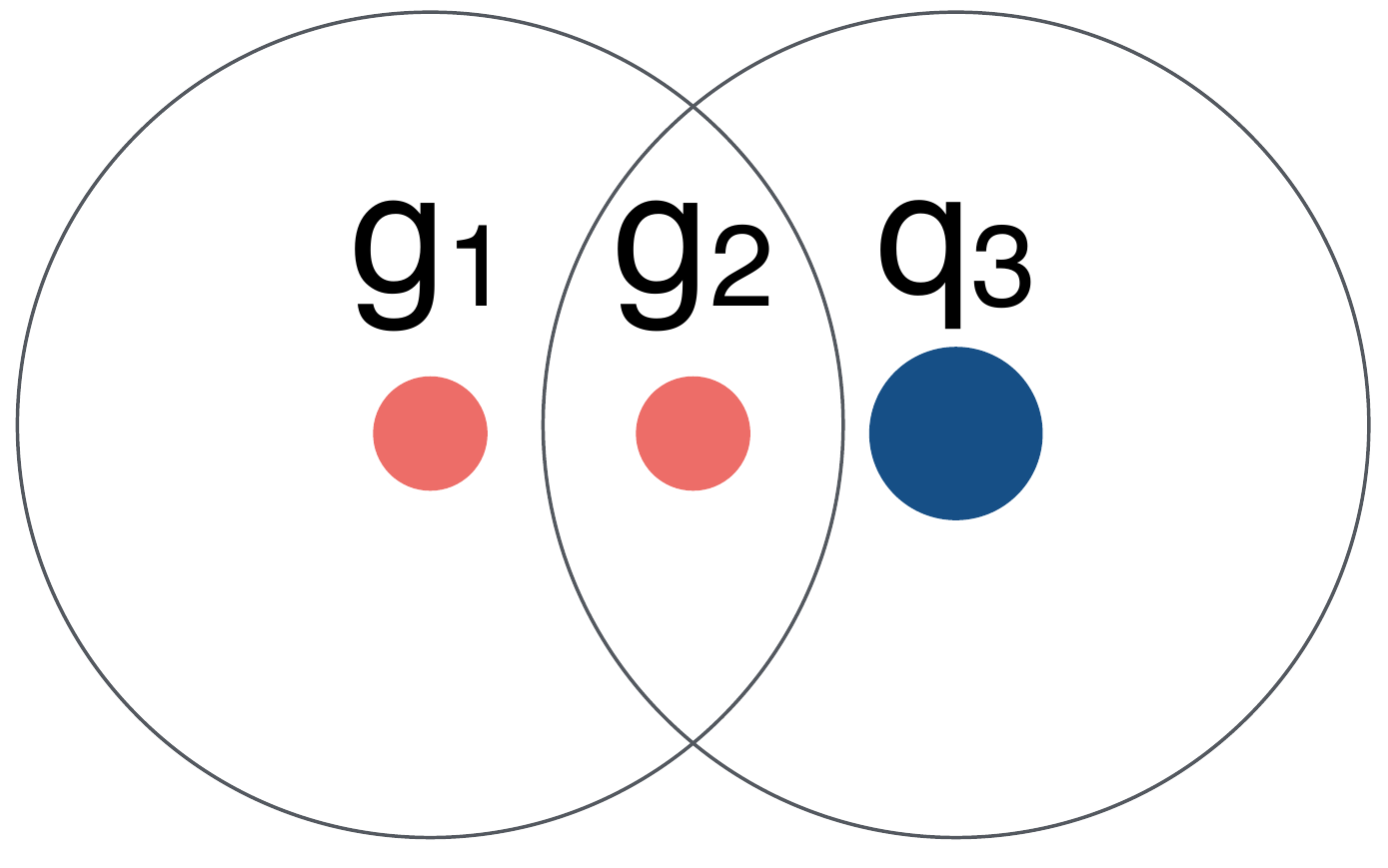} &\; \includegraphics[width=0.15\textwidth]{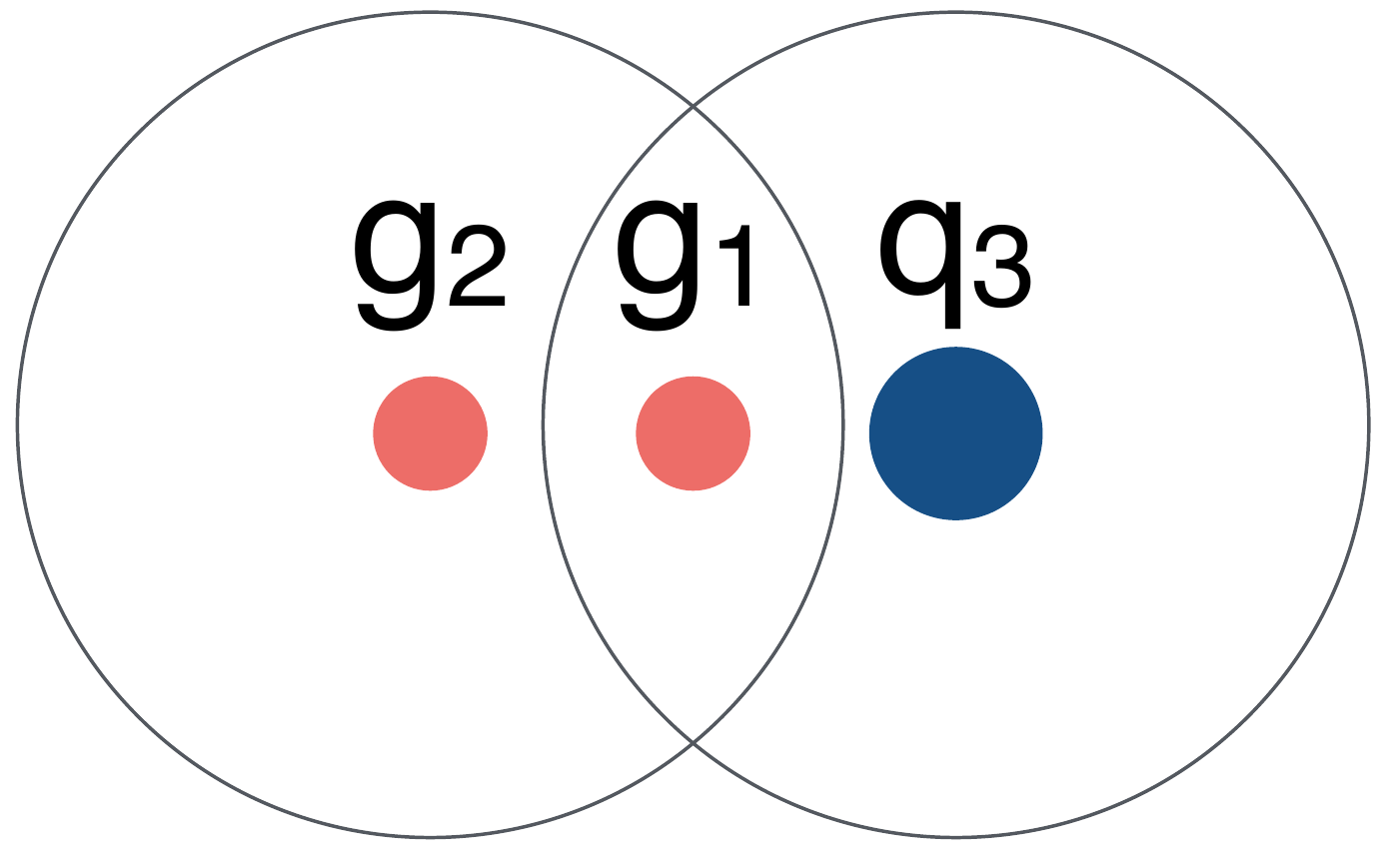} &\; --\\
\hline
$\dNp$ &\; \includegraphics[width=0.15\textwidth]{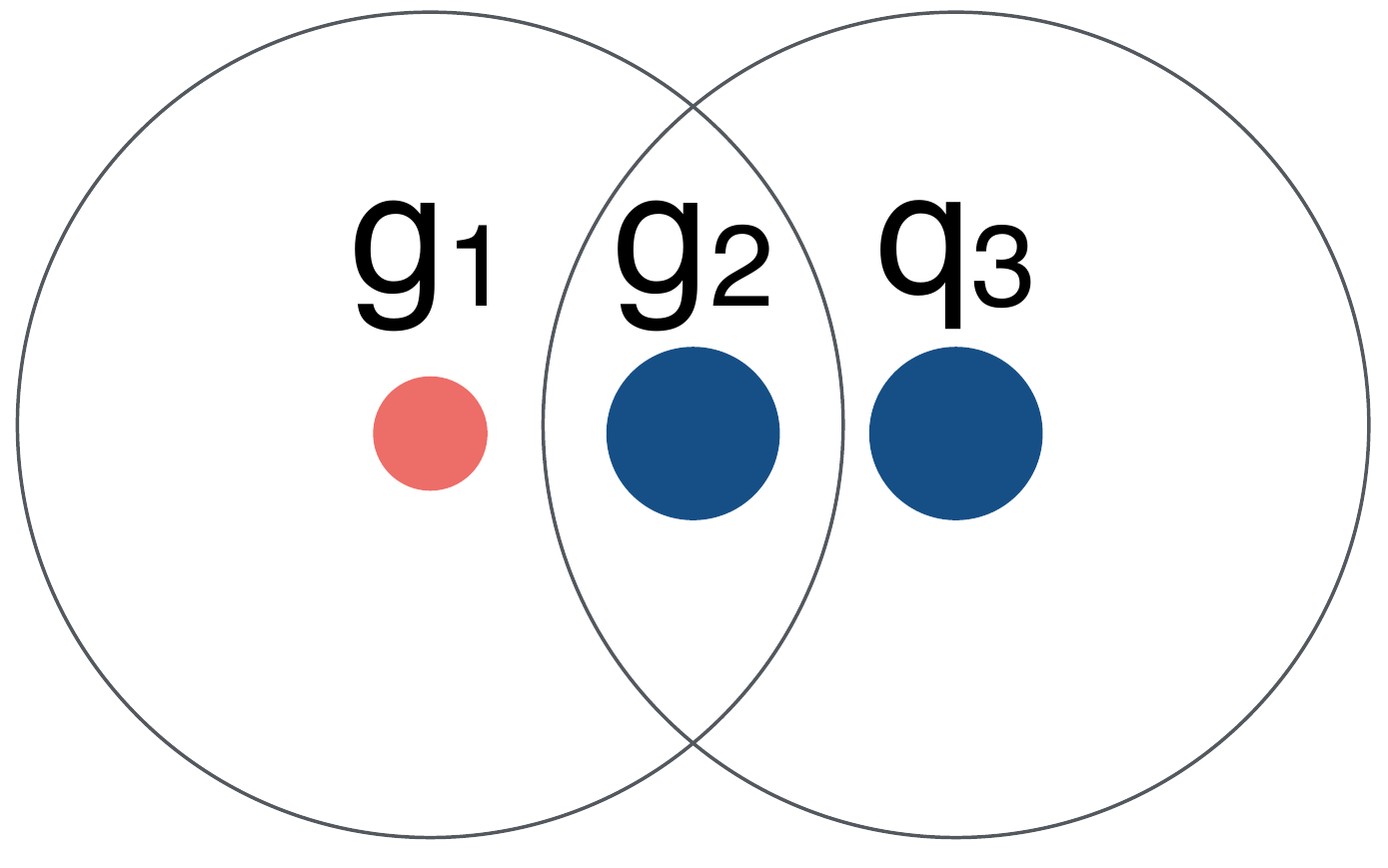} &\; \includegraphics[width=0.15\textwidth]{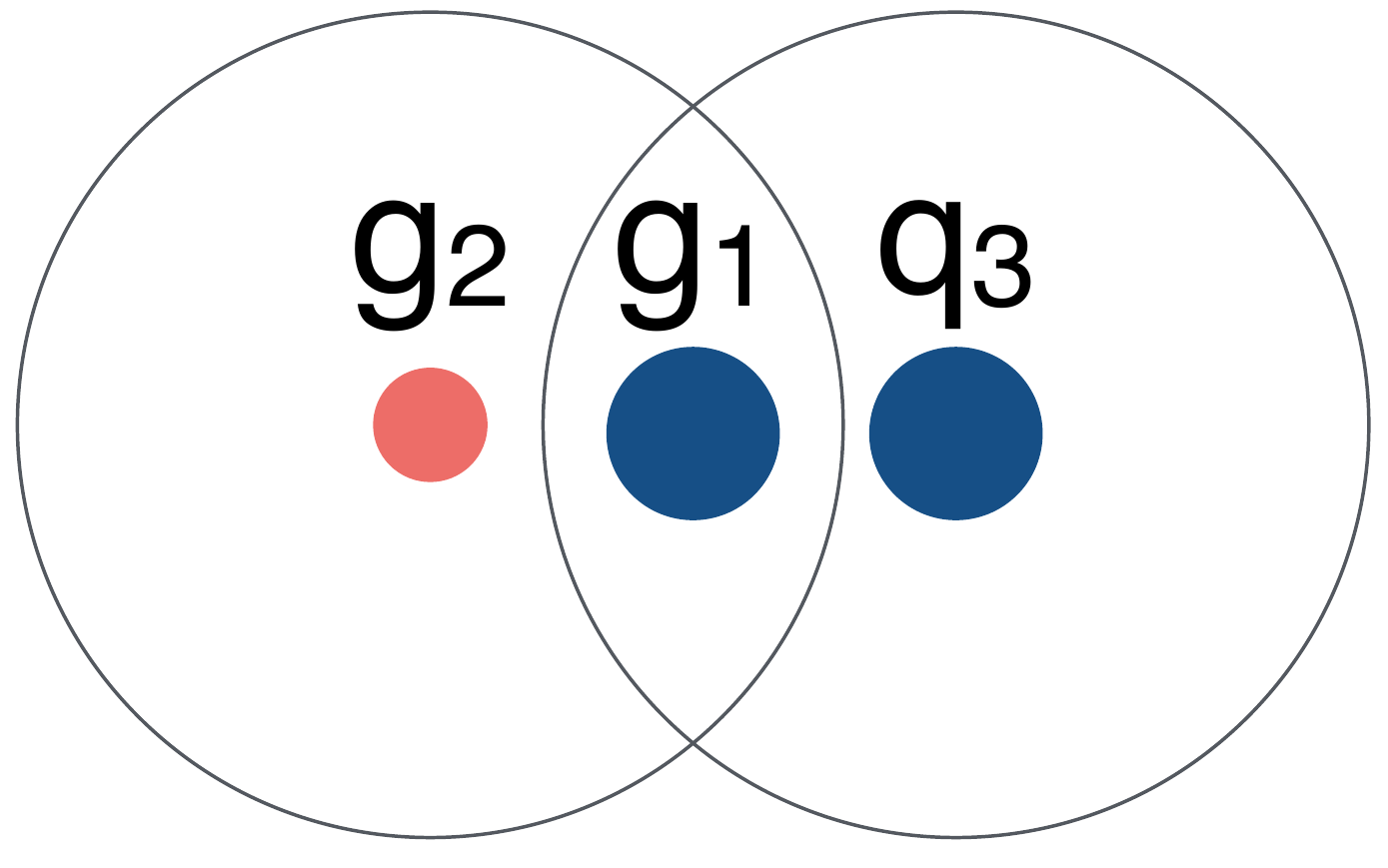} &\;
\includegraphics[width=0.15\textwidth]{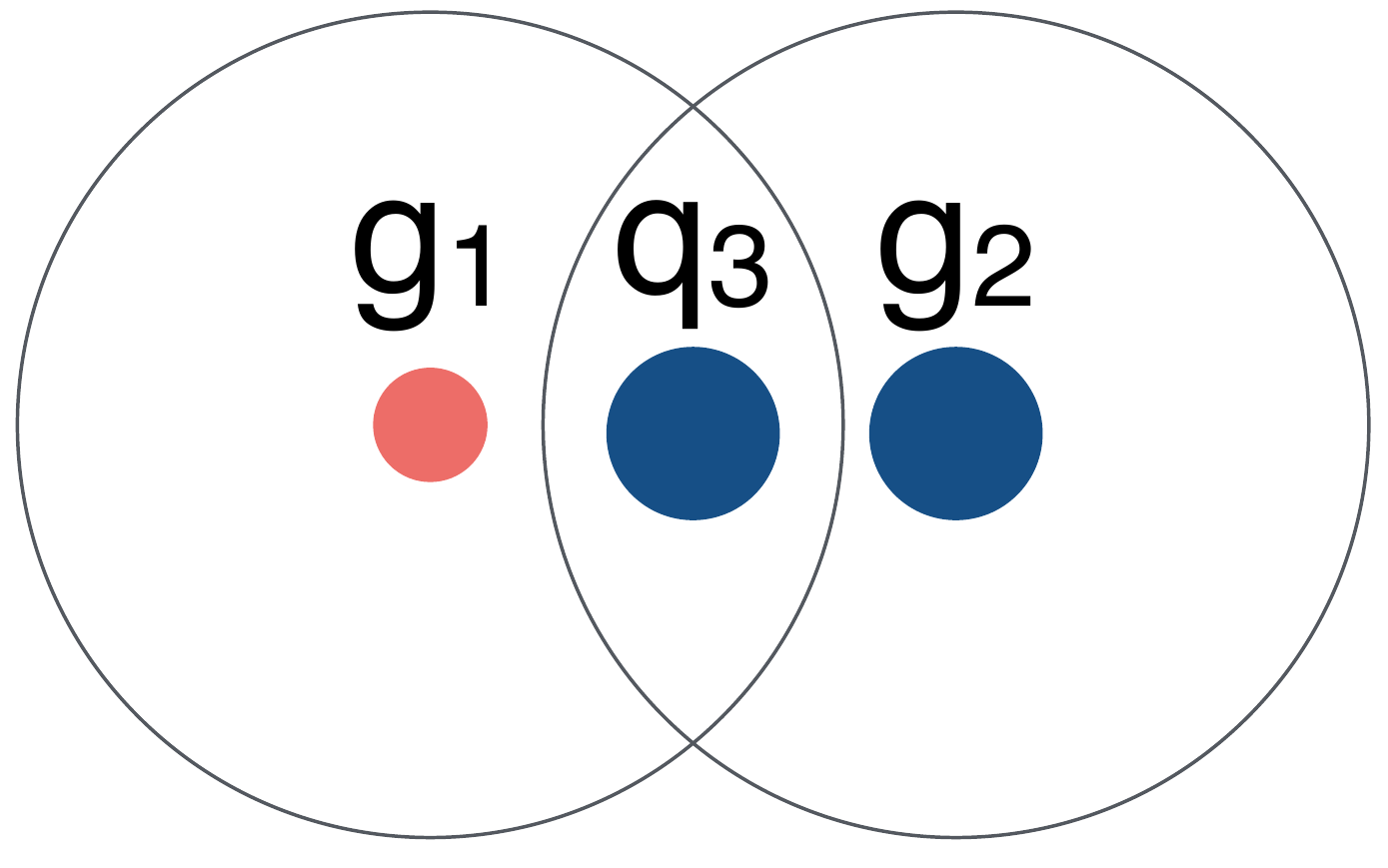}\vspace{0.1cm}
\includegraphics[width=0.15\textwidth]{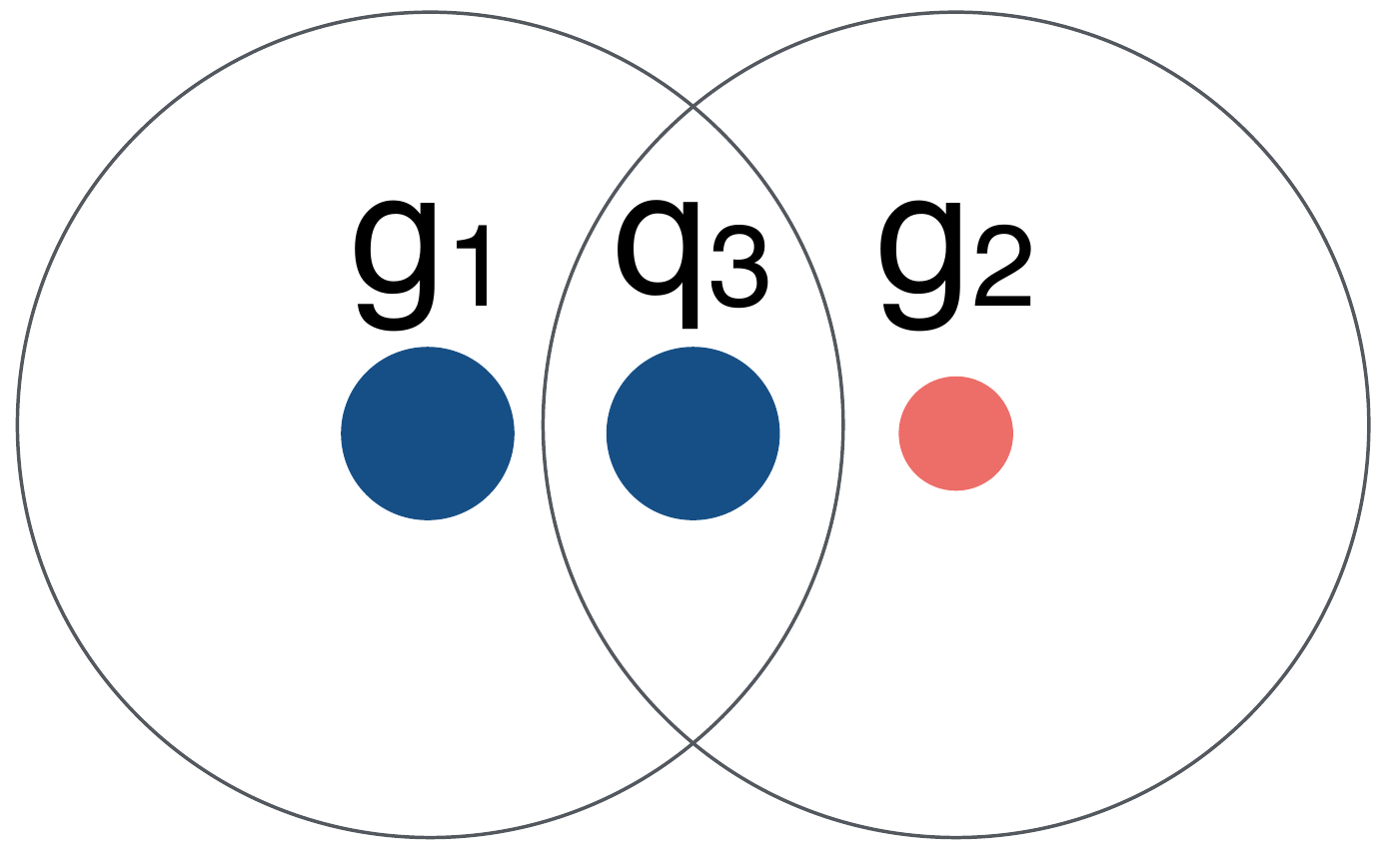}\vspace{0.1cm}
\includegraphics[width=0.15\textwidth]{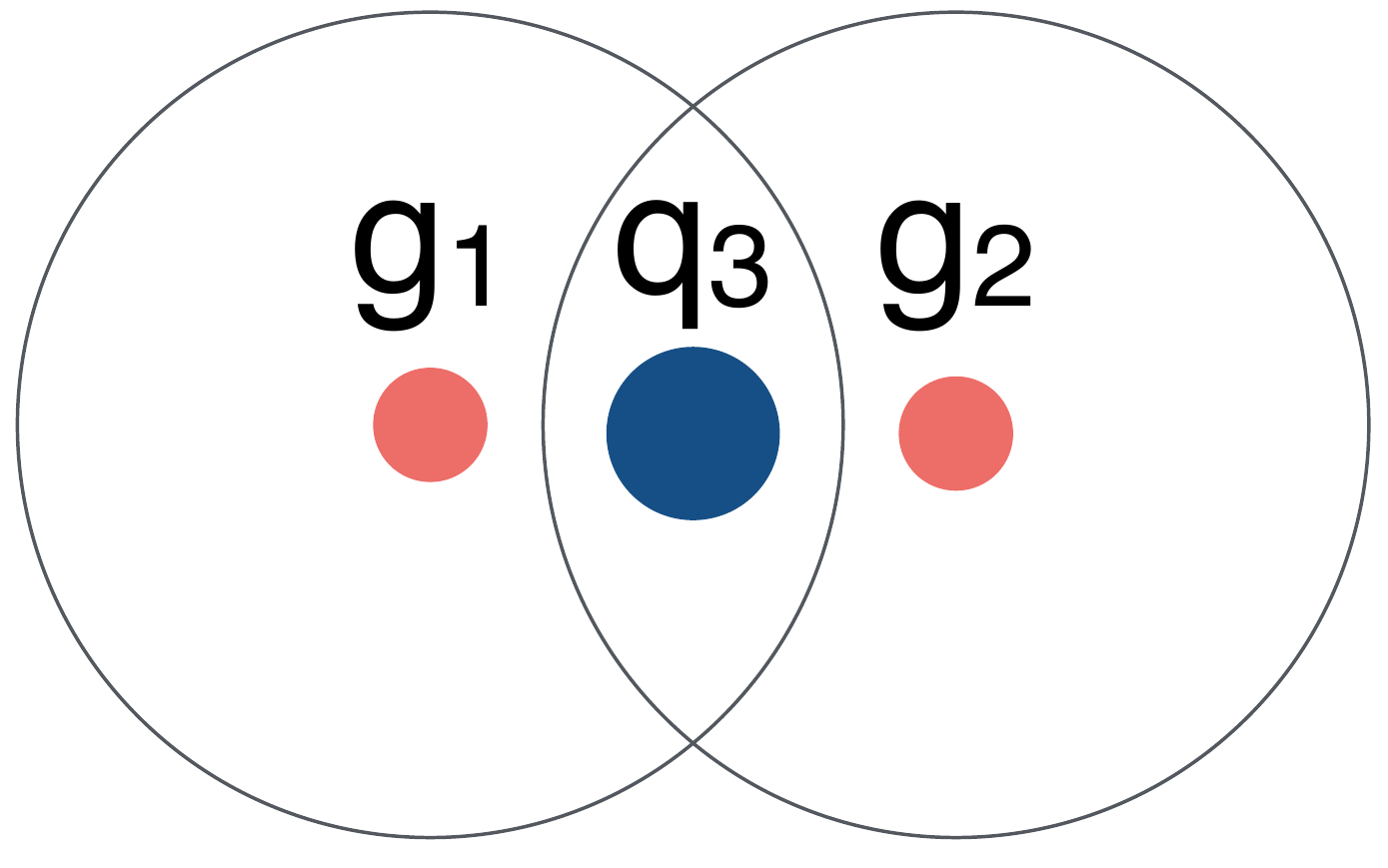}\\
\hline
$\dNtm$ &\; \includegraphics[width=0.15\textwidth]{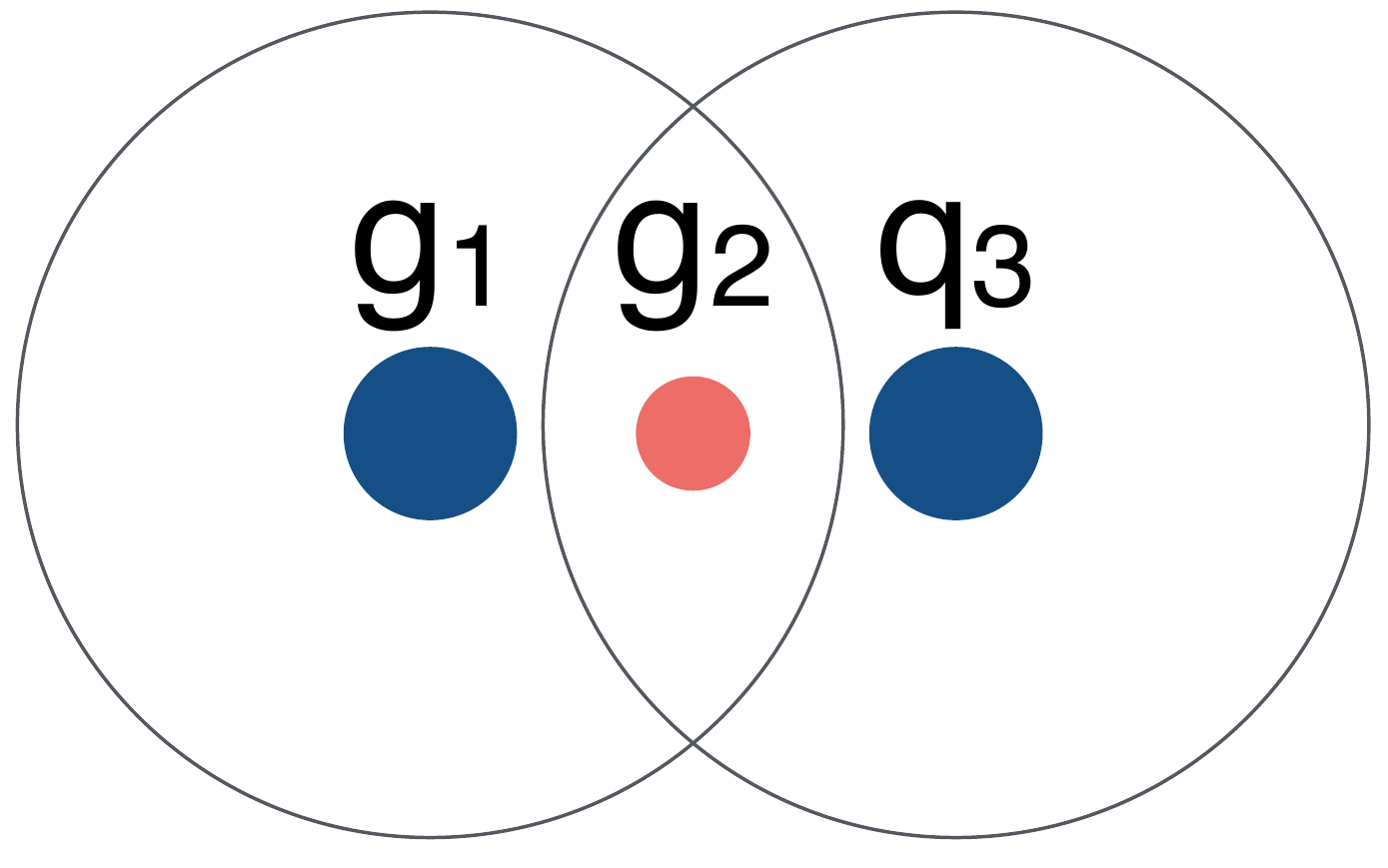} &\; \includegraphics[width=0.15\textwidth]{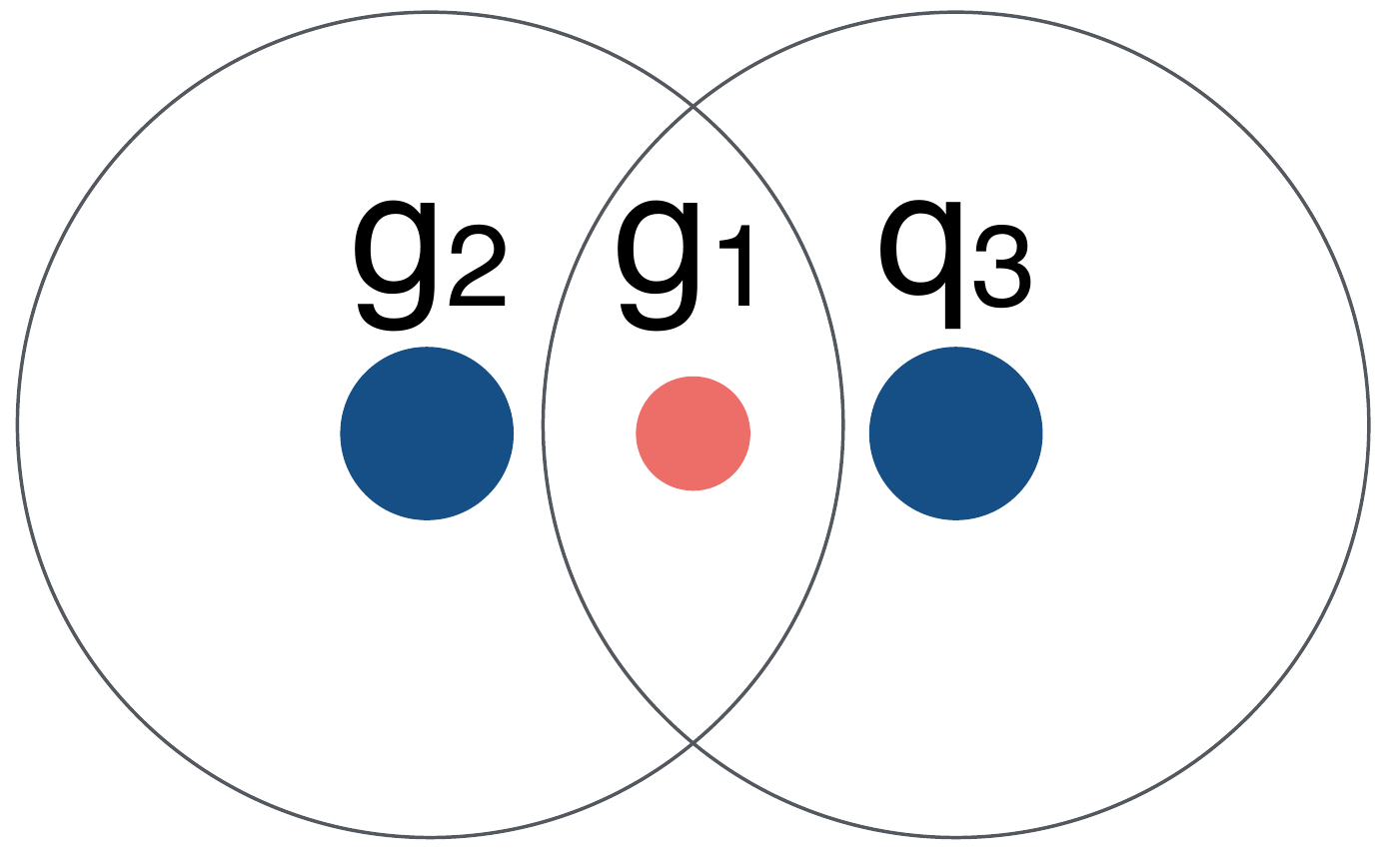} &\; --\\
\hline\hline
\end{tabularx}
\end{center}
\caption{Representation of the phase space configurations contributing to the near-integer jet multiplicities  $\dNm$, $\dNp$, and $\dNtm$.  Compared to \Fig{fig:phasespace}, the value of $\Njet$ is 1 unit higher, because the event contains an additional isolated parton (not shown).  For each observable, we show the corresponding contributions from different angular regions and soft limits. Circles represent cones of radius $R$, large blue dots represent energetic partons, small red dots soft partons with $z<z_\text{cut}$. The angular regions $\cR_A$, $\cR_B$, and $\cR_C$ are defined in \Eq{eq:regions}.  }
\label{tab:PhaseSpace}
\end{table}

\begin{table}[p]
\begin{center}
\setlength{\tabcolsep}{4pt}
\begin{tabularx}{\textwidth}{c|c||c|c|c}
\hline\hline
\; Observable \;&\; Region \;&\; Expression \;&\; Limit \;&\; Cuts \\
\hline \hline
$\dNm$ \;&\; $\cR_A$ \;&\; $z_1 z_2$ \;&\; 1, 2 soft  \;&\; $z_1 + z_2 < z_\cut$ \; \\
$\dNm$ \;&\; $\cR_B$ \;&\; $z_1 z_2$ \;&\; 1, 2 soft  \;&\; $z_1 + z_2 < z_\cut$ \; \\
\hline
$\dNp$ \;&\; $\cR_A$ \;&\; $z_1 (1 - z_2^2) / z_2$ \;&\; 1 soft \;&\; $z_2 > z_\cut$ \; \\
$\dNp$ \;&\; $\cR_B$ \;&\; $z_2 (1 - z_1^2) / z_1$ \;&\; 2 soft \;&\; $z_1 > z_\cut$ \; \\
$\dNp$ \;&\; $\cR_C$ \;&\; $z_1 z_2 (2 - z_2) / (1 - z_2)$ \;&\; 1 soft \;&\; -- \; \\
$\dNp$ \;&\; $\cR_C$ \;&\; $z_1 z_2 (2 - z_1) / (1 - z_1)$ \;&\; 2 soft \;&\; -- \; \\
$\dNp$ \;&\; $\cR_C$ \;&\; $2 z_1 z_2$ \;&\; 1, 2 soft \;&\; -- \; \\
\hline
$\dNtm$ \;&\; $\cR_A$ \;&\; $z_2 [1 - z_1(1 - z_1)]/[z_1 (1 - z_1)]$ \;&\; 2 soft  \;&\; $z_1 > z_\cut$ \; \\
$\dNtm$ \;&\; $\cR_B$ \;&\; $z_1 [1 - z_2(1 - z_2)]/[z_2 (1 - z_2)]$ \;&\; 1 soft  \;&\; $z_2 > z_\cut$ \; \\
\hline\hline
\end{tabularx}
\end{center}
\caption{Near-integer behavior of $\Njet$, shown for various phase space regions as depicted in \Tab{tab:PhaseSpace}.  In each case, the expression for the observable is given along with the relevant limits and phase space cuts.}
\label{tab:observables}
\end{table}

In \Tabs{tab:PhaseSpace}{tab:observables} we collect contributions to $\dNpm$ in the relevant soft limits and regions of phase space.  At this order, we also get a contribution to $\dNtm=3-\Njet$, and we will carry out the calculation for $\dNtm$ as well.

We now gather the relevant pieces to evaluate \Eq{eq:sigmadefinition} in the collinear limit.  There are three relevant regions of phase space:
\begin{align}
\label{eq:regions}
\cR_A&=\Theta(\angle_{13}>R) \, \Theta(\angle_{23}<R) \, \Theta(\angle_{12}<R),\\
\cR_B&=\cR_A(1\leftrightarrow 2),\\
\cR_C&=\Theta(\angle_{13}<R) \, \Theta(\angle_{23}<R) \, \Theta(\angle_{12}>R),
\end{align}
where $\angle_{ij}$ labels the opening angle between partons $i$ and $j$.   In \Tab{tab:PhaseSpace} we show a schematic representation of the angular phase space regions and soft limits which are explicitly listed in \Tab{tab:observables}.
From \Tab{tab:observables}, the measurement functions are given by
\begin{align}
\mathcal{F}(\dNm,\Phi_4)&=\mathcal{F}_{\dNm}^{\,ss},\\
\mathcal{F}(\dNp,\Phi_4)&=\mathcal{F}_{\dNp}^{\,sc}+\mathcal{F}_{\dNp}^{\,cs}+\mathcal{F}_{\dNp}^{\,ss},\label{eq:FdNp} \\
\mathcal{F}(\dNtm,\Phi_4) &=\mathcal{F}_{\dNtm}^{\,sc}+\mathcal{F}_{\dNtm}^{\,cs},
\end{align}
where
\begin{align}
\mathcal{F}_{\dNm}^{\,ss} &=\delta(\dNm-z_1z_2)\,\Theta(z_1+z_2<z_\text{cut})\,\cR_A+(1\leftrightarrow 2) , \\
\begin{split}
\mathcal{F}_{\dNp}^{\,sc}&=\,\delta\left[\dNp-z_1\left(1/z_2-z_2\right)\right]\,\Theta(z_2>z_\text{cut})\cR_A\\
&\quad+\delta [\dNp-z_1z_2(2-z_2)/(1-z_2)]\,\cR_C,\label{eq:FdNpsc}
\end{split}\\
\mathcal{F}_{\dNp}^{\,cs}&=\mathcal{F}_{\dNp}^{\,sc}(1\leftrightarrow 2),\label{eq:FdNpcs}\\
\mathcal{F}_{\dNp}^{\,ss}&=\delta (\dNp-2z_1z_2)\,\cR_C,\label{eq:FdNpss} \\
\mathcal{F}_{\dNtm}^{\,sc}&=\delta\left(\dNtm-z_1 [1 - z_2(1 - z_2)]/[z_2 (1 - z_2)]\right)\,\Theta(z_2>z_\text{cut})\cR_B,\\
\mathcal{F}_{\dNtm}^{\,cs}&=\mathcal{F}_{\dNtm}^{\,sc}(1\leftrightarrow 2).
\end{align}
The single- and double-soft limits for the $1 \to 3$ matrix elements are
\be
\label{eq:Tab}
\cT_{\rm ab}(q\to g_1 g_2 q_3)\simeq
\begin{cases} 
\cT_{\rm ab}^{sc}=\dfrac{4(4\pi\as\mu^{2\e})^2}{E_J^4}\dfrac{2 - 2 z_2 + z_2^2 (1-\e)}{\angle_{13}^2\angle_{23}^2z_1^2z_2^2(1-z_2)}, & \mbox{(1 soft, 2 collinear)} \vspace{1ex} \\
\cT_{\rm ab}^{cs}=\cT_{\rm ab}^{sc}(1\leftrightarrow 2), & \mbox{(1 collinear, 2 soft)} \vspace{1ex} \\
\cT_{\rm ab}^{ss}=\dfrac{8(4\pi\as\mu^{2\e})^2}{E_J^4}\dfrac{1}{\angle_{13}^2\angle_{23}^2z_1^2z_2^2}, & \mbox{(1 soft, 2 soft)}
\end{cases}
\ee
where we have included a symmetry factor of $1/2!$ for identical gluons. Notice that in the double-soft limit, the Abelian matrix element simply reduces to the product of two eikonal factors. In the collinear regime, the 3-body phase space can be written as~\cite{GehrmannDeRidder:1997gf}
\be
\df\Phi_3^{\rm coll}=\frac{E_J^{4-4\e}}{(4\pi)^{4-2\e}\Gamma(1-2\e)} \, \df\Phi_\Omega\, \df\Phi_z,
\ee
where the angular phase space $\df\Phi_\Omega$ is given by
\be
\df\Phi_\Omega=\df\angle_{13}^2 \df\angle_{23}^2 \df\angle_{12}^2\Theta(\Delta)\Delta^{-1/2-\e},\quad\Delta=4\angle_{13}^2\angle_{23}^2-(\angle_{12}^2-\angle_{23}^2-\angle_{13}^2)^2,
\ee
and the energy phase space $\df\Phi_z$ is given by
\be
\df\Phi_z\simeq
\begin{cases}
\df\Phi_z^{sc}=\Theta(z_1<\infty)\Theta(z_2<1)[z_1z_2(1-z_2)]^{1-2\e} \df z_1 \df z_2, & \mbox{(1 soft, 2 collinear)}\\
\df\Phi_z^{cs}=\df\Phi_z^{sc}(1\leftrightarrow 2), & \mbox{(1 collinear, 2 soft)}\\
\df\Phi_z^{ss}=\Theta(z_1<\infty)\Theta(z_2<\infty)(z_1z_2)^{1-2\e} \df z_1 \df z_2. &  \mbox{(1 soft, 2 soft)}
\end{cases}
\ee

As discussed in \Sec{rapidity}, for certain contributions to the observable, these energy integrals give rise to rapidity divergences which require special regularization to make each term well defined.  We use the rapidity regulator to do this, and the sum of all contributions is regulator independent.  Specifically, the contribution to $\dNp$ from $\cR_C$ is the only one with both single- and double-soft limits (see \Tabs{tab:PhaseSpace}{tab:observables}), and hence the only contribution which requires this extra regulator.  While $\dNm$ does receive contributions from the double-soft region of phase space, the constraints on the gluon energies from $\zcut$ implies that both gluons must be soft, and hence the single-soft limits do not contribute.  The limit on the gluon energies also implies there is a kinematic limit on the observable, $\dNm < (\zcut / 2)^2$.

By combining the measurement function with the proper limits of the matrix elements, we get an analytic expression for the Abelian contribution to the $\dNpm$ and $\dNtm$ distributions at leading power in $\lambda$.  Including the identical contribution from $\bar{q} \to \bar{q} gg$ yields
\begin{align}
\left[\frac{\df\sigma}{\df\dNm}\right]_{\rm ab}&= C_F^2\left(\frac{\as}{\pi}\right)^2 \cI_\Omega \bigl\{ -2\mathcal{L}_1(\dNm) + 4\ln z_\text{cut}\,\mathcal{L}_0(\dNm) \bigr\} \Theta( \dNm < \zcut^2 / 4 ) \,, \label{eq:ab1}\\
\left[\frac{\df\sigma}{\df\dNp}\right]_{\rm ab}&= C_F^2\left(\frac{\as}{\pi}\right)^2 \cI_\Omega \biggl\{ -\frac{14}{5}\mathcal{L}_1(\dNp) + \biggl(-\frac{57}{10} + \frac{14}{5}\ln 2 - 2\ln z_\text{cut}\biggr)\mathcal{L}_0(\dNp) \biggr\} \,, \label{eq:ab2}\\
\left[\frac{\df\sigma}{\df\dNtm}\right]_{\rm ab}&=C_F^2\left(\frac{\as}{\pi}\right)^2 \cI_\Omega \biggl( -\frac{3}{2} - 2\ln z_\text{cut} \biggr) \mathcal{L}_0(\dNtm) \,, \label{eq:ab3}
\end{align}
where the distributions $\cL_n$ are the usual logarithmic plus distributions, defined in \App{convs}.  The integral over the angular phase space is given by\footnote{We have written the common angular coefficient $\cI_\Omega$ in terms of the phase space region $\cR_A$.  The same angular coefficient appears for regions $\cR_A$, $\cR_B$, and $\cR_C$ due to an unexpected symmetry in the angular integrals.  This symmetry is only present for certain angular integrals in the non-Abelian case.}
\be
\label{eq:IOmega}
\cI_\Omega=\frac{4}{\pi}\int\frac{\df\Phi_\Omega}{\angle_{13}^2\angle_{23}^2}\cR_A = \frac{5\pi^2}{54} \,.
\ee

\Eqss{eq:ab1}{eq:ab2}{eq:ab3} represent the leading order distributions in the small $R$ and $z_\text{cut}$ limits. Notice that at this order there is no dependence on $R$, as the $1/\theta_{ij}^2$ factors in the denominator of the angular integral provide a logarithmic scaling which can be used to explicitly remove the $R$ dependence.  Since we are focused on the non-integer behavior of $\Njet$, we have suppressed contributions proportional to $\delta(\dN{n\pm})$ in our results (see, however, \Eq{N2contribnear}).

The calculation of the $C_FC_A$ and $C_F T_R n_f$ contributions are very similar to the Abelian case, albeit more involved as the matrix elements do not simply factorize into separate angular and energy functions as in \Eq{eq:Tab}.  In \App{deets}, we give the full set of results for the different color structures.

In the $\ord{\as^2}$ calculation performed above, double logarithms of $\dN{n\pm}$ (appearing as $\cL_1 (\dN{n\pm})$) arose when both emitted gluons became soft, and single logarithms ($\cL_0 (\dN{n\pm})$) arose when a single gluon became soft.  This correspondence holds true with more soft emissions: each soft emission results in a single logarithm of $\dN{n\pm}$ (in certain regions of phase space).  Therefore, in general the largest logarithm at $\ord{\as^k}$ appears when all emissions are soft gluons, and the contribution to the cross section is of the form $\as^k \cL_{k-1} (\dN{n\pm})$.

\subsection{Nonsingular Contributions from \eventtwo}
\label{event2}

To go beyond leading power in $\lambda$, we need to incorporate the $\ord{\as^2}$ matrix elements from full QCD.  For fixed-order calculations of $e^+ e^- \to\text{partons}$ at low multiplicity, the program \eventtwo is a particularly useful tool~\cite{Catani:1996jh,Catani:1996vz}.  \eventtwo performs next-to-leading order calculations of $e^+e^- \to$ 2 and 3 partons, so it includes the needed $e^+e^- \to 4$ parton tree-level matrix elements.  Conveniently, it allows for the decomposition of its results order-by-order in $\as$  and by color structure.  Crucially, \eventtwo can probe the far infrared regions of phase space, meaning the (singular) logarithmic terms of $\dNpm$ and $\dNtm$ are enhanced.  This allows us to perform a robust comparison and cross check of our splitting function calculations above.  Having verified the singular logarithmic contributions to $\dNpm$ and $\dNtm$, we can then extract the nonsingular $\ord{\as^2}$ contributions directly from \eventtwo.

\begin{figure}[p]
\begin{center}
\includegraphics[width=\textwidth]{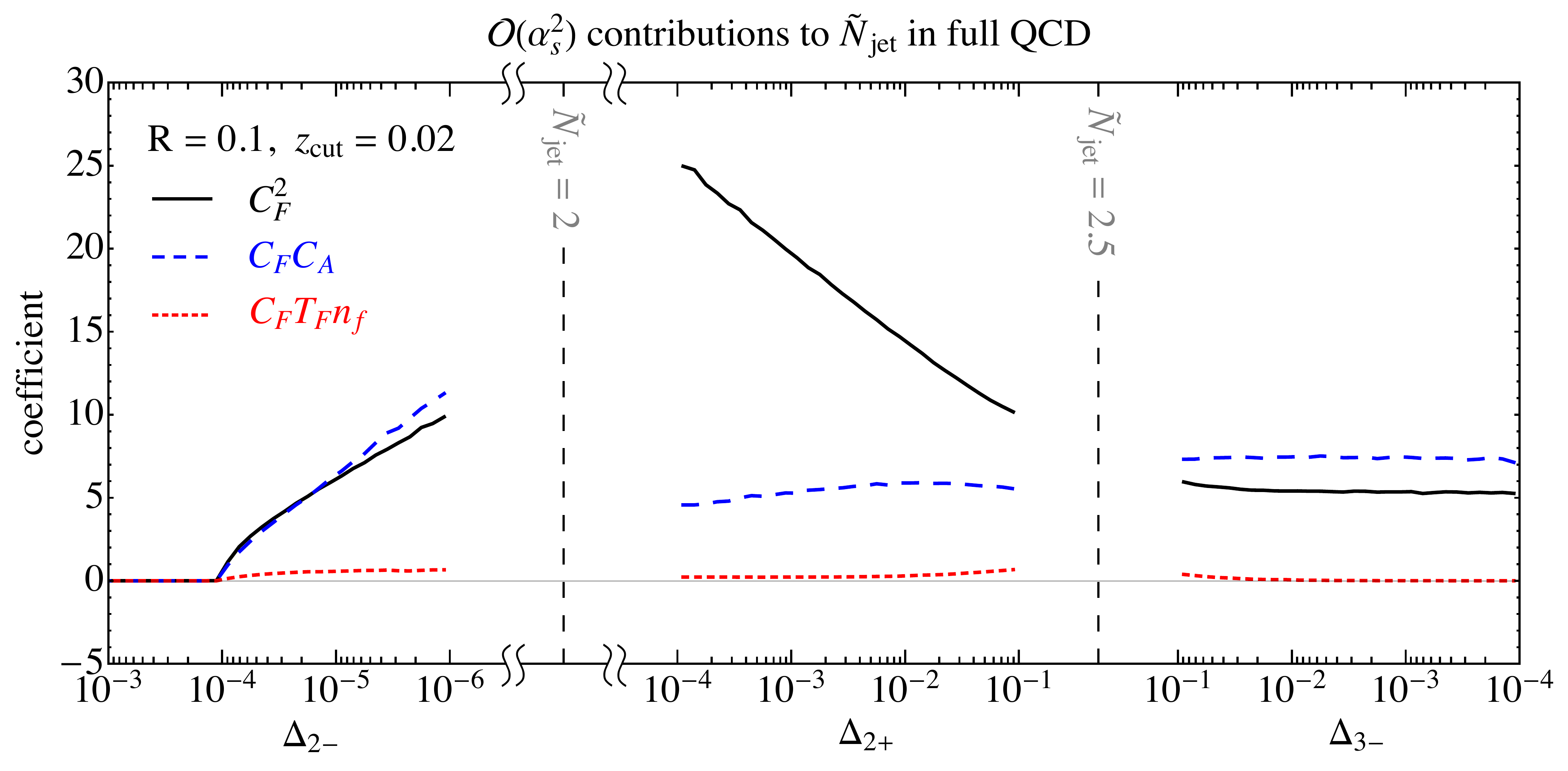}
\caption{The $\dNpm$ and $\dNtm$ distributions extracted from \eventtwo.  Shown are the separate $C_F^2$, $C_F C_A$, and $C_F T_F n_f$ contributions to the cross section, plotted as the coefficient of $(\as / 2\pi)^2 \, C$, where $C$ is the relevant color factor.  Note that the $\dNm$ and $\dNtm$ axes run backwards.}
\label{event2full_triptych}
\end{center}
\end{figure}

\begin{figure}[p]
\begin{center}
\includegraphics[width=\textwidth]{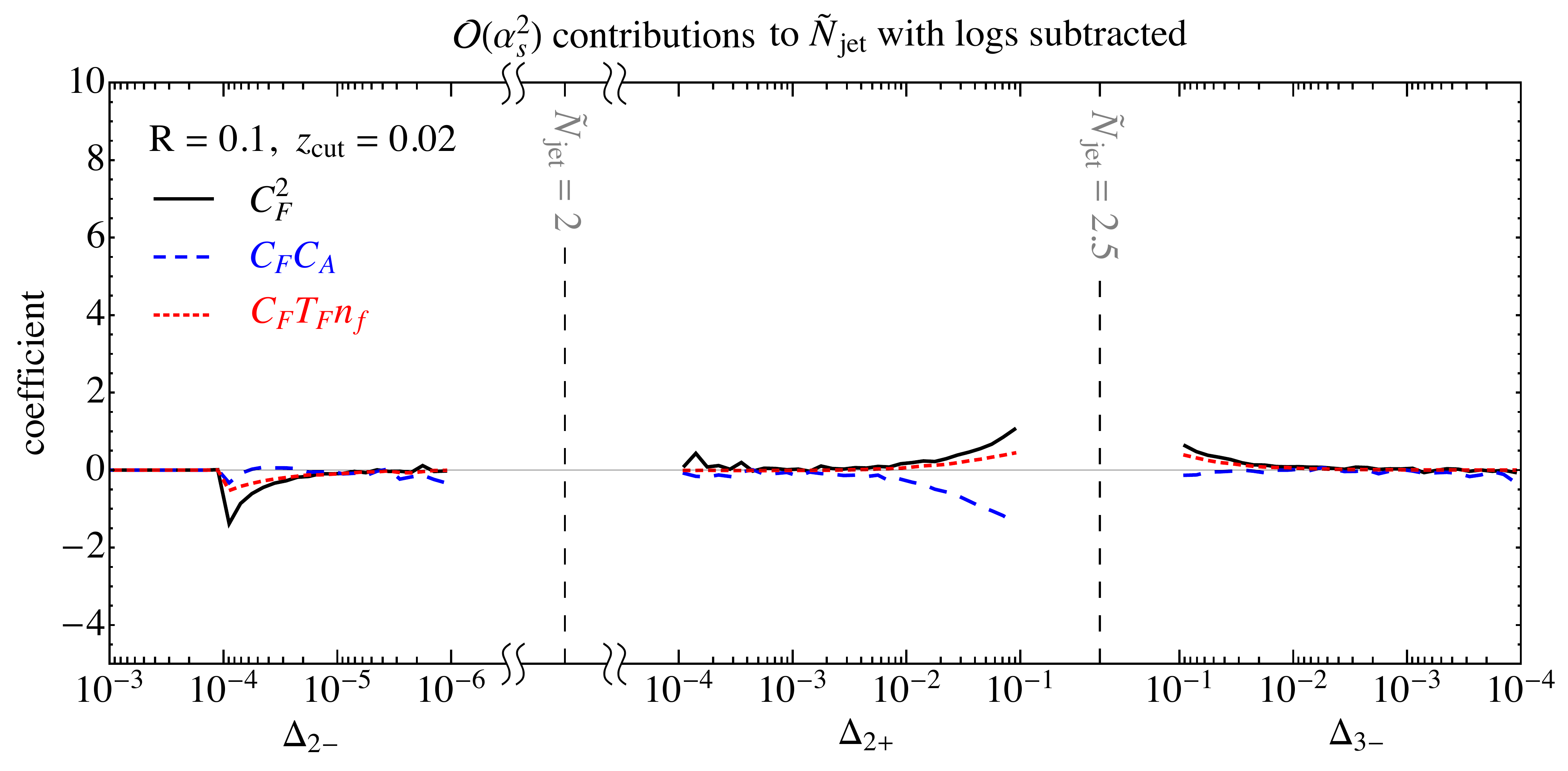}
\caption{Same as \Fig{event2full_triptych}, but subtracting our calculations for the singular contributions in each region.  The residuals vanish in the logarithmic ($\dN{n\pm} \to 0$) regime, indicating that the only remaining terms are nonsingular.  }
\label{event2sub_triptych}
\end{center}
\end{figure}

In \Fig{event2full_triptych}, we show the $\ord{\as^2}$ contributions to $\dN{2\pm}$ and $\dN{3-}$ extracted from \eventtwo.  We plot the coefficients of the $(\as / 2\pi)^2 \, C$ terms in the cross section as a function of $\ln \dN{n\pm}$, where $C$ is the relevant color factor ($C_F^2$, $C_F C_A$, or $C_F T_F n_f$).  To enhance the logarithmic contributions and minimize the power corrections in $R$ and $\zcut$, we choose the small values $R = 0.1$ and $\zcut = 0.02$.  Plotted this way, double logarithms $\bigl(\cL_1 (\dN{n\pm})\bigr)$ appear as lines of constant non-zero slope,\footnote{Note that these are not Sudakov double logarithms, since they appear at $\ord{\as^2}$, not $\ord{\as}$.} single logarithms $\bigl(\cL_0 (\dN{n\pm})\bigr)$ appear as lines with zero slope and non-zero offset, and nonsingular contributions vanish as $\dN{n\pm} \to 0$.  It is clear that logarithmically-enhanced terms are indeed present in the full QCD result from \eventtwo.

To make sure the logarithmic behavior from \eventtwo matches our analytic calculations in \Sec{singular} and \App{deets}, we can extract the nonsingular contribution to the cross section, which are the residual fixed-order terms after the logarithmic contributions are subtracted:
  \be
  \label{eq:nonsingdefinition}
\frac{\df\sigma_{\rm ns}}{\df \dN{n\pm}} = \frac{\df\sigma_{\rm full}}{\df \dN{n\pm}} - \frac{\df\sigma_{\rm sing}}{\df \dN{n\pm}} \,.
\ee
These are shown in \Fig{event2sub_triptych}, again separated by color structure, and confirm that our splitting function calculation, which includes the leading contributions in a small $R$ and $z_\text{cut}$ expansion, captures the leading-order near-integer behavior of $\Njet$ correctly.\footnote{Power suppressed terms of the form $\ord{R^2, \zcut} \cL_{0,1} (\dN{n\pm})$ are generically present in the nonsingular terms, though their contributions are negligible in \Fig{event2sub_triptych} since we have chosen both $R$ and $\zcut$ small.}

For $\Njet < 2$, the singular terms tend to dominate over the nonsingular ones.  The reason is that the $\dNm$ distribution has a kinematic endpoint at $(\zcut / 2)^2 \ll 1$ at this order, so the distribution only has support in the singular region of phase space.  

For $2 < \Njet < 3$, there is an ambiguity in determining the nonsingular terms in the cross section, related to the fact that singular terms exist at multiple points in the $\Njet$ spectrum.  In this range, there are singular terms from $\dNp$ that dominate near $\Njet = 2$ and singular terms from $\dNtm$ that dominate near $\Njet = 3$.  In \Fig{event2sub_triptych}, we defined the nonsingular term by removing the singular terms from the nearest singular point in the $\Njet$ distribution, using $\Njet = 2.5$ as an arbitrary dividing line.\footnote{This corresponds to the nonsingular definition
$$
\frac{\df\sigma_{\rm ns}}{\df\Njet} (2 < \Njet < 3) = \frac{\df\sigma_{\rm full}}{\df\Njet} - \biggl[ \frac{\df\sigma_{\rm sing}}{\df\dNp} \Theta(\Njet < 2.5) + \frac{\df\sigma_{\rm sing}}{\df\dNtm} \Theta(\Njet > 2.5) \biggr] \,.
$$
}
Alternatively, we could define the nonsingular term by removing \emph{both} sets of singular terms over the entire spectrum.\footnote{This would correspond to the nonsingular definition
$$
\frac{\df\sigma_{\rm ns}}{\df\Njet} (2 < \Njet < 3)= \frac{\df\sigma_{\rm full}}{\df\Njet} - \biggl[ \frac{\df\sigma_{\rm sing}}{\df\dNp} + \frac{\df\sigma_{\rm sing}}{\df\dNtm}\biggr] \,.
$$
}
This is a valid approach as well, since the singular $\dNp$ and $\dNtm$ distributions are governed by different soft limits of phase space, hence there is no double-counting by including both singularities.

\begin{figure}
\begin{center}
\includegraphics[width=0.7\textwidth]{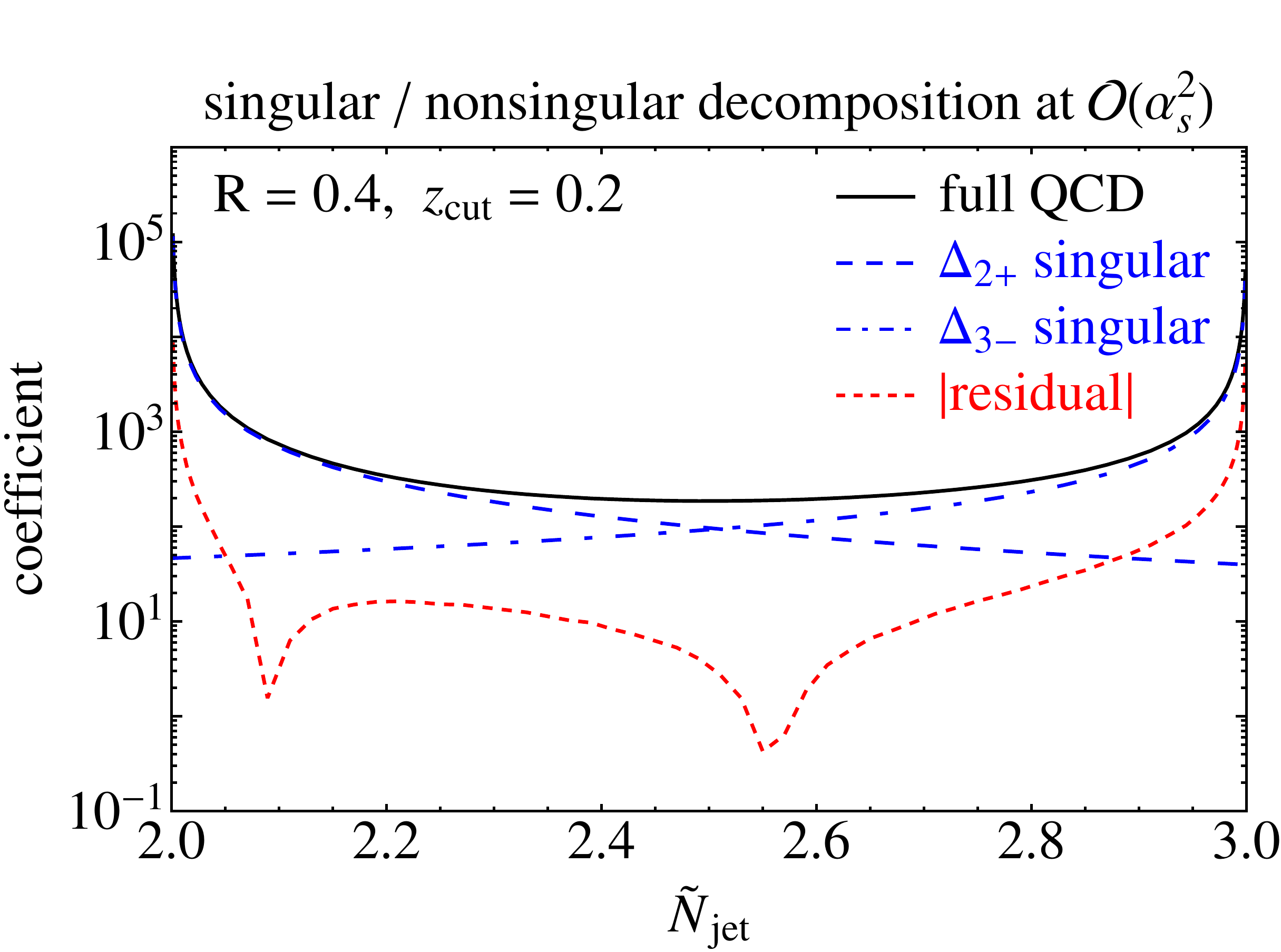}
\caption{The $2 < \Njet < 3$ distribution in full QCD (black, solid), decomposed into singular contributions from $\dNp$ and $\dNtm$ (blue, dashed/dot-dashed) and residual nonsingular contributions (red, dotted).  The coefficient of $(\as / 2\pi)^2$ is plotted for each contribution.  Note that the sum of singular contributions is numerically dominant over the entire range.}
\label{fig:singularVSnonsingular}
\end{center}
\end{figure}

Since we are mainly interested in describing the behavior in the vicinity of $\Njet = 2$, though, for the rest of the paper we will simply adopt the nonsingular definition in \Eq{eq:nonsingdefinition}, even in the vicinity of $\Njet = 3$:
\be
\label{eq:def2p_nonsingular}
\frac{\df\sigma_{\rm ns}}{\df\Njet} (2 < \Njet < 3) = \frac{\df\sigma_{\rm ns}}{\df \dNp} = \frac{\df\sigma_{\rm full}}{\df\Njet} - \frac{\df\sigma_{\rm sing}}{\df\dNp} \,.
\ee
That said, this non-singular term turns out to be dominated by the singular $\dNtm$ piece.  Writing the nonsingular term as
\be
\label{eq:def2p_residual}
\frac{\df\sigma_{\rm ns}}{\df \dNp} = \frac{\df\sigma_{\rm sing}}{\df\dNtm} + \frac{\df\sigma_{\rm res}}{\df\dNp}, 
\ee
the residual term is quite small, even in the vicinity of $\Njet = 2.5$.  This is shown in \Fig{fig:singularVSnonsingular}, where the full QCD result between $2 < \Njet < 3$ is decomposed into the $\dNp$ singular, $\dNtm$ singular, and residual terms for $R = 0.4$ and $\zcut = 0.2$.  The fact that the $\dNp$ nonsingular term is nearly saturated by the $\dNtm$ singular term suggests that higher-order logarithmic terms can play an important role in determining the shape of the $\Njet$ distribution, even for large deviations from integer values of the observable.

\subsection{Complete Results to Order $\as^2$}
\label{sec:completeResultFO}

Summarizing the results from \Secs{singular}{event2}, the full $\ord{\as^2}$ result for the near-integer $\dNpm$ distributions can be written as
\be
\label{eq:full}
\frac{\df\sigma}{\df\dNpm}=\underbrace{\left[\frac{\df\sigma}{\df\dNpm}\right]_{\rm ab}+\left[\frac{\df\sigma}{\df\dNpm}\right]_{\rm nab}+\left[\frac{\df\sigma}{\df\dNpm}\right]_{C_FT_Rn_f}}_{\rm singular} \, + \; \frac{\df\sigma_{\rm ns}}{\df \dNpm}.
\ee
The first three terms in \Eq{eq:full} are the singular contributions to the cross section.  The Abelian terms are given in \Eqs{eq:ab1}{eq:ab2}, while the contributions of non-Abelian and $C_FT_Rn_f$ color structures can be found in \App{deets}.  The last term in \Eq{eq:full} is the nonsingular contribution that we extract numerically from \eventtwo as discussed above.

As discussed in \Eq{eq:def2p_nonsingular}, the entirety of the $\dNtm$ singular distribution in \Eq{eq:ab3} has been absorbed into the $\dNp$ nonsingular distribution.   Thus, when we plot $\dNtm$ in \Sec{MC}, we are in effect plotting the prediction for $\dNtm = 1 - \dNp$.

\section{Towards a Factorization Theorem}
\label{Njet}

In the previous section, we have seen that soft and collinear matrix elements govern the near-integer behavior of $\Njet$ in fixed-order QCD.  In this section, we explore the all-orders behavior at and near $\Njet = 2$, building a candidate factorization theorem to describe the logarithmically-enhanced contributions to the cross section.

Near $\Njet = 2$, energetic collinear modes must be confined to two jet regions of diameter $R$, such that no two energetic particles are more than $R$ apart.  This suggests that the collinear radiation may be described by jet functions.  Additional radiation must be sufficiently soft so as not to create an additional jet, meaning there is a local veto of size $\zcut$ outside of the primary jets.  This suggests that additional radiation may be described by a soft function.  However, while one might hope that a factorization theorem of the form
\be
\label{eq:badfact}
\sigma (\dNpm) \sim \sigma_0 \, H_{q\bq} \Bigl[ J_q (\dNpm) \otimes J_{\bq} (\dNpm) \otimes S_{q\bq} (\dNpm) \Bigr]
\ee
would hold, it fails on several fronts. First, this standard factorization picture is challenged by both the non-additivity of the observable and soft-collinear non-factorization (discussed in \Sec{NGLs} and examined in more detail below). Additionally, the logarithms of $\dNpm$ are non-global, and may not be summed with the above factorization into jet and soft functions. Furthermore, the cross section at $\Njet = 2$ $(\dNpm = 0)$ behaves similarly to a dijet cross section from a standard discrete jet algorithm, for which a convolution structure does not apply.

We will discuss these issues below, en route to a candidate ``local'' factorization theorem which applies in the small $R$ limit. This factorization theorem will take the form
\be \label{factform}
\frac{\df\sigma}{\df\dNpm} = \sigma(\Njet = 2) \Bigl[ C_q (\dNpm) \otimes C_{\bq} (\dNpm) \Bigr] \,,
\ee
where $\sigma(\Njet = 2)$ is the cross section exactly at $\Njet = 2$ (see \Sec{integerNjet}) and $C_{q,\bq}$ are ``collinear functions'' (see \Sec{sec:collinear_functions}).  This form satisfies a number of plausibility checks, but a formal proof of its validity is beyond the scope of the present paper.

\subsection{The Cross Section at $\Njet = 2$}
\label{integerNjet}

We begin by considering the cross section at exactly $\Njet = 2$.  At $\ord{\as}$, the constraints to obtain $\Njet = 2$ are the same as for obtaining two jets from a discrete jet algorithm, namely that either two of the three partons must lie within a mutual radius of $R$ or one of the partons must be below $\Ecut$.  At leading power, the phase space restrictions on the $q \overline{q} g$ final state are
\be
\Theta( \theta_{qg} < R) + \Theta( \theta_{\bq g} < R) + \Theta( \theta_{qg} > R)\Theta( \theta_{\bq g} > R) \Theta (z_g < \zcut) \,.
\ee

At higher orders, the phase space restrictions differ from discrete jet algorithms (which also differ among themselves), and the $\Njet = 2$ cross section has unique features.  Consider an $\ord{\as^2}$ configuration contributing to the near-integer behavior, as discussed in the \Sec{singular}.  A concrete example is the $\dNp$-$\cR_C$ configuration in \Tab{tab:PhaseSpace}, which has a collinear $q\to qg$ splitting with another soft gluon (labeled $s$) emitted such that
\be \label{configscc}
\theta_{qg} < R \,, \quad \theta_{qs} > R \,, \quad \theta_{gs} < R \,.
\ee
This event will give a non-integer contribution to $\Njet$, and its contribution to the cross section is of the form
\be \label{N2contribnear}
\sim \as^2 \biggl[ \frac{1}{\epsilon^2} \delta(\Njet - 2) + \zeta \delta(\Njet - 2) + \kappa_0 \, \cL_0 (\dNp)  + \kappa_1 \, \cL_1 (\dNp) \biggr] \,.
\ee
The plus distribution terms were already calculated in \Eq{eq:ab2}.  The $\delta(\dNp)$ terms were suppressed in our previous discussion (since we were focused on the non-integer behavior), but both the $1/\epsilon^2$ term and the $\zeta$ term follow directly from the calculations in \Sec{singular}.

The structure of \Eq{N2contribnear} reflects the hybrid jet algorithm/event shape behavior of $\Njet$.  The plus distribution terms have no support at $\Njet = 2$, since the plus function prescription removes any divergence there.\footnote{To see this, we can use the definition of the plus function in \Eq{eq:plus_definition}.  Integrating $[q(x)]_+$ against $\delta(x)$ gives the value at 0,
\be
[q(0)]_+ = \int \df x \, [q(x)]_+ \, \delta(x) = 0 \,.
\ee
}
This feature is similar to other event shape variables with a singular limit, such as thrust, where there is zero cross section in the singular limit at higher orders.  By contrast, the delta function terms behave more like a jet algorithm, as anticipated in \Eq{eq:deltafunctioninteger}, with a finite cross section at exactly $\Njet = 2$.  The IR pole from the real radiation at $\Njet = 2$ will cancel divergences from virtual matrix elements, such that the $\Njet = 2$ cross section is IR finite.\footnote{It is straightforward to see that at $\ord{\as^2}$ the cross section at $\Njet = 2$ must be IR finite.  Since non-integer values of $\Njet$, as well as $\Njet = 3$ and 4, have IR finite cross sections, and since the inclusive cross section is IR finite, the remaining piece, the cross section at $\Njet = 2$, must be as well.}  The additional finite $\zeta$ term at $\Njet = 2$ implies that the non-global structure in the near-integer part of the cross section is also contributing at $\Njet = 2$.

\subsection{Soft-Collinear Non-Factorization}
\label{softcollnonfact}

The non-global nature of the near-integer behavior of $\Njet$ is suggestive of a non-standard picture of factorization.  In fact, the $\Njet$ observable does not obey soft-collinear factorization, meaning separate soft and collinear functions cannot be easily (or usefully) defined.  This was foreshadowed in \Sec{NGLs} and can be illustrated with a simple example at $\ord{\as^2}$.

Consider the same phase space configuration as \Eq{configscc}, which has a single soft gluon and a pair of collinear (energetic) partons.  The soft gluon is in the region of phase space where it is in the jet region of only one of the collinear partons, and the value of the observable is dependent on this fact.  This implies that the observable receives contributions that intrinsically depend on the soft and collinear modes in a non-factorizable way; the measurement function cannot be separated into separate soft and collinear pieces that do not depend on each other (see \Fig{fig:phasespace}).  Therefore, soft-collinear factorization does not hold.  We note that, unlike other cases where soft-collinear factorization is not straightforward (see, e.g.,~\cite{Walsh:2011fz,Tackmann:2012bt}), here the \emph{leading} non-integer behavior does not factorize.

Because the non-global structure of $\dNpm$ feeds into the $\Njet = 2$ cross section through the $\zeta$ term in \Eq{N2contribnear}, the same non-factorization is also true of the integer value.  Of course, for the exact $\Njet = 2$ case, non-factorization of the soft and collinear modes happens only at $\ord{\as^2}$.  Therefore, one can write the $\Njet = 2$ cross section as the sum of terms, one with global contributions that can consistently be resummed and one with non-factorizing contributions.  Such a form is
\begin{align}
\label{eq:n2quasi_factorization}
\sigma(\Njet = 2) &= \sigma_0 H_{q\bq} (Q, \mu) J_q (Q, R, \zcut, \mu) J_{\bq} (Q, R, \zcut, \mu) S_{q\bq} (R, \zcut, \mu) \nn \\
& \quad + \sigma_2^{\rm non-fac} (Q, R, \zcut, \mu) \,.
\end{align}
The factorized part of the cross section is similar to dijet (or, generally, exclusive jet) cross sections.  In $e^+e^-$ collisions, such cross sections are known to contain non-global logarithms that spoil a standard effective theory picture of the dynamics.  These non-global logarithms span the jet and soft functions, affecting the RG evolution in nontrivial ways.  Similar to the non-factorizing terms, the non-global contributions start at $\ord{\as^2}$.

\subsection{Introducing Collinear Functions}
\label{sec:collinear_functions}

Let us now try to simultaneously describe the near-integer and exact-integer behavior of $\Njet$.  We have established that the logarithms of $\dNpm$ are purely non-global, and there is no standard soft-collinear factorization.  As mentioned in \Sec{NGLs}, however, the contributions from each jet region are independent, meaning they can be separated:
\be
\label{eq:convolution_basic}
\frac{\df\sigma}{\df\dNpm} \sim \sigma_0 \, C_q (\dNpm) \otimes C_{\bq} (\dNpm) \, ,
\ee
where $\sigma_0$ is an overall prefactor and $\otimes$ refers to the standard convolution in \Eq{eq:convolution_def}.  Here, we have introduced  ``collinear functions'' $C_q$ and $C_{\bq}$ for the two jet regions, which give the contribution to $\dNpm$ for each jet region separately. 

The collinear functions contain only the singular terms and have the general form
\begin{align}
\label{eq:Cq}
C_{q,\bq} (\dNpm) &= \delta(\dNpm) + \sum_{n = 2}^{\infty} \Bigl( \frac{\as}{\pi} \Bigr)^n \sum_{k = -1}^{n-1} \Bigl[ \kappa_{k,+}^{(n)} \, \cL_k (\dNp) + \kappa_{k,-}^{(n)} \, \cL_k (\dNm) \Bigr] \,,
\end{align}
where $\kappa_{k,\pm}^{(n)}$ are coefficients which in general depend on $\zcut$ (except for the leading coefficient $\kappa_{n-1,\pm}^{(n)}$).  Recall that $\cL_{-1} (x) = \delta(x)$, and these delta functions are needed to describe the $\Njet = 2$ cross section.  Since $\dNpm$ describe different regimes of the same observable ($\Njet$ just above/below 2), each term in the expansion has support for one term or the other.

The convolution between the collinear functions  in \Eq{eq:convolution_basic} mixes the distributions for $\dNp$ and $\dNm$.  Therefore, we require not only the usual convolutions between distributions of either $\dNp$ or $\dNm$ (which we refer to as convolutions of one-sided distributions, e.g.\ $\cL_0 (\dNp) \otimes \cL_0 (\dNp)$), but also convolutions between distributions of $\dNp$ and $\dNm$ (which we refer to as convolutions of two-sided distributions, e.g.\ $\cL_0 (\dNp) \otimes \cL_0 (\dNm)$).  Details about the definitions of one- and two-sided distributions and convolutions between them are presented in \App{convs}.

We can now reinterpret the calculation in \Sec{singular} directly as a calculation of $C_q$ and $C_{\bq}$ to $\ord{\as^2}$ (see \App{collinear}). The convolution between $C_q$ and $C_{\bq}$ in \Eq{eq:convolution_basic} then gives part of the higher-order $\ord{\as^4}$ cross section.  In addition, we can estimate the $\ord{\as^4}$ structure of an individual collinear function by performing na\"ive (Abelian) exponentiation to capture some of the multiple emission terms.  That is, we assume that each of the collinear functions in \Eq{eq:Cq} has the form
\be
\label{eq:Cq1}
\begin{split}
C_{q,\bq} (\dNpm)&=\delta(\dNpm)+ \Bigl( \frac{\as(\mu)}{\pi} \Bigr)^2\Bigl[\mathcal{K}_+(\dNp)+\mathcal{K}_-(\dNm)\Bigl] \\
&\quad+\frac{1}{2}\Bigl( \frac{\as(\mu)}{\pi} \Bigr)^4\Bigl[\mathcal{K}_+(\dNp)+\mathcal{K}_-(\dNm)\Bigl]\otimes\Bigl[\mathcal{K}_+(\dNp)+\mathcal{K}_-(\dNm)\Bigl] \,,
\end{split}
\ee
where
\begin{align}
\mathcal{K}_+(\dNp)&=\kappa_{0,+}^{(2)}\cL_0(\dNp)+\kappa_{1,+}^{(2)}\cL_1(\dNp) \,, \\
\mathcal{K}_-(\dNm)&=\kappa_{0,-}^{(2)}\cL_0(\dNm)+\kappa_{1,-}^{(2)}\cL_1(\dNm) \,.
\end{align}
The convolution terms at $\mathcal{O}(\as^4)$ are those coming from na\"ive Abelian exponentiation in one jet region.   Note that this does not fully capture the correct higher-order terms (for example, we do not get any term at $\mathcal{O}(\as^3)$ from this exponentiation), but we will see in \Sec{MC} that it is enough to reproduce some of the higher-order effects observed in parton shower Monte Carlo generators.  In defining \Eq{eq:Cq1}, we have used the fact that one can absorb corrections to the $\Njet=2$ cross section into the $\sigma_0$ prefactor (see further discussion below).

\subsection{A ``Local'' Factorization Theorem}
\label{sec:local_fact}

Via \Eq{eq:convolution_basic}, we can capture the impact on $\Njet$ of soft and collinear emissions within the two jet regions.  But soft emissions away from the jet regions can still give logarithmically-enhanced contributions to the cross section, even if they do not change the value of $\Njet$.   For example, a soft gluon well-separated from the jets with energy fraction less than $\zcut$ will not change $\Njet$, but it will contribute to the cross section.  For $e^+ e^- \to q \bq g$, this wide-angle soft radiation is part of the $\Njet = 2$ phase space.  Going to higher orders, there is a contribution to both the exact-integer and near-integer cross sections from wide-angle soft radiation, and up to power corrections for small $R$, both contributions are identical.

This logic implies that the $\Njet = 2$ cross section should \emph{multiply} the near-integer contributions, leading to the candidate factorization theorem from \Eq{factform}, repeated for convenience:
\be \label{factform_again}
\frac{\df\sigma}{\df\dNpm} = \sigma(\Njet = 2) \Bigl[ C_q (\dNpm) \otimes C_{\bq} (\dNpm) \Bigr] \,. 
\ee
This is the same structure as \Eq{eq:convolution_basic}, but we have identified $\sigma_0$ with $\sigma(\Njet = 2)$.  As discussed in \Eq{eq:n2quasi_factorization}, $\sigma(\Njet = 2)$ itself has its own quasi-factorization theorem.  If this candidate factorization theorem is indeed true, then the definition of $C_{q,\bq} (\dNpm)$ in \Eq{eq:Cq1} should be revised such that the coefficient of the $\delta(\dNpm)$ piece is always 1 (i.e.~the sum over $k$ should start at $0$ instead of $-1$).

We stress here that \Eq{factform} is only a candidate factorization theorem, and we have not proven that such a form exists.  In particular, we do not have an operator definition of the collinear functions $C_{q,\bq}$, and without such a definition, the extraction of $C_{q,\bq}$ is ambiguous from fixed-order calculations alone.  The reason is that the structure of $C_q (\dNpm) \otimes C_{\bq} (\dNpm)$ is single-logarithmic order-by-order, without any definite relation between coefficients, so higher-order terms can absorb corrections from lower-order ones.  For example, the leading nontrivial terms in $\sigma(\Njet = 2)$ (the $\ord{\as}$ term) and $C_q (\dNpm) \otimes C_{\bq} (\dNpm)$ (the $\ord{\as^2 \cL_1}$ terms) will contribute to $\kappa_{1,\pm}^{(3)}$ (the $\ord{\as^3} \cL_1 (\dNpm)$ terms), but cannot unambiguously determine that coefficient.

Despite these limitations, our candidate factorization theorem does describe important effects (like convolutions between the two jet regions) that go beyond a simple perturbative, log series expansion, which is why we will use \Eq{factform} in our comparison studies in \Sec{MC}.  One thing we can say unambiguously is that if the factorization theorem in \Eq{factform} is valid, then the collinear functions must be RG independent.  The reason is that the prefactor $\sigma(\Njet = 2)$ is itself a cross section, so it must be RG independent, and therefore $C_q \otimes C_{\bq}$ must be RG independent.  Similarly, the modes that contribute to $C_{q}$ and $C_{\bq}$ are completely disjoint, so there is no possibility that RG-scale-dependence could cancel between $C_{q}$ and $C_{\bq}$.  This does not rule out, however, a further factorization of the collinear functions.

\subsection{Complete Results}
\label{sec:completeResult}

Let us summarize our final prediction for the $\dNpm$ distributions using the candidate factorization theorem in \Eq{factform}.  Beyond the calculation summarized in \Sec{sec:completeResultFO}, we can include two higher-order effects.

First, as described in \Sec{sec:collinear_functions}, the collinear function approach allows us to capture $\ord{\as^4}$ terms coming from convolutions.  There are two types of convolutions: convolutions between $C_q$ and $C_{\bq}$ given by the factorization theorem, and convolutions within an individual collinear function coming from na\"ive Abelian exponentiation.  Though the na\"ive exponentiation can in principle give results to all orders in $\as$, we will truncate the convolutions to order $\ord{\as^4}$, especially since there are known $\ord{\as^3}$ terms missed by this approach.

Second, we can include running coupling effects by evaluating $\as(\mu)$ at an energy scale $\mu= Q\sqrt{\dNpm}$.  To see why this is the correct scale, note that in \Sec{singular}, the fixed-order expansion in dim reg had a prefactor of
\be
\biggl( \frac{\mu^2}{E_J^2 \dNpm} \biggr)^{2\epsilon} \,,
\ee
where $E_J \simeq Q/2$ is the jet energy.  This implies that at higher orders, the fixed-order calculation generates terms like $\ln(\mu^2/ E_J^2 \dNpm)$ where $\mu$ is the renormalization scale, suggesting that $Q\sqrt{\dNpm}$ is a relevant running coupling energy scale.  

As in \Sec{sec:completeResultFO}, the nonsingular contributions to the cross section enter at fixed $\ord{\as^2}$ and can be simply included additively.  Like in \Sec{sec:completeResultFO}, we absorb the entirety of the $\dNtm$ singular distribution into the $\dNp$ nonsingular distribution.  To try to capture some higher-order effects in $\dNtm$, though, we make use of the nonsingular decomposition in \Eq{eq:def2p_residual}, repeated for convenience:
\be
\label{eq:nonsingularagain}
\frac{\df\sigma_{\rm ns}}{\df \dNp} = \frac{\df\sigma_{\rm sing}}{\df\dNtm} + \frac{\df\sigma_{\rm res}}{\df\dNp}.
\ee
We evaluate the singular $\dNtm$ piece at the scale $\mu= Q\sqrt{\dNtm}$ and the small $\dNp$ residual term at a fixed scale $\mu= Q$.  While this approach misses out on genuine $\ord{\as^3}$ fixed-order corrections from 3-jet events, they have an endpoint at $\dNtm = (\zcut/2)^2$ analogous to \Eq{eq:merged_triplet_phase_space}, so we will simply not make a prediction for $\dNtm < (\zcut/2)^2$.

Finally, while the singular terms for $\dNpm$ come with an overall prefactor of $\sigma(\Njet = 2)$, the nonsingular terms do not.  However, since the nonsingular terms start at $\ord{\as^2}$, we can multiply them by $\sigma(\Njet = 2) / \sigma_0$, which introduces corrections beyond the order to which we are working.  This allows us to factor out a global $\sigma(\Njet = 2)$ from our final predictions.

Putting these pieces together, our final analytic prediction for the $\dNpm$ distribution is:
\begin{align}
\label{eq:fullResult}
\frac{\df\sigma}{\df\dNpm} &= \sigma(\Njet = 2) \, \Bigl\{ C_q (\dNpm) \otimes C_{\bq} (\dNpm) + \frac{1}{\sigma_0} \frac{\df\sigma_{\rm ns}}{\df\dN{2\pm}} \Bigr\} \\
&= \sigma(\Njet = 2)\biggl\{\delta(\dNpm)+2\Bigl(\frac{\as(Q \sqrt{\dNpm})}{\pi} \Bigr)^2\Bigl[\mathcal{K}_+(\dNp)+\mathcal{K}_-(\dNm)\Bigl] \nn \\
&\quad+2\Bigl( \frac{\as(Q\sqrt{\dNpm})}{\pi} \Bigr)^4\Bigl[\mathcal{K}_+(\dNp)+\mathcal{K}_-(\dNm)\Bigl]\otimes\Bigl[\mathcal{K}_+(\dNp)+\mathcal{K}_-(\dNm)\Bigl] \nn \\
& \quad + \frac{1}{\sigma_0} \frac{\df\sigma_{\rm ns}}{\df\dN{2\pm}} \biggl\} \,. \nn
\end{align}

\begin{figure}
\begin{center}
\includegraphics[width=\textwidth]{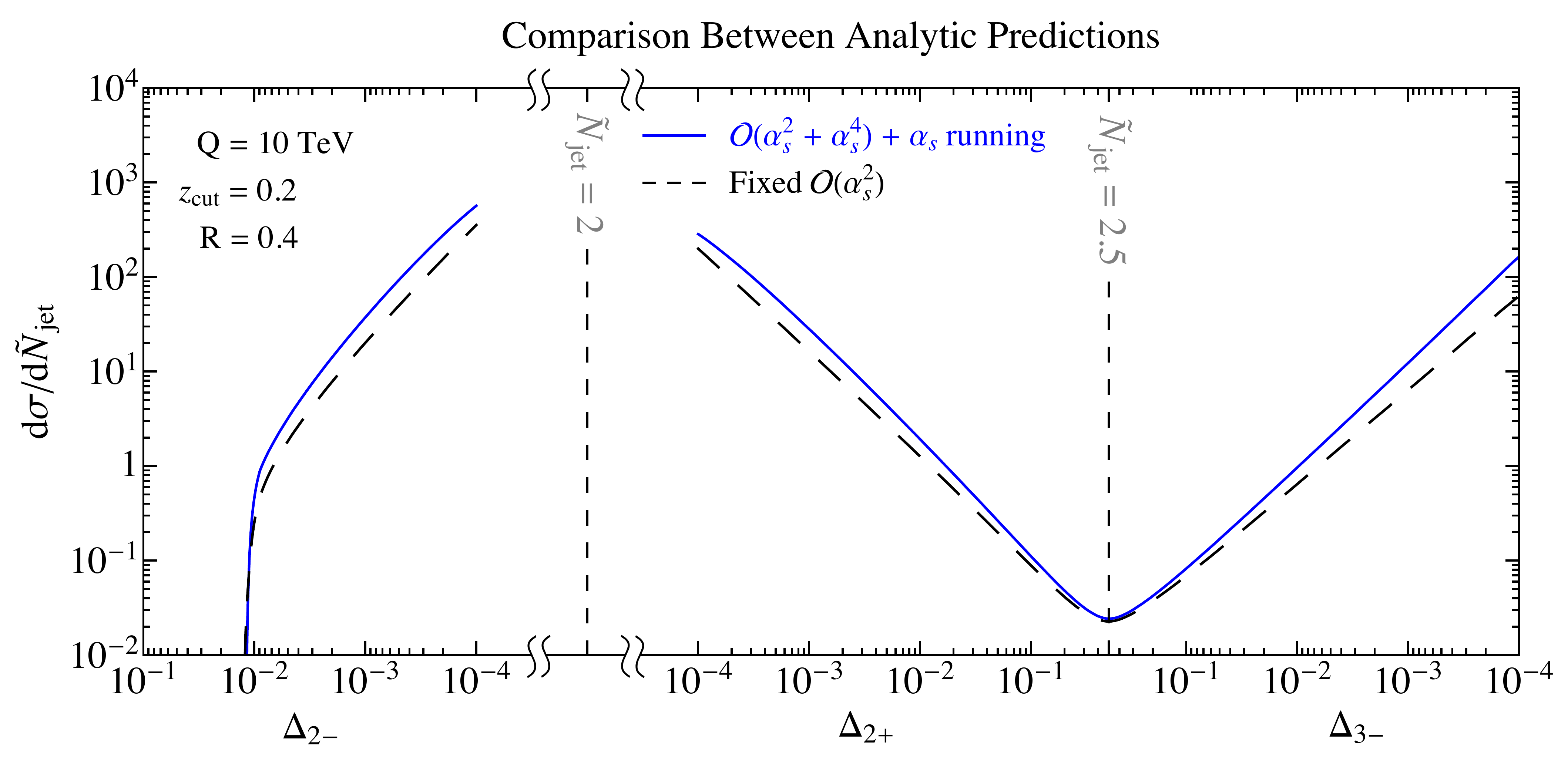}
\caption{A comparison between analytic predictions for the $\Njet$ distribution.  Shown in dashed black are the fixed $\ord{\as^2}$ calculations from \Eq{eq:full}.  Shown in solid blue are our final predictions from \Eq{eq:fullResult}, which include an estimate of higher-order $\ord{\as^4}$ terms as well as $\dNpm$-dependent running of $\as$.  Note that the $\dNm$ and $\dNtm$ axes run backwards.}
\label{compFOresum}
\end{center}
\end{figure}

In \Fig{compFOresum}, we compare the pure fixed $\ord{\as^2}$ distributions for $\dNpm$ and $\dNtm$ from \Eq{eq:full} with the final prediction from \Eq{eq:fullResult}.  At small values of $\dN{n\pm}$, the dominant differences come from the higher-order logarithms included in our final prediction, while at larger values of $\dN{n\pm}$ the running effects from our $\dN{n\pm}$-dependent scale choice give most of the difference from the fixed-order prediction.  Since we do not have a candidate factorization theorem for the $\Njet = 3$ region of phase space, the $\dNtm$ distribution is really just \Eq{eq:fullResult} evaluated at $\dNtm = 1 - \dNp$.

\section{Looking Towards the LHC}
\label{sec:LHC}

Given the interesting features of $\Njet$ and the wide-ranging and robust measurements of jet substructure at the LHC \cite{Abdesselam:2010pt,Altheimer:2012mn,Altheimer:2013yza}, a natural consideration is the $\Njet$ distribution in hadronic collisions.  In this case, it is more convenient to use the original definition of $\Njet$ from \Eq{eq:Njetintro} based on transverse momenta and rapidity-azimuth distances (instead of energies and angles):
\be
\label{eq:NjetLHC}
\Njet(\ptc,R) = \sum_{i\in\ev}\frac{p_{T i}}{p_{T i,R}}\, \Theta(p_{T i,R}-\ptc),
\ee
where $p_{Ti,R}=\sum_j p_{Tj}\,\Theta(R-R_{ij})$ and $R_{ij}=\sqrt{\Delta y_{ij}^2+\Delta\phi_{ij}^2}$.  

At first glance, the calculation of $\Njet$ at the LHC would seem to be much more difficult than at an $e^+e^-$ collider.  After all, $\Njet$ depends sensitively on soft radiation, and soft QCD is notoriously complicated at a hadron collider.  However, $\Njet$ only depends on soft radiation in a region of size $\lesssim 2R$ around energetic partons.  To the extent that we can exploit color coherence at small $R$, we can make predictions for $\Njet$ by simply knowing the collinear functions for quark- and gluon-initiated jets, $C_q$ and $C_g$.  In particular, the $1\to3$ splitting function formalism that we used in \Sec{singular} to calculate the collinear function for quarks in $e^+e^-$ may also be used to calculate the quark and gluon collinear functions for LHC processes.

Conveniently, when switching between the $e^+ e^-$ and hadronic definitions of $\Njet$, the collinear functions $C_{q,g}$ only differ by terms of $\mathcal{O}(R)$.  For small jet radii, one can therefore neglect those corrections and use the $e^+ e^-$ collinear functions calculated in this paper also for the hadronic case, though one has to be careful to match the $\zcut$ value to the outgoing parton momentum.  To see why this is the case, notice that 
\be
z_i=\frac{E_i}{E_J}\simeq\frac{p_{Ti}}{p_{TJ}}[1+\mathcal{O}(R)], \qquad \zcut = \frac{\Ecut}{E_J}\simeq\frac{\ptc}{p_{TJ}}[1+\mathcal{O}(R)], 
\ee
where $p_{TJ}$ is the scalar sum of transverse momenta in a jet region.  Thus, the energy integrals in $C_{q,g}$ would differ at most by $\mathcal{O}(R)$ terms.  The angular phase space regions in \Eq{eq:regions} would be written as constraints on $R_{ij}$ instead of $\theta_{ij}$, but notice that
\be
R_{ij}=\theta_{ij}(\cosh y_i\cosh y_j)^{1/2}\simeq\theta_{ij}\cosh y_J[1+\mathcal{O}(R)],
\ee
where $y_J$ is some characteristic rapidity associated with the jet region, for example the rapidity of the summed jet region momenta.  At leading power, the logarithmic scaling of the angular integrals can be used to remove any dependence on $\cosh y_J$, so that the differences in the angular integrals are again at most an $\mathcal{O}(R)$ effect.

In order to extend the candidate dijet factorization formula in \Eq{factform} to dijet events from hadronic collisions, we have to sum over all relevant partonic channels:
\be
\label{eq:candidatehadronicfactorization}
\frac{\df\sigma}{\df\dNpm} = \sum_{k,\ell} \sigma_{h_1h_2\to k \ell}(\Njet = 2)\,C_{k} (\dNpm) \otimes C_{\ell} (\dNpm) \,,
\ee
where $k, \ell=q$, $\bar{q}$, or $g$, and $h_{1,2}$ represents the colliding hadrons.  In general, $\sigma_{h_1h_2\to k \ell}$, $C_k$, and $C_\ell$ depend on the outgoing parton kinematics, though we have suppressed that dependence in \Eq{eq:candidatehadronicfactorization} for readability.  Note that the incoming beams only create additional high-$p_T$ jets through hard, perturbative emissions.  We can write the total rate $\sigma_{h_1h_2\to k \ell}(\Njet = 2)$ schematically as
\be
\sigma_{h_1h_2\to k \ell}(\Njet = 2)=\sum_{a,b}\int_0^1 \df x_a\,  \df x_b\, f_{a/h_1}(x_a)f_{b/h_2}(x_b)\,\sigma_{ab\to k \ell}(\Njet = 2),
\ee
where $f_{a/h_1}$ and $f_{b/h_2}$ are parton distribution functions for partons $a$ and $b$ carrying momentum fractions $x_{a,b}$ of the initial hadrons $h_{1,2}$, and $\sigma_{ab\to k \ell}(\Njet = 2)$ is the total rate at $\Njet = 2$ for the partonic channel $ab\to k \ell$.  As in the $e^+ e^-$ case, this cross section can have large logarithms of $R$ and $\ptc$ that require resummation to obtain reliable predictions.

Compared to the $e^+ e^-$ case, the main new ingredient is the collinear function $C_g(\dNpm)$ for a gluon-initiated jet.  For completeness, we have calculated the gluon collinear function using the same approach as \Sec{singular}, and we present both $C_q$ and $C_g$ in \App{collinear}.  The structure in \Eq{eq:candidatehadronicfactorization} can be easily extended to handle the $\dN{n\pm}$ cross section for $n$-jet processes by using $n$ collinear functions multiplied by the $\Njet = n$ cross section, appropriately summed over the various partonic channels.

At a hadron collider, there is also a new potential source of non-global logarithms from initial state partons.  In general, non-global logarithms of $\dN{n\pm}$ only appear in the collinear functions $C_{q,g}$ for each final state jet, and are associated with correlated emissions from final state partons.  For events with well-separated jets, the collinear functions are universal and independent of other jets in the event.  However, there are also non-global logarithms of $\zcut$ and $R$ which appear both in the collinear functions and in the exact integer $\Njet$ cross section (e.g.~$\sigma_{h_1h_2\to k \ell}(\Njet = 2)$, see \Eq{eq:n2quasi_factorization}).  For the exact integer cross section at a hadron collider, one also has to include correlated emissions from initial state partons, which may introduce super-leading logarithms \cite{Forshaw:2006fk,Forshaw:2008cq,Keates:2009dn} at sufficiently high $\alpha_s$ order.  Ideally, one would want to understand the resummation of non-global effects in the collinear functions and the exact integer cross section to achieve accurate predictions.

Because a detailed study for the LHC is beyond the scope of this paper, we will not present any results for hadronic collisions.  An important effect to account for in future LHC predictions and comparisons with measurements is hadronization, which we have not considered here.  Additionally, for hadronic collisions, effects from the underlying event and initial state radiation are not present in the $e^+e^-$ case.  To lowest order in $R$, we expect those effects to be captured by color coherence, and one could imagine using the techniques of \Ref{Stewart:2014nna} to understand the $R$ dependence.  A similar concern is pileup contamination, though the closed-form nature of $\Njet$ makes it well-suited to analytic studies using area subtraction \cite{Cacciari:2007fd,Cacciari:2008gn,Soyez:2012hv}.  One could further mitigate the impact of pileup by using a version of $\Njet$ that includes jet trimming \cite{Krohn:2009th} in closed form \cite{Bertolini:2013iqa}, and we expect the collinear function for the trimmed $\Njet$ version could be calculated using the same techniques used here.

\section{Monte Carlo Comparisons}
\label{MC}

We now compare our analytic prediction in \Eq{eq:fullResult} with parton shower Monte Carlo generators.   As discussed in \Sec{sec:local_fact}, the rate $\sigma(\Njet = 2)$ only enters as an overall normalization factor, and it in principle requires resummation of logarithms of $R$ and $\zcut$ to obtain a reliable prediction.  Since we are mainly interested in describing only the near-integer (and not the exact-integer) behavior, we will divide out by an overall normalization factor and perform only a shape comparison.  In all of the plots below, we normalize the cross section to the region $\dNp \in (10^{-4}, 10^{-2})$.

To make an apples-to-apples comparison of our analytic results to \pythia and \herwig, we turn off hadronization in the Monte Carlo generators.  In principle, the collinear functions $C_q$ and $C_{\bq}$ should get hadronization corrections, but the non-factorizing nature of $\Njet$ means that we cannot adopt a standard shape function analysis \cite{Korchemsky:1999kt,Korchemsky:2000kp}.   We run at a sufficiently high energy to allow for a comparison over a wide logarithmic range in $\dNpm$.  The large collision energy we choose, $Q = 10$ TeV, also mitigates the effect of low energy cutoffs ($\simeq 1$ GeV) on the parton shower.  

Because the singular terms in our calculation are numerically dominant for small $\dNpm$, higher-order logarithms may be important in determining the shape of the distribution in this regime.  Some of the $\ord{\as^4}$ effects are captured by the convolution structure in \Eq{eq:fullResult}, but they are not fully reliable.  Therefore, we include an uncertainty in our predictions derived from the addition of $\ord{\as^3 \cL_2 (\dNpm)}$ and $\ord{\as^4 \cL_3 (\dNpm)}$ terms with unknown coefficients that we vary.  These terms represent the leading logarithms at each order, and are an estimate of missing higher-order terms.  The form we use to determine the uncertainty is
\be
\label{eq:uncertainty}
\frac{\df\sigma}{\df\dNpm}\supset \Bigl(\frac{\as}{\pi} \Bigr)^3\kappa^{(3)}_{2,\pm}\cL_2(\dNpm)+\Bigl(\frac{\as}{\pi} \Bigr)^4\kappa^{(4)}_{3,\pm}\cL_3(\dNpm),
\ee
and we vary independently the coefficients in the ranges $\kappa^{(3)}_{2,\pm} \in (-5,5)$ and $\kappa^{(4)}_{3,\pm} \in (-15,15)$.  These coefficients are of similar size as the leading coefficients $\kappa^{(2)}_{1,\pm}$ in our calculation, and the $\ord{\as^4}$ term generated by the na\"ive exponentiation in \Eq{eq:fullResult} has $\kappa^{(4)}_{3,\pm} \sim \ord{10}$.  The uncertainty band is given by the envelope of these variations, including effects both on the cross section normalization (where we just add \Eq{eq:uncertainty}) and on the cross section shape (where we add \Eq{eq:uncertainty} and readjust the normalization in the $\dNp \in (10^{-4}, 10^{-2})$ window).

\begin{figure}
\begin{center}
\includegraphics[width=0.75\textwidth]{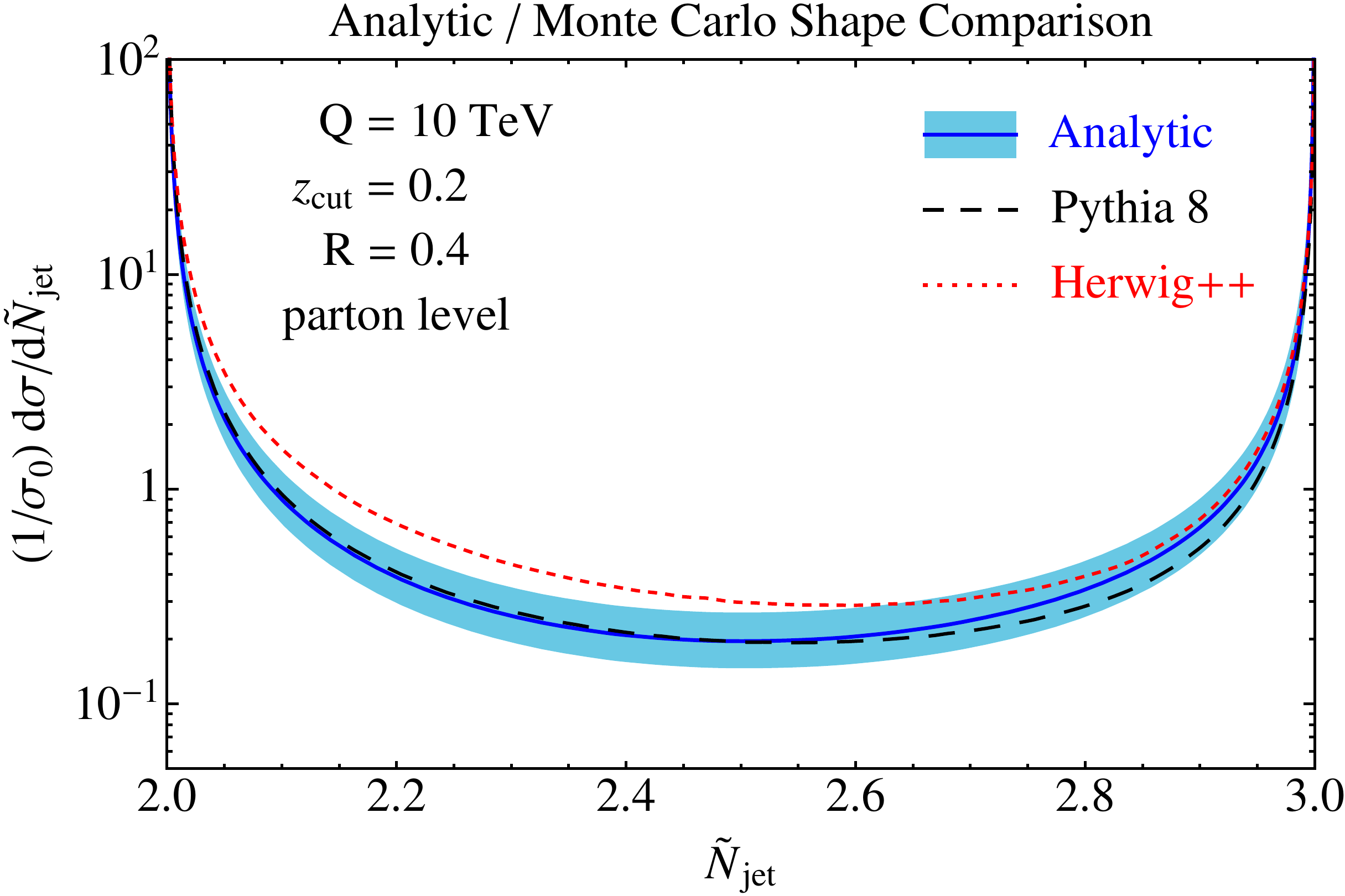}
\caption{A shape comparison between our analytic predictions for the $2 < \Njet < 3$ distribution with the Monte Carlo generators \pythia and \herwig.  Our calculation is summarized in \Eq{eq:fullResult} with uncertainties given in \Eq{eq:uncertainty}.  To make an apples-to-apples comparison, we include showering but not hadronization in the Monte Carlo samples.  All cross sections are normalized to have the same value in the $\dNp \in (10^{-4}, 10^{-2})$ window.}
\label{MCcomp1}
\end{center}
\end{figure}
\begin{figure}
\begin{center}
\includegraphics[width=\textwidth]{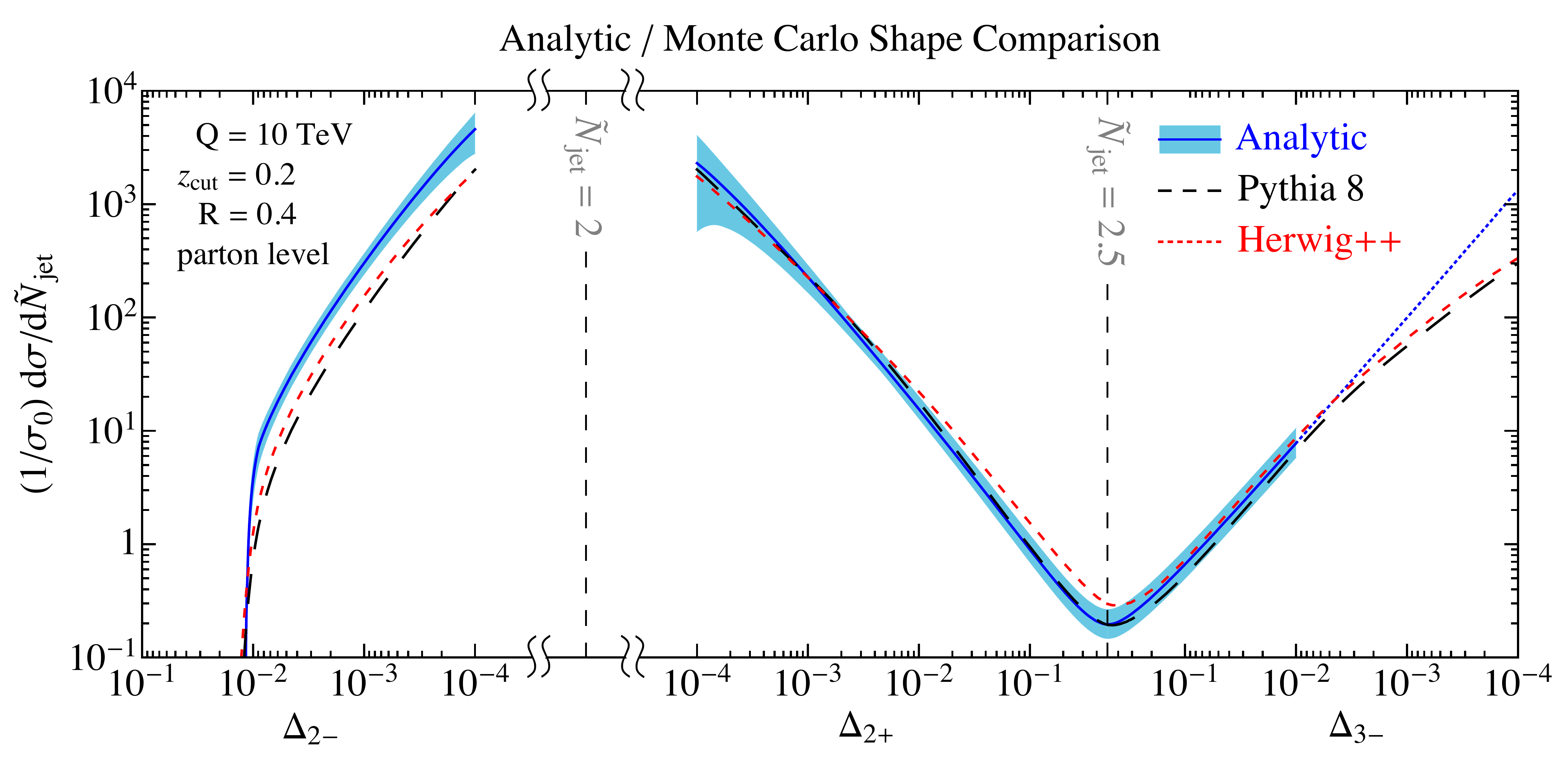}
\caption{Same distributions as \Fig{MCcomp1}, but in triptych form to highlight the near integer behavior.  Because we normalize only to the $\dNp \in (10^{-4}, 10^{-2})$ window, the normalization differences in the $\dNm$ region are accentuated; the shape agreement is much better.  For the $\dNtm$ distribution, we extend our central prediction (with a dotted blue line) to $\dNtm < (\zcut/2)^2$, where genuine 3-jet events, whose contribution we have not calculated, contribute to the observable.  Note that the $\dNm$ and $\dNtm$ axes run backwards.}
\label{MCcomp2}
\end{center}
\end{figure}

In \Fig{MCcomp1}, we show the shape comparison between our result and \pythia/\herwig over the range $2 < \Njet < 3$.  The singular cross section for $\dNtm$ is included as part of the nonsingular correction to the $\dNp$ distribution (see \Eq{eq:nonsingularagain}).  Overall, we find good agreement with the Monte Carlo generators within uncertainties.  It is amusing that there is such close agreement with \pythia in the $\Njet\to 2$ region and with \herwig in the $\Njet\to 3$ region, with the analytic result effectively interpolating between the two.  Given that the $\dNp$ and $\dNtm$ parts of the distribution are dominated by different phase space configurations (see \Tab{tab:PhaseSpace}), it is possible that we are seeing the impact of different shower ordering variables in \pythia ($p_\perp$-ordered) versus \herwig (angular-ordered).  Of course, the theoretical uncertainties in our calculation are too large to make a definitive statement.

In \Fig{MCcomp2}, we compare distributions of $\dNpm$ over the range $(10^{-4}, 1)$, using the triptych format to also see the $\dNtm$ region.   Again, our analytic prediction reproduces the shape of the \pythia and \herwig distributions remarkably well over the whole range.  The Monte Carlo distributions, which include higher-order logarithms of $\dNpm$ through multiple emissions, generally lie within the uncertainty band of our prediction for $\Njet > 2$.  This indicates that we have made a reasonable estimate of the higher-order corrections not included in our calculation.  Because we normalize to the window $\dNp \in (10^{-4}, 10^{-2})$, it is not surprising that there is a normalization discrepancy in $\dNm$, though one can see that the shape agreement for $\Njet < 2$ is excellent.  We suspect that the normalization issues for $\Njet < 2$ may be due to the treatment of correlated soft radiation (and angular ordering) in the Monte Carlo generators, since the $\dNm$ distribution is dominated by double-soft emissions with a $C_F C_A$ color structure.  The overall shape agreement between \pythia and \herwig is quite surprising, given that they exhibit a factor of 3 difference (not shown) in their predicted value of $\sigma(\Njet = 2)$, suggesting that the shape of the non-integer distributions is more robust than the cross section at integer values.

\section{Conclusions}
\label{conclusions}

In this paper, we studied the analytic properties of $\Njet$, a jets-without-jets event shape that can return a fractional value of jet multiplicity.  Focusing on $e^+e^-\to\text{jets}$, we calculated the distribution of $\Njet$ in the vicinity of a dijet configuration at $\ord{\as^2}$.  A fractional number of jets requires at least three collimated partons, such that for $e^+e^-\to\text{jets}$, the first non-trivial contribution requires at least two emissions.  The singular parts of this emission structure can be captured using $1 \to 3$ splitting functions, and we validated this approach (and included nonsingular contributions) using the fixed-order code \eventtwo.  To partially capture higher-order effects out to $\ord{\as^4}$, we included convolutions from different phase space regions and running coupling effects.  We found very good agreement between our calculation and Monte Carlo distributions from both \pythia\ and \herwig.  

Fractional jet multiplicity exhibits unique analytic features that are not shared by other jet observables.  At $\ord{\as^2}$, we showed how rapidity-like divergences, related to the energy sharing between emissions, appear when one or two partons become soft, and we explained how to regulate them.  Beyond the fixed-order result, we proposed a candidate local factorization theorem and used it to predict a hybrid jet algorithm/event shape behavior for the $\Njet$ distribution in the vicinity of an integer.  At exact-integer values, $\Njet$ behaves much like a standard jet algorithm, yielding spikes in the $\Njet$ cross section.  The near-integer behavior of $\Njet$  is more characteristic of event shapes, where towers of higher-order logarithms give rise to (single-logarithmic) suppression of the singular phase space, yielding shoulders in the $\Njet$ cross section.  Finally, as opposed to typical event shapes, collinear emissions do not generate logarithms of $\Njet$, so the shape of the near-integer distribution is entirely determined by soft logarithms.  These soft logarithms are purely non-global, as near-integer values force soft partons to lie in a  restricted angular region of phase space and correlate different emissions (even if generated from an Abelian matrix element).

Beyond our present understanding of $\Njet$, there are three key directions to pursue.  The first is to extend our calculations to $\ord{\as^3}$.  Though the differential cross section shapes were largely within our uncertainty estimates, the normalization differences seen between our analytic calculations and Monte Carlo generators suggest that higher-order terms might be relevant.  While we were able to estimate some $\ord{\as^4}$ effects through convolutions, there are genuine $\ord{\as^3}$ effects that may get a phase space enhancement in the merged jet region (relative to the $\ord{\as^2}$ phase space) to partially overcome the $\as$ suppression.  The second is to attempt an understanding of logarithmic resummation.  Because of the non-global nature of the observable, standard renormalization group methods will not work, but there may be a way to exploit the rapidity-like divergences seen in \Sec{rapidity} or the techniques introduced in \Refs{Larkoski:2015zka,Caron-Huot:2015bja}.  The third is to perform a detailed study for hadron colliders like the LHC, as discussed in \Sec{sec:LHC}, including the important effect of hadronization.  Fractional jets offer a more nuanced understanding of jet formation than is possible with standard jet algorithms, and a measurement of $\Njet$ seems feasible given the increasingly sophisticated approach to jet physics at the LHC.

\acknowledgments{We thank Andrew Larkoski for helpful discussions.  D.B.~and J.R.W.~are supported by the Office of Science, Office of High Energy Physics, of the U.S.~Department of Energy (DOE) under Contract No.~DE-AC02-05CH11231, and J.T.~is supported under grant Contract No.~DE-SC00012567.  J.T. is also supported by the DOE Early Career research program DE-FG02-11ER-41741 and by a Sloan Research Fellowship from the Alfred P. Sloan Foundation.  This research used resources of the National Energy Research Scientific Computing Center, which is supported by the Office of Science of the DOE under Contract No.~DE-AC02-05CH11231.}

\appendix

\section{Results for Non-Abelian Contributions}
\label{deets}

Following the Abelian results in \Sec{singular}, here we show the singular $C_F C_A$ and $C_F T_R n_f$ contributions to the $\dNpm$ and $\dNtm$ cross sections.  The $C_F C_A$ terms are
\begin{align}
\left[\frac{\df\sigma}{\df\dNm}\right]_{\rm nab} &= \left(\frac{\as}{\pi}\right)^2 C_FC_A \biggl\{ -\frac{14}{5} \cI_\Omega \mathcal{L}_1(\dNm) \\
& \qquad\qquad\qquad\qquad + \biggl[ \frac{28}{5} \cI_\Omega \ln \zcut + \cI_\Omega^{(a)} \biggr] \mathcal{L}_0(\dNm) \biggr\} \Theta( \dNm < \zcut^2 / 4 ) \,, \nn \\
\left[\frac{\df\sigma}{\df\dNp}\right]_{\rm nab} &= \left(\frac{\as}{\pi}\right)^2 C_FC_A \biggl\{ \frac25 \cI_\Omega \mathcal{L}_1(\dNp) \\
& \qquad\qquad\qquad\qquad + \biggl[ \biggl( -\frac32 - \frac25 \ln 2 - \frac{14}{5} \ln \zcut \biggr) \cI_\Omega + \cI_\Omega^{(b)} \biggr] \mathcal{L}_0(\dNp) \biggr\} \,, \nn \\
\left[\frac{\df\sigma}{\df\dNtm}\right]_{\rm nab} &= \left(\frac{\as}{\pi}\right)^2 C_FC_A \cI_\Omega \biggl( -\frac{21}{10} - \frac{14}{5} \ln \zcut \biggr) \mathcal{L}_0(\dNtm) \,,
\end{align}
where $\cI_\Omega$ is defined in \Eq{eq:IOmega}, and the remaining angular integrals can only be done numerically:
\be
\label{eq:angfactorAB}
\cI_\Omega^{(a)} = -2.44393 \,, \qquad \cI_\Omega^{(b)} = -0.035397 \,.
\ee
The $C_F T_R n_f$ terms are
\begin{align}
\left[\frac{\df\sigma}{\df\dNm}\right]_{C_F T_R n_f} &= \left(\frac{\as}{\pi}\right)^2 C_F T_R n_f \cI_\Omega^{(c)} \mathcal{L}_0(\dNm) \Theta( \dNm < \zcut^2 / 4 ) \,, \\
\left[\frac{\df\sigma}{\df\dNp}\right]_{C_F T_R n_f} &= \left(\frac{\as}{\pi}\right)^2 C_F T_R n_f \cI_\Omega^{(d)} \mathcal{L}_0(\dNp) \,,
\end{align}
with the angular integrals
\be
\label{eq:angfactorCD}
\cI_\Omega^{(c)} = 0.724689 \,, \qquad \cI_\Omega^{(d)} = 0.251525\,.
\ee
The $C_F T_R n_f$ contribution to $\dNtm$ is purely power suppressed.

\section{Properties of Distributions}
\label{convs}

In this appendix we collect useful formulae for the convolution of one-sided distributions, and then discuss two-sided distributions and their convolutions.  The relations given here are straightforwardly derived from the results and techniques in \Ref{Ligeti:2008ac}.

Throughout this paper, we use the standard plus distribution notation.   For a function $q$,
\be
\label{eq:plus_definition}
[q(x)]_+ = \lim_{\beta \to 0} \frac{\df}{\df x} \bigl[ \Theta(x - \beta) Q(x) \bigr] \,, \qquad Q(x) = \int_1^x \df x' q(x') \,.
\ee
This distribution has a boundary at 1, so that
\be
\int_{-\infty}^1 \df x \, [q(x)]_+ = 0 \,.
\ee
We also use the shorthand
\be
\label{eq:plusfunctiondef}
\cL_n (x) = \biggl[ \frac{\ln^n x}{x} \Theta(x) \biggr]_+ \text{ for } n \text{ an integer } \ge 0 \,, \qquad \cL_{-1} (x) = \delta(x) \,,
\ee
as well as the distribution
\be \label{powerdist}
\bigl[ \Theta(x) \, x^{-1+\beta} \bigr]_+^{[\infty]} = \frac{1}{\beta} \delta(x) + \sum_{n = 0}^{\infty} \frac{1}{n!} \beta^n \cL_n (x) \,.
\ee
The $[\infty]$ notation on the plus distribution indicates that the boundary is at $\infty$ (instead of 1), so that
\be
\int_{-\infty}^{\infty} \bigl[ \Theta(x) \, x^{-1+\beta} \bigr]_+^{[\infty]} = 0 \,.
\ee

\subsection{One-Sided Distributions and their Convolutions}
\label{1sidedconvs}

For many applications, one often makes use of convolutions between plus distributions, especially $\cL_n$.  The convolution is defined as
\be
\label{eq:convolution_def}
(f \otimes g) (x) \equiv \int_{-\infty}^{\infty} \df x' \, \df x'' \, \delta(x - x' - x'') \, f (x') \, g (x'') \,. 
\ee
We can take the Fourier transform $\cF$ to make the convolution multiplicative, and also use the Fourier transform and its inverse to determine a convolution:
\be
\cF \bigl\{ f \otimes g \bigr\} = \cF \bigl\{ f \bigr\} \cF \bigl\{ g \bigr\} \qquad \Rightarrow \qquad f \otimes g = \cF^{-1} \bigl\{ \cF \bigl\{ f \bigr\} \cF \bigl\{ g \bigr\} \bigr\} \,.
\ee

It is straightforward to determine the convolution between two plus functions $\cL_k$ and $\cL_n$ by relating it to \Eq{powerdist}:
\be
\bigl( \cL_k \otimes \cL_n \bigr) (x) = \text{the } \ord{\frac{\alpha^k}{k!} \frac{\beta^n}{n!}} \text{ coefficient of } \Bigl( \bigl[ \Theta(x) \, x^{-1 + \alpha} \bigr]_+^{[\infty]} \Bigr) \otimes \Bigl( \bigl[ \Theta(x) \, x^{-1 + \beta} \bigr]_+^{[\infty]} \Bigr).
\ee
Using the Fourier transform
\be
\cF \bigl\{ \bigl[ \Theta(x) \, x^{-1+\beta} \bigr]_+^{[\infty]} \bigr\} = \Gamma[\beta] (-is)^{-\beta} \,,
\ee
with $s$ as the conjugate variable, it is straightforward to show
\begin{align} \label{onesidedconv}
&\bigl( \cL_k \otimes \cL_n \bigr) (x) = \text{the } \ord{\frac{\alpha^k}{k!} \frac{\beta^n}{n!}} \text{ coefficient of } \frac{\Gamma(\alpha) \Gamma(\beta)}{\Gamma(\alpha + \beta)} \, \bigl[ \Theta(x) \, x^{-1 + \alpha + \beta} \bigr]_+^{[\infty]} \\
&\qquad\qquad= \text{the } \ord{\frac{\alpha^k}{k!} \frac{\beta^n}{n!}} \text{ coefficient of } \frac{\Gamma(\alpha) \Gamma(\beta)}{\Gamma(\alpha + \beta)} \biggl[ \frac{1}{\alpha + \beta} \delta(x) + \sum_{m = 0}^{\infty} \frac{1}{m!} (\alpha+\beta)^m \cL_m (x) \biggr] \,. \nn
\end{align}
This approach is similar to the one used in \Ref{Ligeti:2008ac} to give the general form of the convolution $(\cL_k \otimes \cL_n) (x)$, where an equivalent result to \Eq{onesidedconv} is given.

We give the first few convolutions here for convenience:
\begin{align}
\cL_0 (x) \otimes \cL_0 (x) &= 2 \cL_1 (x) - \frac{\pi^2}{6} \delta(x) \,, \nn \\
\cL_0 (x) \otimes \cL_1 (x) &= \frac32 \cL_2 (x) - \frac{\pi^2}{6} \cL_0(x) + \zeta_3 \delta(x) \,, \nn \\
\cL_1 (x) \otimes \cL_1 (x) &= \cL_3 (x) - \frac{\pi^2}{3} \cL_1 (x) + 2\zeta_3 \cL_0 (x) - \frac{\pi^4}{360} \delta(x) \,.
\end{align}
These distributions only have support for $x > 0$, and appear in many applications. 

\subsection{Two-Sided Distributions and their Convolutions}
\label{2sidedconvs}

In this work, we encountered observables whose cross sections have support for all $x$ and behave like distributions as $x \to 0^\pm$.  We will refer to them as \emph{two-sided distributions} (one may think of the usual distributions as one-sided), and we now discuss convolutions of them.

Consider an observable $x$ with singular behavior for $x\to 0^\pm$.  The fixed-order singular behavior of $x$ is described by distributions $\cL_n ( x_\pm)$, where $x_+ = x$ for $x \ge 0$ and $x_- = -x$ for $x \le 0$, so that
\begin{align}
\cL_n (x_+) &= \cL_n (x) = \biggl[ \frac{\ln^n x}{x} \, \Theta(x) \biggr]_+ \,, \nn \\
\cL_n (x_-) &= \cL_n (-x) = \biggl[ \frac{\ln^n (-x)}{-x} \, \Theta(-x) \biggr]_+ \,.
\end{align}
Convolutions between various $\cL_n (x_\pm)$ will mix the $x_\pm$ distributions.

First, we note that convolutions between two $x_+$ distributions or two $x_-$ distributions are effectively one-sided, meaning the above results can be used without modification.  Convolutions between an $x_+$ and an $x_-$ distribution are the novel ones we derive here, adapting the technique in \App{1sidedconvs} to find the general form of two-sided convolutions.  

Using the Fourier transforms of the $x_\pm$ distributions
\begin{align} \label{FTpmconv}
\cF \bigl\{ \bigl[ \Theta(x_\pm) \, x_\pm^{-1+\beta} \bigr]_+^{[\infty]} \bigr\} = \Gamma(\beta) (\mp is)^{-\beta} \,,
\end{align}
we have
\begin{align}
&\cF \biggl\{ \Bigl( \bigl[ \Theta(x_+) \, x_+^{-1+\alpha} \bigr]_+^{[\infty]} \Bigr) \otimes \Bigl( \bigl[ \Theta(x_-) \, x_-^{-1+\beta} \bigr]_+^{[\infty]} \Bigr) \biggr\} = \Gamma(\alpha) \Gamma(\beta) (-i s)^{-\alpha} (i s)^{-\beta} \\
& \qquad\qquad = \frac{\Gamma(b) \Gamma(1 - (a+b))}{\Gamma(1-a)} \, \Gamma(a+b) (-i s)^{-(\alpha+\beta)} + \frac{\Gamma(a) \Gamma(1 - (a+b))}{\Gamma(1-b)} \, \Gamma(a+b) (i s)^{-(\alpha+\beta)} \,. \nn
\end{align}
We can easily Fourier invert the right-hand side given \Eq{FTpmconv}.  This implies
\begin{align} \label{twosidedconv}
&\bigl( \cL_k (x_+) \otimes \cL_n (x_-) \bigr) (x_\pm) \nn \\
& \qquad = \text{the } \ord{\frac{\alpha^k}{k!} \frac{\beta^n}{n!}} \text{ coefficient of } \\
& \qquad\qquad \frac{\Gamma(b) \Gamma(1 - (a+b))}{\Gamma(1-a)} \bigl[ \Theta(x_+) \, x_+^{-1 + \alpha + \beta} \bigr]_+^{[\infty]} + \frac{\Gamma(a) \Gamma(1 - (a+b))}{\Gamma(1-b)} \bigl[ \Theta(x_-) \, x_-^{-1 + \alpha + \beta} \bigr]_+^{[\infty]} \nn \\
&\qquad = \text{the } \ord{\frac{\alpha^k}{k!} \frac{\beta^n}{n!}} \text{ coefficient of } \nn \\
& \qquad\qquad\qquad \frac{\Gamma(b) \Gamma(1 - (a+b))}{\Gamma(1-a)} \biggl[ \frac{1}{\alpha + \beta} \delta(x_+) + \sum_{m = 0}^{\infty} \frac{1}{m!} (\alpha+\beta)^m \cL_m (x_+) \biggr] \nn \\
& \qquad\qquad\quad\!\! + \frac{\Gamma(a) \Gamma(1 - (a+b))}{\Gamma(1-b)} \biggl[ \frac{1}{\alpha + \beta} \delta(x_-) + \sum_{m = 0}^{\infty} \frac{1}{m!} (\alpha+\beta)^m \cL_m (x_-) \biggr] \,.
\end{align}

Using this relation, the first few nontrivial convolutions for two-sided distributions are
\begin{align}
\cL_0 (x_+) \otimes \cL_0 (x_-) &= \cL_1 (x_+) + \cL_1 (x_-) + \frac{\pi^2}{3} \delta(x) \,, \nn \\
\cL_1 (x_+) \otimes \cL_0 (x_-) &= \cL_2 (x_+) + \frac{\pi^2}{6} \cL_0 (x_+)  \nn \\
& \quad + \frac12 \cL_2 (x_-) + \frac{\pi^2}{6} \cL_0 (x_-) + \zeta_3 \, \delta(x) \,, \nn \\
\cL_0 (x_+) \otimes \cL_1 (x_-) &= \frac12 \cL_2 (x_+) + \frac{\pi^2}{6} \cL_0 (x_+) \nn \\
& \quad + \cL_2 (x_-) + \frac{\pi^2}{6} \cL_0 (x_-) + \zeta_3 \, \delta(x) \,, \nn \\
\cL_1 (x_+) \otimes \cL_1 (x_-) &= \frac12 \cL_3 (x_+) + \frac{\pi^2}{3} \cL_1 (x_+) + \zeta_3 \, \cL_0 (x_+) \nn \\
& \quad + \frac12 \cL_3 (x_-) + \frac{\pi^2}{3} \cL_1 (x_-) + \zeta_3 \, \cL_0 (x_-) + \frac{7\pi^4}{180} \delta(x) \,.
\end{align}

\section{Quark and Gluon Collinear Functions}
\label{collinear}

We summarize here results for the collinear functions for quark- and  gluon-initiated jets through $\ord{\as^2}$.  We express our results for dijet observables ($\dNpm$ and $\dNtm$), but they can be applied to the more general $n$-jet case of $\dN{n\pm}$ as well.  The quark collinear function can be derived from the calculations in \Sec{singular} and \App{deets}.   The gluon collinear function calculation proceeds in analogy with the quark case, and there is a strong relation between the quark and gluon results for nearly all coefficients.

To $\ord{\as^2}$, the quark and gluon collinear functions contribute to $\dNpm$ and $\dNtm$, with
\begin{align}
C_i (\dNm) &= \delta(\dNm) + \Bigl(\frac{\as}{\pi} \Bigr)^2 \cK^{(i)}_{2-} (\dNm) \,, \\
C_i (\dNp) &= \delta(\dNp) + \Bigl(\frac{\as}{\pi} \Bigr)^2 \cK^{(i)}_{2+} (\dNp) \,, \nn \\
C_i (\dNtm) &= \Bigl(\frac{\as}{\pi} \Bigr)^2 \cK^{(i)}_{3-} (\dNtm) \,, \nn
\end{align}
where $i = q, g$ for the different parton types, and the subscripts on $\cK^{(i)}$ indicate the contribution of an individual jet region to $\Njet$.  The $\ord{\as^2}$ terms are given by
\begin{align}
\cK^{(q)}_{2-} (\dNm) &= \biggl\{ \biggl( - C_F^2 - \frac75 C_F C_A \biggr) \cI_\Omega \cL_1 (\dNm) + \biggl( 2 \ln \zcut C_F^2 + \frac{14}{5} \ln \zcut C_F C_A \biggr) \cI_\Omega \cL_0 (\dNm) \nn \\
& \qquad + \biggl( \frac12 \cI_\Omega^{(a)} C_F C_A + \frac12 \cI_\Omega^{(c)} C_F T_R n_f \biggr) \cL_0 (\dNm) \biggr\} \Theta( \dNm < \zcut^2 / 4)  \,, \nn \\
\cK^{(g)}_{2-} (\dNm) &= \biggl\{ \biggl( -\frac{12}{5} C_A^2 \biggr) \cI_\Omega \cL_1 (\dNm) + \biggl( \frac{24}{5} \ln \zcut C_A^2 \biggr) \cI_\Omega \cL_0 (\dNm) \nn \\
& \qquad + \biggl( \frac12 \cI_\Omega^{(a)} C_A^2 + \frac12 \cI_\Omega^{(c)} C_A T_R n_f \biggr) \cL_0 (\dNm) \biggr\} \Theta( \dNm < \zcut^2 / 4) \,,
\end{align}
and
\begin{align}
\cK^{(q)}_{2+} (\dNp) &= \biggl( -\frac75 C_F^2 + \frac15 C_F C_A \biggr) \cI_\Omega \cL_1 (\dNp) \nn \\
& \qquad + \biggl[ \biggl(  - \frac{57}{20} + \frac75 \ln 2 - \ln \zcut \biggr) C_F^2 + \biggl( -\frac34 - \frac15 \ln 2 - \frac75 \ln \zcut \biggr) C_F C_A \biggr] \cI_\Omega \cL_0 (\dNp) \nn \\
& \qquad + \biggl( \frac12 \cI_\Omega^{(b)} C_F C_A + \frac12 \cI_\Omega^{(d)} C_F T_R n_f \biggr) \cL_0 (\dNp) \,, \nn \\
\cK^{(g)}_{2+} (\dNp) &= \biggl( -\frac{6}{5} C_A^2 \biggr) \cI_\Omega \cL_1 (\dNp) \nn \\
& \qquad + \biggl[ \frac65 \biggl( -\frac{11}{3} + \ln 2 - 2\ln \zcut \biggr) C_A^2 + \frac23 C_F T_R n_f + \frac23 \cdot \frac75 C_A T_R n_f \biggr] \cI_\Omega \cL_0 (\dNp) \nn \\
& \qquad + \biggl( \frac12 \cI_\Omega^{(b)} C_A^2 + \frac12 \cI_\Omega^{(d)} C_A T_R n_f \biggr) \cL_0 (\dNp) \,,
\end{align}
and
\begin{align}
\cK^{(q)}_{3-} (\dNtm) &= \biggl[ \biggl(-\frac34 - \ln \zcut \biggr) C_F^2 + \biggl( -\frac34 - \ln \zcut \biggr) \frac75 C_F C_A \biggr] \cI_\Omega \cL_0 (\dNtm) \,, \\
\cK^{(g)}_{3-} (\dN{3-}) &= \biggl[ \frac65 \biggl( -\frac{11}{6} - 2\ln \zcut \biggr) C_A^2 + \frac23 \cdot \frac75 C_F T_R n_f - \frac23 \cdot \frac15 C_A T_R n_f \biggr] \cI_\Omega \cL_0 (\dN{3-}) \,. \nn
\end{align}
The angular integral factor $\cI_\Omega$ is defined in \Eq{eq:IOmega} and $\cI_\Omega^{(a,b,c,d)}$ are given in \Eqs{eq:angfactorAB}{eq:angfactorCD}.

\bibliographystyle{JHEP}

\bibliography{NtildeCalc}

\end{document}